
\magnification=1200
\hsize=13cm
\headline={\hfil}
\footline={\ifnum\count0 = 1 \hfil \else\hss\tenrm\folio\hss \fi}
\topskip10pt plus30pt
\def\<{{<}} \def\>{{>}} \def\dsize{\displaystyle} \def\text{\hbox}
\def\newline{\hfil\break}
\def\tr{\mathop{\rm tr}}
\def\ent#1#2#3{{\rm ent}_{#1}(#2\,|\,#3)}
\def\app#1#2#3{{\rm app}_{#1}(#2\,|\,#3)}
\abovedisplayskip=3pt plus 1pt minus 1pt
\belowdisplayskip=3pt plus 1pt minus 1pt
\def\proclaim#1#2{\medskip\noindent{\bf #1}\quad \begingroup #2}
\def\endproclaim{\endgroup\medskip}
\def\A{{\cal A}} \def\B{{\cal B}} \def\H{{\cal H}} \def\sect{\S}
\def\F{{\cal F}} \def\SM{{\cal SM}} \def\V{{\cal V}}
\def\implies{\Rightarrow}
\font\Bbb =msbm10
\def\Minkowski{{\hbox{\Bbb M}}}
\def\blacksquare{\vrule height4pt width3pt depth2pt}
\def\hcrh{\hfill \cr \hfill} \def\crh{\cr \hfill} \def\hcr{\hfill \cr}
\font\smc =cmcsc10
\font\brm=cmbx12  \def\til{\lower 1.1 ex\hbox{\brm \char'176}}

\vbox to 2cm{}
\centerline{\bf Progress in a Many-Minds Interpretation of Quantum
Theory.}
\vbox to 1.5cm{}

\centerline{\bf Matthew J. Donald}
\vbox to 1.5cm{}

{\bf \hfill The Cavendish Laboratory,  Madingley Road,  Cambridge 
CB3 0HE, 

\hfill Great Britain.}

\vbox to 0.75cm{}

{\bf \hfill e-mail:\quad matthew.donald@phy.cam.ac.uk}

\bigskip

{\bf \hfill web site:\quad  http://www.poco.phy.cam.ac.uk/\til mjd1014}

\vbox to 2cm{}

\proclaim{abstract}{}  In a series of papers, a many-minds
interpretation of quantum theory has been developed.  The aim in
these papers is to present an explicit mathematical formalism which
constitutes a complete theory compatible with relativistic quantum
field theory.  In this paper, which could also serve as an
introduction to the earlier papers, three issues are discussed. 
First, a significant, but fairly straightforward, revision in some
of the technical details is proposed.  This is used as an
opportunity to introduce the formalism.  Then the probabilistic
structure of the theory is revised, and it is proposed that the
experience of an individual observer can be modelled as the
experience of observing a particular, identified, discrete
stochastic process.  Finally, it is argued that the formalism can be
modified to give a physics in which no constants are required. 
Instead, ``constants'' have to be determined by observation, and are
fixed only to the extent to which they have been observed.
\endproclaim
\vbox to 1cm{}

{\bf \hfill	                                               April 1999
	
\hfill	                                 
   
\hfill             quant-ph/9904001}
\vfill \eject

\centerline{\smc CONTENTS}
\smallskip
  
{\bf \parindent=54pt
\item{1.\quad} 	Introduction.

\item{2.\quad}  An abstract model of a finite information-processing
structure. 

\item{3.\quad}	Constraints and temporal development for switching 
structures.

\item{4.\quad}	Manifestations of switching structures in spacetime.

\item{5.\quad}	Life is like a game of chance.

\item{6.\quad}	Observing quantum probabilities.

\item{7.\quad}	Elementary models.

\item{8.\quad}	A model for private probabilities.

\item{9.\quad}	Physics without any physical constants.

\item{10.\quad}	Conclusion.

\item{Appendix} A Hypothesis.

\item{}	References.}
\bigskip

\proclaim{1. Introduction.}
\endproclaim

Following von Neumann, the conventional formalism for quantum
mechanics postulates that two quite distinct processes are involved in the
time evolution of the wave-function of a quantum system.  Process I --
``collapse of the wave-packet'' -- is an abrupt change which supposedly
occurs as a result of measurements.  With appropriate probabilities, it
replaces the wave-function by one of the eigenfunctions of some operator
which is being measured.  Process II, on the other hand, which is supposed
to govern the system at all other times, is deterministic evolution under
some version of the Schr\"odinger equation.  However, no satisfactory
account of process I has ever been given.  In particular, it has never made
been clear how a ``measurement'' should be distinguished from any other
physical process, nor, given a measurement, how the ``operator being
measured'' is to be defined, and nor how the ``abrupt'' change is to be made
to conform with special relativity.  Everett, therefore, made the brilliant
suggestion that process I simply did not occur.  Instead, he argued that
process II was sufficient, and that the apparent occurrence of process I was
entirely due to the way in which process II inevitably affected the physical
structure of observers.  Once again however, a detailed account was
required.  In particular,  what is the physical structure of an observer?

	For many years now, I have been attempting to answer this question with
the aim of providing a firm technical foundation for Everett's ideas.  There
should be no doubt that such a foundation is required, even in
the context of decoherence theory -- the latest and most sophisticated
version of the lesson that, as far as correlations to quasi-classical states of a
given macroscopic object are concerned, an entirely quantum universe will,
under normal circumstances, appear, for all practical purposes, to behave
classically at the macroscopic level (Giulini et al.~(1996)). When
that lesson is absorbed, the problems remain: firstly to
characterize unambiguously structures within such a universe which
could provide a foundation for the correlations which we
individually actually experience; and secondly to describe how those
structures themselves can develop with time.

The details of my work have so far appeared in three papers (Donald 
(1990, 1992, 1995)).   In Donald (1995) in particular, I proposed a
``mathematical characterization of the physical structure of an observer''. 
This characterization forms part of a hypothesis, the current version of
which is presented in the appendix to this paper. The hypothesis consists of
a completely explicit set of definitions for the structure of an observer and
for the set of possible future extensions of such a structure, together with
definitions for the a priori probability of existence of an observer and for
the probability of moving from a given structure to a given extension.  In
Donald (1990), with a detailed neurophysiological analysis, I began the
work of arguing that humans do have the proposed type of structure and
that the only entities which can with significant a priori probability possess
such structure at the human level of complexity are those which we would
be prepared to believe might be physical manifestations of consciousness. 
In Donald (1992), I gave an axiomatic definition of a priori probabilities for
localized observers in quantum theory, in such a way as to be compatible
with special relativity and quantum field theory.

As a whole, the hypothesis constitutes a complete interpretation of quantum
theory.  According to this interpretation, an observer has a particular type
of physical structure and the probabilities of changes in that structure
(corresponding to new experiences) are as defined.  Our observations of
the physical world are explained if we have the type of physical structure
suggested and if the suggested probabilities are empirically justified.  If
arguments show that these goals are not achieved, then it may be possible
to save the interpretation by improvements in the hypothesis.  In this
paper, I shall present several such improvements.  Starting in section 2, I
shall review the hypothesis, step by step, and in sections 3 and 4, I shall
explain the changes which I feel need to be made in the non-probabilistic
parts of the hypothesis.  Some elementary models are presented in section
7 in order to elucidate certain aspects of the interpretation.   Although this
paper could be read independently, much of the hypothesis is unchanged
from Donald (1995), and explanation of some of the details is to be found
there.  Wherever possible, the parts of the hypothesis in this paper carry
the same labels as the corresponding parts of the hypothesis in Donald
(1995).

The second purpose of this paper is to revise the probabilistic structure of
the theory.  This is a more interesting and fundamental revision as the
nature of probability in many-minds interpretations has been a matter of
some controversy (Lockwood et al.~(1996), Donald (1997)).  Given the
framework already constructed, the solution proposed here is conceptually
quite simple.  In Donald (1992, 1995), a number was associated with each
observer.  Those numbers were referred to as ``a priori probabilities'', but I
do not now believe that, by themselves, those numbers can form an
adequate basis for a theory of observation.  In section 5 of this
paper therefore, I shall modify the theory in such a way that it becomes
possible to model the experience of an individual observer as being
equivalent to the experience of observing a particular, identified, stochastic
process.  In other words, life is like a game of chance.

Once the stochastic process has been defined, it is necessary to show
that its probabilities can be expected to be compatible with observed
statistics.  In section 6, the nature of probability in a many-minds
interpretation is discussed.  It is then argued that, with probabilities
defined by the stochastic process, a typical modern human observer will be
aware of a world in which quantum theory is accepted and in which its
detailed theoretical predictions are confirmed.  In sections 7 and 8,
attention is turned to the observation of individual events with uncertain
outcomes.  After a preliminary discussion in section 7, the full complexity
of human neural processing is addressed in section 8.

	It will be clear from a glance at the appendix that the proposed
hypothesis is technically sophisticated.  It is also speculative.  Some 
elements are more speculative than others.  In my opinion, the details are
important, not merely because of the possibility that they might be correct,
but because it is only by accepting the discipline of trying to construct a
complete mathematical framework that one can come to understand the
conceptual ramifications of the many-minds idea and test whether that
idea is coherent.  The overall aim is just to provide one theory which
constitutes a complete mathematical framework for a many-minds
interpretation and which is compatible with modern theoretical physics
and, of course, with observation.  The complexity of the hypothesis and its
speculative nature may in themselves cast light on the nature of
many-minds interpretations.

Even the most speculative of the elements of the hypothesis is fiercely
constrained by the other elements; by the goal of compatibility with
quantum field theory, with special relativity, and with observation; and by
the goal of being able to describe the existence of human observers in the
framework of universal quantum theory.  In my view, the difficulty of
working out the details has been widely underestimated in the stream of
preliminary sketches for interpretations which have gained popularity
over the years (cf.~Bacciagaluppi, Donald, and Vermaas (1995),
Dowker and Kent (1996)).  It has been my experience in the
development of the present theory that apparently minor problems
when pursued have usually turned into major problems.  However, such
problems become opportunities, when they force the refinement of the
interpretation.  Although it seems unlikely this process will be
sufficient to lead to a theory which is unique in all its details;
it can perhaps lead us further than might be expected.  Except where
grounds are given for modifying the details of Donald (1995), the
many versions of the hypothesis discarded during its development to
date are absent from this work.  These failures and the difficulty
in moving beyond them give me no doubt, not only that the present
hypothesis could also be shown to be unsatisfactory, but even that
unless there is some truth in the many-minds idea, then any version
of it will ultimately be shown to be flawed in its detail.

If the version of the many-minds idea presented here is true, then there
is a law of nature which states that to any conscious entity there
corresponds a physical structure characterized by the hypothesis.  In other
words, the hypothesis is proposed as a physical law in the same way that
the standard model in particle physics might be so proposed.  The only way
to find a physical law is to try to describe the physical world in some
particularly elegant fashion.  In the case of the hypothesis, the idea is to
provide a high-level abstract description of the functioning of a human
brain; a description that is sufficiently abstract to be definable entirely
within the mathematics of axiomatic quantum field theory.  The elegance
comes in trying to find the most simple such description sufficiently
complete to express everything about the physical structure of the
observer which is relevant to his or her mental life.  Nevertheless, it should
be realized that even if a physical law holds, there is no guarantee that it is
sufficiently elegant to be distinguished from all alternative possibilities --
either minor variations or radical replacements. 

The search for an elegant description, refined by specific problems, is
demonstrated in sections 3 and 4.  Here, for example, we meet a problem
in individuating observers.  This is solved by a mathematical translation of
the fact that individual human brains are (usually) spatially separated
from each other.  The solution is elegant in that it fits naturally into the
overall mathematical structure, but variations in the details of the solution
are certainly possible.

At present, the formalism is compatible with a fixed Lorentz invariant
quantum field theory on Minkowski space which satisfies the
Haag-Kastler-Schroer axioms.   In section 9, I shall speculate about further
changes to the formalism and argue that, while remaining consistent with
observation, one could allow for observation-dependent variations in
the underlying quantum field theory.  This suggests the possibility
of a physics without physical constants and may be a preliminary step
towards compatibility with more sophisticated physical theories. 

Eventually, I hope that some version of the present formalism might be
useful in the development of axiomatic structures for theories like
noncommutative geometry or quantum gravity or string theory.  There are
two general principles at the heart of my work which may make it useful in
this context.  These principles are that the formalism should define a {\sl
minimal }structure for an observer and that an observer should be
specified in entirely {\sl abstract }terms as a developing pattern of
information.  The result is that the formalism places minimal constraints
upon the governing physical theory and is potentially adaptable to be
compatible with the most exotic mathematical framework; requiring
ultimately only the possibility of defining the structures used in minimal
descriptions of individual observers.  

As our physical theories have become more sophisticated, it has constantly
been found that in order to accomodate apparently esoteric empirical
results, or even just theoretical advances, technical concepts which had
once been considered fundamental have come to be seen as derived.  If our
idea of the world was based on the idea of a collection of particles with
definite properties, then our entire philosophy may have been put at risk
by the realization that particles do not have definite properties.  In
mathematical quantum field theory, not only the definite properties, but
even the particles themselves have become derived concepts useful only in
certain observational situations; including of course, scattering
experiments.   Wave-functions and eigenfunctions also cannot be
considered as fundamental, unless at the level of the entire universe.  It is
thus to the advantage of the present formalism that it depends on
generalizable concepts like patterns of spacetime relations, continuous
paths, local geometry, and quantum states in the mathematical sense
(rather than wave-functions).  In terms of such high-level concepts it may
be possible, (and will perhaps be required by quantum gravity), to allow
even for time itself to appear only as a derived concept at the level of the
observer.  Allowing observation-dependent quantum field theory, as is
proposed in section 9, may be a hint of the kind of formal flexibility which
may be needed for the implementation of this idea.

In Donald (1997), the goals of this paper were announced and an
introduction to ``many-minds'' interpretations was provided, together with
discussions of and references to some alternative formulations.

\proclaim{2. An abstract model of a finite information-processing
structure.}
\endproclaim

	The first step in the interpretation is the modelling of an observer
as a finite information-processing structure.  This step is based on the
preparatory intuition that human cognition works through patterns of
neural firing developing over time.  As a first approximation, a neuron can
be thought of as a two-status entity: either a signal is passing along the
neuron and it is ``firing'', or no signal is passing and the neuron is
``quiescent''.  A human brain contains a finite number of neurons (around
$10^{11}$) and the preliminary idea would be that a {\sl complete}
description of everything relevant to the workings of the brain as the
embodiment of a mind at a given moment, could be given by describing the
pattern of neural firings in that brain over the lifetime of the person
involved, up until the moment in question.  

	There are many problems with this idea.  From a physical point of view,
neurons are macroscopic objects.  Whether or not they only have two
statuses as far as information processing is concerned, the family of
quantum mechanical states which they visit over time is much too complex
to express a simple binary opposition.  To deal with this problem, it is
necessary to focus attention not on neurons, but on substructures within
neurons which do have a pair of quantum-mechanically simple statuses,
directly tied to the firing or quiescent status of the neuron in they are
contained.  I call such substructures ``switches''.  Much attention is devoted
in Donald (1990, 1995) to identifying and characterizing possible switches. 
An example might be a small piece of neural wall, whose electrical
polarization will express the status of the encompassing neuron.

It is the switch statuses which are the ``classical facts'' or the ``measured
observables'' of the present interpretation.  In the hypotheses of the
previous papers of this series, the status of each switch was required to be
``observed'', or specified, or determined, exactly once per change of status
and so, with a rather sloppy use of language, it was possible to blur the
distinction between determinations of switch status and changes of switch
status.  The word ``switching'' was used to refer to both.  However, as will
be explained in section 6 of the present paper, it now appears necessary, in
order to achieve a satisfactory probabilistic structure, to allow the
possibility of switch status being determined more than once per change. 
For this reason, the distinction can no longer be blurred, and the
fundamental events will now be referred to as ``determinations of switch
status'' or ``determinations'' rather than as ``switchings''.  It should be
stressed that this is merely a linguistic amendment; no change is actually
being proposed in the nature of the fundamental events.  Patterns of
switch status determinations will continue to be referred to as ``switching
patterns'' or ``switching structures''. 

Many-minds interpretations are conceptually radical, so it should not be
surprising that words must be used with some care.  For example, a
determination that a switch has status $C$ in some region is not something
chosen by the observer.  ``A determination'' is being used in the sense of ``a
finding'' rather than ``a decision''.  In fact it is very difficult entirely to
avoid using language which gives an impression of an observer's future
possibilities as being either, on the one hand, chosen or shaped
by the observer; or, on the other hand, merely an expression of the
observer's ignorance about a pre-determined unique future.  Many-worlds
theory rejects the latter idea because of the difficulty of constructing a
theory of physical ``wave-packet collapse'' compatible, for example, with
the experimental evidence for special relativity and for the violation of
Bell's inequalities.  The former idea should also be rejected if possible,
because it makes the causal effects of an observer unphysical.  The present
theory is an attempt to show that both of these ideas can consistently be
rejected; leaving us instead with the idea of the possible immediate futures
of an observer at any moment, as being given by a well-defined set of
possible extensions of his current history, each of which will occur with its
own well-defined probability. 

Determinations are defined precisely by the hypothesis, but in essence they
are the elementary facts from which the observer's world (or ``history'') is
constructed.  The facts are of the form that it appears to the observer as if
a certain switch had a definite given status in a region of spacetime which
is related in a given way to the other regions in which switch status is
determined.  At the neural level, a sequence of determinations of the
polarization of a patch of neural membrane might be described in the form
$LLHLLL$, where $L$ indicates that the electric potential inside the
membrane is lower than that outside, and $H$ the opposite.  This
sequence would be indicative of one period of neural firing (the
$H$), preceded and followed by periods of quiescence.  For an
abstract switch, a corresponding set of six switch determinations
might be described by $CCOCCC$ with $C$ for ``closed'' and $O$ for
``open''.

For the complete description of the mental processes of an
observer, we shall require more geometrical information about a set of
switch determinations than mere time-ordering.  This brings us back to the
idea of ``a pattern'' of neural firing (or more precisely ``a pattern'' of 
switch determinations).  As will be explained below, it is essential for the
understanding of probability that only a finite amount of information be
required to describe an observer.  The definition of a pattern which I
introduce and discuss in Donald (1995) amounts to a listing, called a
``docket'', of the spacetime relations between the spacetime sets in which
switch statuses are determined.  Thus a docket is a geometrical
structure in spacetime defined as an equivalence class of ordered sequences
$(A_i)_{i=1}^M$ of suitable spacetime sets.  Two such sequences
$(A_i)_{i=1}^M$ and $(B_i)_{i=1}^M$ will have the same docket if they have
the same spacetime, or causal, arrangement -- in other words, if, for every
pair $i$, $j$,  $B_i$ is in the past of/spacelike to/in the future of
$B_j$ exactly when $A_i$ is in the past of/spacelike to/in the future of
$A_j$ -- and if one sequence can be continuously deformed into the other
while the arrangement is essentially unaltered.  With $M$ finite, only
finitely many dockets are possible for $M$ sets.

These ideas are sufficient to define an abstract ``pattern of switch
determinations''.  This provides us with part A of the
hypothesis in the appendix, according to which the structure of a
mind is defined by an ordered sequence of $N$ switches and
$M$ determinations of status, by a docket $d$ which defines the spacetime
relations between the determinations, and by a function $\varphi$ which
defines the switch $\varphi(m)$ which has its status determined in the
$m^{th}$ determination.   The status of switch $n$ is determined $K_n$
times and its $k^{th}$ determination is determination number $j_n(k)$. 
Taken together, $M$, $N$, $d$, and $\varphi$ define a ``minimal ordered
switching structure $SO(M, N, d, \varphi)$''.  In part B3 of the hypothesis,
irrelevant labels are removed from $SO(M, N, d, \varphi)$ to define a
``minimal switching structure $S(M, N, [d, \varphi])$''.  All sorts of
philosophical questions arise with the claim that such a definition is
adequate as a description of everything which is relevant about the
behaviour of a brain in its functioning as the embodiment of a mind.  These
questions are discussed at length in Donald (1995, 1997).  The only
shortcut I can offer to that philosophical discussion is to emphasize the
following points:
\smallskip

\item{a.}  Such an abstract pattern is an abstract analog of a
pattern of neural firing and human cognition does seem to function through
such a pattern.
\smallskip

\item{b.}  A satisfactory theory of probabilities depends on only a {\sl
finite} number of distinct future possibilities being available to any
observer within a finite period. 
\smallskip

\item{c.} In the context of an interpretation of quantum mechanics without
a physical process of wave-packet collapse, there is no direct way
to identify the sequence of quantum states occupied by a warm wet human
brain.  It is, for example, extremely difficult to see how a natural ``preferred
basis'' of brain wave-functions could be identified.  Because of this, we
cannot make the assumption,  inevitable in classical physics, that a brain
is simply something given; ``out there''; existing in itself; a ready-made
vessel for the mind.  Thus quantum mechanics requires us to characterize
the essential structure of a brain before we can identify the physical
manifestations of such structures.

\proclaim{3. Constraints and temporal development for switching 
str\-uctures.}
\endproclaim

Even if it is accepted that for suitable $M$, $N$, $d$, and $\varphi$, the
definition of \newline $S(M, N, [d, \varphi])$ gives an adequate ``description
of everything which is relevant about the behaviour of a brain in its
functioning as the embodiment of a mind'', it is not the case that every
possible switching structure $S(M, N, [d, \varphi])$ is a description of
something which we could interpret as such an embodiment.  Indeed most
switching structures are random and presumably meaningless.  A
satisfactory many-minds theory should define a priori probabilities so that
the futures of human minds are not dominated by randomness.  In order to
do this, various constraints are imposed in the hypothesis on the physical
manifestations of switching patterns.  This disallows some manifestations
which are not like individual functioning human brains; for example,
because they are spatially disconnected.  This section introduces some of
these constraints.  The temporal development of structures is also
discussed.

Pruning the apparently vast set of ``observer-like'' structures may be the
most difficult part of developing a many-minds theory.  There is always a
temptation to abandon this task; to suggest that quantum theory should be
viewed simply a theory of correlations (Wheeler (1957), Saunders (1996),
Mermin (1998)), and that every interaction between different physical
systems is a species of observation.  One drawback of this suggestion is that
it provides no hope of understanding our own personal physical structures
as observers.  After all, the information-processing structure of the human
brain appears not to depend on anything like the full complexity of the
physical interactions of the atoms of which the brain is formed; we usually
think of most of that complexity as mere thermal noise.  Moreover, at least
in relativistic quantum field theory, it is not all clear how ``physical
systems'' are to be identified; even the idea of ``an electron'' is not without
ambiguity.  The most serious problem, however, in a proposal which at
each moment, provides an infinite variety of ``observer-like'' entities, is
that it is not clear that anything remains of the concept of temporal
succession except for a vague idea about similarity between past and
future (cf.~Bell (1976, 1981), Butterfield (1996)).

Given the problems with the idea that there are no rules constraining the
definition of the physical structure of observers, we must either abandon
the many-minds interpretation or propose some rules.  Of course, such a
proposal will only be correct, if the rules are facts about reality or physical
laws, rather than mere invention.  The main test of whether or not they
could be factual will consist of asking whether or not they are consistent
and allow us to explain belief in quantum theory.  If they pass this test,
then they can annex all the empirical evidence usually taken to support the
(hopelessly inconsistent) conventional theory.  However, as mentioned in
the introduction, physical laws are also frequently required to be, in some
sense, ``elegant'', or ``simple'', or even ``beautiful''.  In the present context,
this might mean something like ``unfussy'' or ``not subject to special
conditions'', and, in these terms, it could be suggested that the constraints
introduced in this section are ugly and that the hypothesis might be, at
best, accurate phenomenological description rather than fundamental law.  
That is for the reader to decide, and to improve on if s/he can.  It should be
noted, however, that in compensation for these constraints, not only does
one gain a possibly complete and consistent interpretation of quantum
theory, but also, as we shall see in section 9, one may be able to reduce,
perhaps to zero, the vast amount of information which would be
conventionally required to specify exact initial conditions.  Although initial
conditions and physical laws are neatly separated in conventional physics,
this does not mean that they should be ignored in the total count of the
complexity or fussiness of a theory.

In Donald (1995), I mentioned the possibility of using the details of the
theory with its allowance for fluctuations in switch states to construct
arbitrary switching structures with a priori probability of around
$(1.14)^{-M}$ and I suggested that these structures were unimportant
because their a priori probability was so low.  These structures were
referred to as ``artificial'' because they are not reflections of patterns of
causal correlations implicit in the universal quantum state.   
Unfortunately, I now think, that with the version of the hypothesis
presented in that paper, the future of a structure
$S(M, N, [d, \varphi])$ which did describe a human brain would be
dominated by the large number of possible ways in which small numbers
of new short-term artificial switches could arise.  This brings us to the first
of the new constraints to be imposed in this paper: a requirement of
homogeneity between switches, which will make negligible the total
probability of such occurrences.  The details of this requirement will be
discussed in the next section.

Another type of possible switching structure which would satisfy the
definition of $S(M, N, [d, \varphi])$ could be formed by the amalgamation of
the switching structures describing several individuals and would describe
the brains of all those individuals simultaneously.  This would mean, for
example, Alice and Bob existing not only as separate individuals: as ``Alice''
and as ``Bob'', but also as the joint individual ``Alice-n'-Bob''.  Now the
answer to the question ``What would it be like to be ``Alice-n'-Bob''?'' is
simply that it would be like being Alice and Bob separately but in the same
``world''.  At first sight, one might then think that ``Alice-n'-Bob'' would be
a better description of reality than the individual-minds theory in which
Alice has her world and Bob separately has his.  However, inevitably one is
led by this idea to try describe the entire human race by one vast joint
structure and I do not believe that this can be done satisfactorily.  There
would be correspondingly vast ambiguities in such a structure;  there
would be an uncertainty of at least a twenty-fifth of a second (the light
travel time to the antipodes) in the specification of an instant in time for
such a structure;  and the problem of whether humans also exist as
individuals would arise.  In order to rule out such joint structures (at least
with significant probability),  I therefore propose in this paper a
modification of the hypothesis which will force a switching structure to be
manifested by a connected set of switches.  This also will be discussed in
detail in the next section.

I shall refer to the problem of the extension of a switching structure from
one individual to more than one as the ``breeding problem''.  Another
problem relating to the existence of superfluous structures was introduced
in Donald (1995).  This was the ``trimming problem'' which arises from the
possibility of taking a switching structure describing a single individual
and removing a few of the switches to leave a structure which may have
higher a priori probability but very little less information.  The problem is
to decide which of these structures should be associated with a given
person.  The trimming problem will be resolved in this paper by modifying
the order in which structures are allowed to develop and the definition of a
priori probability so that trim structures will tend to become fuller.

The development of structures is governed by part B of the hypothesis,
which defines what it means for one structure to be an immediate successor
to another.  Two different ways in which a structure can be extended are
allowed by part B.  Either there is a single new determination of the status
of an existing switch (B1), or a new switch is introduced (B2).  However, it
is required by A5 that, for a switch to be part of the structure, it needs to
have opened and closed at least twice.  Thus any new switch must have at
least four determinations of status.  If new switches could be added with
fewer determinations, then all the constraints imposed by parts C to F of
the hypothesis would not be expressed and an unrestricted increase in
switch numbers would be possible.  Because four determinations are
needed in B2, the introduction of a new switch cannot be instantaneous. 
This means that, although in general part of the meaning of a human
switching structure will include a  ``psychological present moment'',
successors of such a structure may involve new determinations of status
localized before that moment.  Otherwise, the entire structure would have
to be reduced for a certain interval to a single switch.  In the case of B1,
such extensions back in time are also permitted.  The single new
determination can be made at any time during the history of the switch;
subject to the later constraint (C13) which prevents a pile-up of
re-determinations of status.

 In general, the possibility of extension back in time resolves the trimming
problem by allowing growth to occur wherever it is possible.  This means
that status determinations of high a priori probability can become part of
the switching structure before earlier determinations of low a priori
probability, but that those earlier determinations can nevertheless still
eventually be included.  However, a fundamental arrow of time remains
built into the theory both by the forward direction of switch paths (part C)
and by the inductive definition of a priori probability (part G). 
In section 8, the existence of the ``psychological present'' will be
explained as an empirical consequence of the theory.

\proclaim{4. Manifestations of switching structures in spacetime.}
\endproclaim

  Once switching structures and their successors have been defined, the
remaining parts of the hypothesis are concerned with the definition of 
probabilities.  This requires the introduction of the set of ``physical
manifestations'' of a structure.  In parts C to F, individual manifestations
are considered.  In part G, the probability of existence of an abstract
structure is defined by maximizing the probabilities of its individual
manifestations.  

An individual manifestation, defined in G1 as a pair 
$((\sigma_m)^M_{m=1}, W)$, can be thought of as a quantum world in
which an appropriate pattern of switch behaviour exists.  It is a world
defined by a sequence of quantum states ($(\sigma_m)^M_{m=1}$ in part
F) and by a sequence of geometrical information ($W$ in part C).  Taken
together, these describe the switches moving through spacetime and
changing in status in such a way as to model the abstract structure defined
in parts A and B.	In developing a many-minds interpretation of quantum
theory, it is natural to construct such possible models of an observer's
world.  They correspond closely to conventional ideas about reality in
quantum physics and use both of von Neumann's processes --
deterministic evolution in the definition of the Heisenberg picture states,
and ``collapse'' in the sequential change from one state to another. 
Nevertheless, ultimately, these conventional ideas break down.  Each
switching structure has continuously many possible individual
manifestations, each of which describes how the world might be for the
observer whose experiences are constituted by the given structure.  But
there is no way to choose between these possibilities.  Only the structure is
experienced.  All of the possible worlds enter into the universal quantum
state in similar ways.  They are all indistinguishable as far as the observer
is concerned.  All will be used here in the calculation of a priori
probability.  They are not ``real'' worlds, neither are they ``worlds'' in the
Everett or DeWitt sense (DeWitt and Graham (1973)); they are just
mathematical constructs which mirror the unique experience of a single
observer characterized by one completely-defined abstract switching
structure.  It is, of course, no more than a hypothesis that the total set of all
such possibilities  -- the equivalence classes of individual manifestations
-- should be relevant to a complete interpretation of quantum theory. 
Whether that hypothesis is successful is for the reader to judge.

Part C of the hypothesis is concerned with the idea of a switch as a
two-status object moving through spacetime and having determined status
in specified regions.  Thus, in part C, for each individual manifestation, a
specific sequence of sets $(A_i)_{i=1}^M$ is identified which has the docket
$d$ defined in part A.  For $n = 1, \dots, N$, a spacetime path
$x^n(t)$ is defined on a parameter interval $[0, T_n]$.  This is the path
along which switch $n$ moves, although it is only ``in operation'' in a
subinterval $[S_n, T_n]$.  The switch status is determined at parameter
times $t_{nk}$ for $k = 1, \dots, K_n$.  At parameter time
$t$, switch $n$ occupies the spacetime region $\Lambda_n(t)$. 
$\Lambda_n(t)$ is a Poincar\'e transform, defined by $x^n(t)$ and by a
path $L^n(t)$ in the restricted Lorentz group,  of a set $\Lambda$. 
$\Lambda$ is the set, common to all the switches, at which comparisons are
made.  C6 demands that the path
$x^n(t)$ followed by switch $n$ and the boosts imposed on that switch by
$L^n(t)$ be consistent. 

The requirement of connectedness which is the fundamental step in making
joint switching structures like ``Alice-n'-Bob'' of negligible probability is
imposed by part C10 of the hypothesis.  This requires that between any
two spacelike separated points within regions where switch status is
determined, there is a spacelike path lying entirely within the spacetime set
traced out by switch paths.  By itself, however, C10 will not solve the
breeding problem.  No constraint is imposed on the shape of the set
$\Lambda$.  By giving $\Lambda$ very long fine hairs, it would be possible
to connect arbitrarily distant switches without any significant loss of a
priori probability.  This possibility is ruled out by C11 which requires that,
at any moment, only a maximum number ($C$) of other switches can
contact a given individual switch.  

$C$ -- the ``contact number'' -- may be the most arbitrary part of the
entire hypothesis.   Some value has to be given to $C$, not only for the
breeding problem, but also to prevent switches piling up on top of each
other.  In Donald (1995), only the latter problem was considered, and $C$
was defined to be zero.  Now, however, I propose that $C$ should be
defined to be thirteen.  Thirteen is one more than the maximum number of
identical spheres which can make simultaneous contact with another sphere
of the same size in three dimensions.  Thus thirteen allows for a switch to
have a maximal number of nearest neighbours, and a replacement. 

The fact that observers are localized is fundamental in the proposed
interpretation.  An observer only interacts with and has information about
limited aspects of the universe.  At the level of an individual geometrical
manifestation $W$, these aspects are defined by a limited set $\B(W)$ of
observables (bounded, but not necessarily self-adjoint operators) on which
the quantum states of the switches are defined.  The sets $\B(W)$ are
defined by parts D1 and D2 of the hypothesis, while in part D3, quantum
states on general sets of bounded operators are defined.  As required for a
theory of localized entities, the states in D3 are not necessarily pure.

The quantum state of a switch when it is in the spacetime region
$\Lambda$ will be on the local algebra $\A(\Lambda)$ -- the algebra of all
observables measurable within $\Lambda$ (Haag (1992)).  The Poincar\'e
transformation $(x^n(t), L^n(t))$ is used to transform the state of a switch
in $\Lambda_n(t)$ back to $\Lambda$.   The two possible statuses (``open''
and ``closed'') of switch $n$ are defined by two orthogonal projections $P_n$
and $Q_n$ in
$\A(\Lambda)$.   $\B(W)$ is generated from transforms of the $P_n$ and
$Q_n$ and from the algebras $\A(\Lambda_n(t))$ for $t \in [S_n, T_n]$.

In parts E and F the quantum states of the switches are characterized.  The
first four items in part F are a formal expression of the idea, introduced in
Donald (1990), that ``A switch is something spatially localized, the quantum
state of which moves between a set of open states and a set of closed
states, such that every open state differs from every closed state by more
than the maximum difference within any pair of open states or any pair of
closed states.''  These open and closed states are the states in which the
switch status is determined.  They will be referred to as ``determination
states''. 

The final item of part F requires that the determination states cannot
change arbitrarily between different switches.  This is the homogeneity
requirement referred to in the previous section.  New switches are only
allowed if they have determination states which are close to the
corresponding states of some other contemporary switch in the structure. 
The requirement is sufficiently loose to permit some variation in
determination states -- as is neccessitated, for example, by changes in
neural temperature -- but it is also tight enough to disallow artificial
switches.  The loss of a priori probability caused by a single arbitrary
collapse sufficient to mimic switching can be as low as $(1.14)^{-1}$, but
only if the switch mimicked is of a very simple type.  However, the most
likely ways in which structures of human complexity can arise is through
the causal correlations within a functioning brain in which the switches are
something like patches of neural membrane.  With the homogeneity
requirement, the determination states of the switches are no longer
independent.  Artifical switches, imposed in a functioning brain, must also
involve collapses to patches of neural membrane.  These will be of
extremely low a priori probability unless the membrane required is
already in existence and has, in conventional terms, some possibility of
behaving in the required fashion.

  In part E, the identity of the switches over time is considered.  As has
already been mentioned, identity over time has frequently been seen as a
problem in many-worlds and many-minds interpretations of quantum
theory (Bell (1976, 1981), Butterfield (1996)).  Part E defines the path
followed by a switch through spacetime as being that path along which the
local quantum state changes least.

	In parts E and F, finite sequences of quantum states are considered.  This
is a reflection of the fundamental problem of the interpretation of quantum
mechanics that although quantum states can be found at any instant which
provide apparently accurate descriptions of any physical system,
compatible with all that is observed about the system at that instant, that
compatibility property is not preserved over time.  This is the
Schr\"odinger cat problem -- a quantum state can be found to describe
the cat as it is observed at the beginning of the experiment, but, by the end
of the experiment, the time propagation of that state is sure to be some
sort of superposition or mixture, which is neither compatible with a live
cat, nor with a dead cat.  

	Similarly, in the human brain, a switch can have a well-defined quantum
state at one instant, but thermal and other fluctuations will soon result in
that quantum state describing a mixture of switch statuses.  The quantum
state must be then replaced (or ``collapsed'') in order (loosely speaking) to
purify the mixture.  In order to avoid an analogue of the quantum Zeno
paradox (Donald (1992)), the number of such replacements must be limited,
and it is natural to allow one replacement per determination of switch
status -- ``one collapse per measurement''.  Each such change of state
should allow the path $x^n(t)$ of switch $n$ to change direction.  However,
in Donald (1995) the path of switch $n$ was only allowed to alter in
response to a determination of the status of switch $n$ itself.  This was a
mistake and resulted in an inadequate hypothesis, because it is possible for
the observer to gain information from only a part of his brain which would
affect how he sees his entire brain moving. 

The mistake is corrected in the hypothesis given in the appendix to this
paper, by introducing in part C a second sequence of parameter times on
path $x^n(t)$: $(t'_{nm})_{m=m_n^i}^{m_n^f}$ such that for  $m = m_n^i,
\dots,m_n^f$, $t'_{nm}$ is the time at which the $m^{th}$ change of state
affects switch $n$ and the local state changes from $\sigma_{m-1}$ to
$\sigma_{m}$. By C8, the spacetime point $x^n(t'_{n m})$ is required to be
in the closure of the causal future of at least $m$ of the points
$x^{n'}(t_{n'k'})$ so that the
$m^{th}$ change of state cannot occur except as an effect of at least $m$
prior status determinations.

\proclaim{5.  Life is like a game of chance.}
\endproclaim

The indeterministic nature of quantum theory indicates that we are
sometimes confronted with a world in which any one of several possible
different futures may occur.  According to the present theory, this happens
practically all the time, because of the stochastic nature of neural
processing and the level of detail in which it is analysed.  The different
futures referred to here involve different switching structures and
therefore different experiences, rather than the many alternative
manifestations of a single switching structure discussed in the previous
section.  In a many-minds interpretation, all such different futures have
similar claim to reality.  Some possibilities may be more likely than others,
but we know from experience that the most likely does not always
happen.  We do, however, seem to see a fairly typical world.  Indeed, we
can only believe that we have evidence about probability if we believe the
assumption that we do see such a world.  Such beliefs do appear to be
consistent.  Why they should be true is the mystery at the heart of
probability, and the more one thinks about it, the more mysterious it
seems to become (Papineau (1995)).  Nevertheless, this mystery can be left
unexplained.  The goal for physics is to tell us what the world is like.  In
this paper, I propose that, for us, the world like a discrete stochastic
process.  However mysterious probability may ultimately be, we have no
problem in understanding what it is like to observe such a process.  Games
of chance have surrounded us from infancy.  Board game players, for
example, know what it is like to wait until two dice show doubles in order
to be released from some trap.  Poker players know the rarity of a really
good hand.  We all know what it means to ``know the odds''.

At the core of this proposal are three essential elements, which indeed are
central to the entire hypothesis.  The first is the idea, discussed further in
Donald (1997), that the stochastic process should be discrete, which
requires that at any moment the future possibilities can be enumerated.  It
is only because of this discreteness that the process could be simulated;
that it is like a game of chance; that it is the sort of process with which we
are familiar.  The second essential element is that the stochastic process
should be well-defined.  The probabilities, in other words, are objective. 
They are facts which have exact numerical values, and it is the purpose of
physical theory to find the physical laws by which they are defined.  In
this section, a possible definition will be put forward. The third essential
element is that the stochastic process should be able to explain our
observations.  Study of this element will occupy sections 6, 7, and 8. 

According to the present theory, everything which is relevant about the
functioning of a brain in its embodiment of a mind can be encoded in a
finite switching structure.  In part B of the hypothesis, the immediate
successors of a given structure are defined.  In order to define a discrete
Markov process,  it is only necessary to define the probabilities for moving
from a given structure to its immediate successors.  This will be done in
three steps, corresponding to G2, G3 -- G7, and G8 of the hypothesis.  In G3
-- G7, a definition is given for the a priori probability of a minimal
switching structure $S(M, N, [d, \varphi])$, and in G8 the jumping
probability for moving from a given structure to a given successor is
defined to be proportional to the successor's a priori probability.  However,
before considering these definitions, it is necessary to recall the
mathematical function defined in G2, which lies at the heart of the
definition of a priori probability.  Here the relation between that function
and the idea of decoherence will be emphasized as this will be fundamental
to the analysis of observations in subsequent sections.   

The definition of a priori probability is developed in Donald (1986, 1992,
1995), where it is based on the introduction of a function
$\app{\B}{\sigma}{\rho}$ of a set of operators $\B$ and two quantum
states (in the sense of definition D3) $\sigma$ and $\rho$ on $\B$.  This
function is interpreted in Donald (1986, 1992), consistent both with an
axiomatic definition and with a broad range of appropriate properties, as
giving ``the probability, per unit trial of the information in $\B$, of being
able to mistake the state of the world on $\B$ for $\sigma$, despite the fact
that it is actually $\rho$''.  In other words, $\app{\B}{\sigma}{\rho}$
provides a probability for the observation on $\B$ of the appearance of a
generalized ``collapse'' from local state $\rho$ to local state $\sigma$.  The
collapse is generalized because we are working at the level of a
macroscopic but localized observer whose state will inevitably be mixed
rather than pure. For such an observer, different possibilities will arise as
different approximately-decoherent components of the local mixed state
rather than as eigen-components of a wave-function.  

Perhaps the most important property of $\app{\B}{\sigma}{\rho}$ is that it
generalizes the idea that the a priori probability of seeing a state
$\sigma$ in a mixture of the form $$\rho = p\sigma + (1-p)\sigma_d
\eqno{(5.1)}$$  where $\sigma_d$ is disjoint from $\sigma$, is the
coefficient $p$ of $\sigma$.  Thus

\item{5.2)}  Suppose that, on a set $\B$,  $\rho = p\sigma +
(1-p)\sigma_d$  for some $0
\leq p \leq 1$, in other words, that 
$\rho(B) = p\sigma(B) + (1-p)\sigma_d(B)$ for all $B \in \B$,
 and suppose that there exists a projection  $Q \in \B$ such that
$\sigma(Q) = 1$,  $\sigma_d(Q) = 0$.

Then, in Donald (1992), it is proved, from the definition, that  
$$\app{\B}{\sigma}{\rho} = \rho(Q) = p.  \eqno{(5.3)}$$

An approximate form of this result also holds, in the sense that if 
 $\sigma(Q) \sim 1$ and $\sigma_d(Q) \sim 0$, then  
$$\app{\B}{\sigma}{\rho} \sim \rho(Q) \sim p,  \eqno{(5.4)}$$ where the
$\sim$ are given an explicit definition in (5.1) of Donald (1992).

(5.3) equates three different expressions for quantum probability:  
$\app{\B}{\sigma}{\rho}$;  the ``expected value'' $\rho(Q)$ of the projection
$Q$ in the state $\rho$; and the coefficient $p$ in (5.2).  The ``expected
value'' of an appropriate projection in an appropriate state is the
expression in terms of which empirical results and textbook quantum
mechanics are most easily and most often calculated.  However, to use this
idea it is necessary to choose ``appropriate'' projections and states.  It is
here that one looks  to ``the classical world'' or ``the macroscopic level'' or
``the measured operator''; and that, because none of these concepts can be
precisely defined, conventional quantum mechanics becomes an art rather
than a science.  The paradox of ``Wigner's friend'' exemplifies the
importance of not using ``inappropriate'' projections.  The goal in this paper
is to explain how ``appropriate expectations'' $\rho(Q)$ can be underpinned
by precisely-defined probabilities. 

(5.2) expresses an idea of local ``decoherence''.  This is one of the most
important themes in the foundations of quantum theory, but it is a theme
which is prey to a well-known confusion.  On one hand, there is the idea of
adding a classical probabilistic structure to a quantum theory, so that one
can speak of the (proper) mixture $p\sigma + (1-p)\sigma_d$ as being the
state $\sigma$ with probability $p$ and the state $\sigma_d$ with
probability
$1-p$.  On the other hand, there is similar mathematics which can be used
to describe the restrictions of states to localized systems.  It is this
mathematics which is often applied through the idea that a pure state is an
expression of complete information about a local system, but that when
such a system has interacted with an environment, complete information is
usually not locally available and the local state can be expressed as an
(improper) mixture of decohering possibilities.  We deal here with these
improper mixtures.  The aim is to express the fundamental probabilistic
structure of quantum theory in terms of the mathematics of (5.2) -- (5.4). 
By relating improper decoherent mixtures to genuine probabilities of
precisely specified events, the traditional confusion will be explained and
the problems of the ambiguity and inexactness of such mixtures will be
solved. 

On a technical level, there are several ways in which the idea of
decoherence of quantum states might be expressed.  For example,

\item{5.5)} With (5.2), we might say that
$\sigma$ is decoherent in $\rho$ on $\B$ and that $\sigma$ and
$\sigma_d$ are mutually decoherent in $\rho$ on $\B$.  In the
circumstances of (5.4), we might say that $\sigma$ and $\sigma_d$ are
approximately mutually decoherent in $\rho$ on $\B$. 

\item{5.6)}  (5.2) implies that, for all $B \in
\B$ such that $QB, BQ \in \B$, $\rho(QB) = \rho(BQ) = \rho(Q) \sigma(B)$
and we might say that $Q$ is a decohering projection for $\rho$ on $\B$
which, with probability $\rho(Q) = p$, reduces $\rho$ to $\sigma$ on $\B$. 
As long as $p = \rho(Q) > 0$, (5.2) also implies that there is an extension
$\rho'$ of $\rho$ such that $\sigma =
\dsize{Q \rho' Q \over \rho'(Q)}\Big|_\B$.

\item{5.7)} The opposite of decoherence is also important.  This is the idea
that a pure state is an expression of complete information about a physical
system.
\medskip

The properties of $\mathop{\rm app}$ are such that much of this can be
expressed by taking $\app{\B}{\sigma}{\rho}$ to measure (or to define) the
probability of $\sigma$ as a decoherent part of $\rho$ on $\B$.  As far as 5.7
is concerned, the set $\B$ defines the localized observables about which
information is available.  The coherence of pure states is expressed by the
following property, which holds for any projection $P \in \B$:
$$\rho(P) = 0 \text{ and } \app{\B}{\sigma}{\rho} > 0 \implies \sigma(P) =
0. \eqno{(5.8)}$$ This implies that if we have total information -- if $\B =
\B(\H)$ (the set of all observables) -- and if $\rho = |\Psi\>\<\Psi|$ is pure,
then
$\app{\B(\H)}{\sigma}{\rho}$ is zero unless $\sigma = |\Psi\>\<\Psi|$.  This
has a generalization for any C$^*$-subalgebra ${\cal C}$ of $\B(\H)$: if
$\rho$ and $\sigma$ are states on $\B$ with ${\cal C} \subset \B \subset
\B(\H)$, and if
$\rho|_{\cal C}$ (the restriction of $\rho$ to ${\cal C}$) is pure, then
$\app{\B}{\sigma}{\rho}$ is zero unless $\sigma|_{\cal C} = \rho|_{\cal C}$.
\medskip

Now we turn to the definition for the a priori probability of a minimal
switching structure $S(M, N, [d, \varphi])$, which was developed in Donald
(1992, 1995).

 Corresponding to a structure $S(M, N, [d, \varphi])$, B3 of the hypothesis
gives us a set of ordered switching structures $SO(M, N, d', \varphi')$ and
for each of these orderings,  part C gives us a set $GSO(M, N, d', \varphi')$
of possible geometrical structures $W$, where each $W$ is a sequence of
the form
$$W = (x, \Lambda, \theta, (T_n, (t_{nk})^{K_n}_{k=1}, (t'_{nm})_{m =
m_n^i}^{m_n^f}, x^n(t), L^n(t), P_n, Q_n)^N_{n=1})$$ corresponding to the
elements defined in part C.  Finally, for each geometrical structure, parts E
and F of the hypothesis give us a set ${\cal N}(W)$ of sequences
$(\sigma_m)^M_{m=1}$ of restrictions of quantum states to a set of
observables $\B(W)$ defined in part D.

The aim now is to use the function $\app{\B}{\sigma}{\rho}$ to define the a
priori probability of the sequences $(\sigma_m)^M_{m=1}$.  These are
sequences of quantum states, and the move from $\sigma_m$ to
$\sigma_{m+1}$ can be interpreted as the appearance of a generalized
collapse in precisely the sense discussed above.  It is then natural to
assume that successive collapses should be treated as independent events
and to impose a product structure on the developing probabilities.  There
are, however, two problems.  One is to know where to start the sequence, 
and the other is to know on which set of operators we have information at
any given collapse.  The initial quantum state, we shall denote by
$\omega$.  This may be thought of as the state corresponding to the
``universal wave-function'' of Everett.  All a priori probabilities are defined
relative to this state.  As we shall see in section 9, it is plausible to think of
$\omega$ as being an essentially featureless background, which carries
only the texture of the observed universe.  Nevertheless, this texture has to
be imposed on all the states in any sequence of collapses.  For this to occur,
the mathematics, explained in section 9 of Donald (1992), requires that the
same set of operators be used at each collapse throughout any given
sequence.

This leads us to G3 of the hypothesis in which the a priori probability of a
sequence of states $(\sigma_m)^M_{m=1}$ on a set of operators $\B$ given
an initial state $\omega$ is defined by the function
$\app{\B}{(\sigma_m)^M_{m=1}}{\omega}$ which satisfies
$$\app{\B}{(\sigma_m)^M_{m=1}}{\omega}  =
{\textstyle\prod\limits^M_{m=1}}
\app{\B}{\sigma_m}{\sigma_{m-1}} \eqno{(5.9)}$$ where  $\sigma_0 =
\omega$.

$\app{\B}{(\sigma_m)^M_{m=1}}{\omega}$ gives us a definition for the a
priori probability of individual sequences of local quantum states.  In G4--
G6, the a priori probability of a geometrical structure $W$ is defined
inductively by looking for the maximum a priori probability to which a
sequence of elements of ${\cal N}(W)$ can approximate, given that the
sequences of initial portions of those elements also approach maximal a
priori probability.  The inductive nature of this definition imposes a causal
structure according to which the most likely states for a given
determination of switch status are influenced by states earlier in sequence,
but not by later states.  

In G7, the a priori probability of a switching structure $S(M, N, [d,
\varphi])$ is defined by taking the supremum over the a priori
probabilities of the elements $W \in GSO(M, N, d', \varphi')$ and over the
possible labellings $SO(M, N, d', \varphi') \in S(M, N, [d, \varphi])$.
\medskip

Finally, in G8, the probability of moving from $S(M, N, [d, \varphi])$ to an
immediate successor $S(M', N', [d', \varphi'])$ is defined to be proportional
to the a priori probability of that successor. The constant of normalization
in G8 is significant.  For any given structure
$S(M, N, [d, \varphi])$, there is a finite set of possible immediate successors,
denoted by $\Xi(M, N,  d, \varphi)$, and each of the elements of this set has
finite a priori probability.  The number 
$$\xi = \sum \{\app{}{S(M', N', [d', \varphi'])}{\omega}:  S(M', N', [d',
\varphi']) \in \Xi(M, N,  d, \varphi)\}$$ is thus finite.  $\xi$ is the sum of the
a priori probabilities of the possible immediate successors of $S(M, N, [d,
\varphi])$.  If $\xi$ was equal to
$\app{}{S(M, N, [d, \varphi])}{\omega}$ then $S(M, N, [d, \varphi])$ could
be thought of as splitting into its distinct successors and it would be natural
to define the jumping probability to any one of these successors to be 
$$\app{}{S(M', N', [d', \varphi'])}{\omega}/\app{}{S(M, N, [d,
\varphi])}{\omega}.$$   However, the analysis of a priori probability
provided in Donald (1992) gives no reason to believe that $\xi$ should
equal  $\app{}{S(M, N, [d, \varphi])}{\omega}$.  If $\xi > 1$ then the
successors more than exhaust the present and it is reasonable to choose the
constant of normalization for the jumping probability to be $1/\xi$.  On the
other hand, if $\xi < 1$,  I speculate that it is appropriate to normalize the
jumping probabilities by
$1/\app{}{S(M, N, [d, \varphi])}{\omega}$  and to introduce a probability of
extinction equal to $1 - \xi/\app{}{S(M, N, [d, \varphi])}{\omega}$.  This is
speculation because there can be no direct empirical evidence.  The
proposed theory considers each individual observer separately, and no
observer can know of, let alone report, his own extinction.  Nevertheless, I
do not think that a many-minds theory can be plausible without such an
extinction probability.  Indeed, without extinction, ``most'' switching
structures would be very large and individually very improbable and we
might wonder why we should observe ourselves to be comparatively small
and probable.

\eject

\proclaim{6.  Observing quantum probabilities.}
\endproclaim

All the definitions have now been reviewed.  The hypothesis defines a
stochastic process on a set of entities.  That process could, in principle, be
simulated, using a suitable lattice quantum field theory.  Given an initial
structure, the probability of going to any other structure, or of hitting any
set of structures, is well defined.  What remains is to establish the
connections between this process and our observations.  The complexities of
the hypothesis are needed because of the difficulty of giving a conceptually
and mathematically complete formulation of the foundations of quantum
mechanics.   But it is also necessary to demonstrate the relationship
between such a complete formulation and the empirical adequacy for all
practical purposes of conventional quantum mechanics.  

This is a complicated issue.  There should be no doubt that the hypothesis
is conceptually radical.  Nothing is ``real'' except the switching structures of
individual observers (each considered separately), the initial condition
$\omega$, the underlying quantum field theory, and the objective
probabilities defined by the hypothesis.  Out of these ``elements of reality'',
each separate observer must construct his experiences and learn to guess
at what his future may bring.  This is done by the observer being aware of
his structure as awareness of an ``observed world''.  How this might be
possible is considered in Donald (1995, 1997).  It is a sophisticated form of
the doctrine that one is aware of the external world entirely through being
aware of the history of one's own brain.  

The ``observed world'', the world we see about us, is only a mental
representation.  In order to understand that representation, we  make
further representations of it.  Much of our mental processing is involved
with such representations of representations.  These representations have
their biological explanation in terms of their survival value; ``survival
value'', of course, implying enhanced relative probability.  Among the most
sophisticated representations available to the modern human observer, are
the theoretical representations of modern physics.  These theoretical
representations of his observed world are used by the observer to predict,
at least statistically, his future observations.  The purpose in this section is
to begin to explain, within the framework of the hypothesis, why these
predictions are usually successful.  This may seem a rather limited goal.  A
more thorough analysis would involve a much more careful and
philosophical discussion of the nature of awareness and of representations,
as well as an investigation of more ordinary aspects of the observed world. 
However, although this will not be provided in this paper, it is my belief
that what is provided is sufficient to indicate the essential outlines of such
an analysis.  Understanding of the quantum level gives a foundation for
conventional ideas about the reduction of the everyday to that level. 

The relationship between the probabilistic predictions of
conventional quantum theory and those of the hypothesis has already
been discussed at some length in my previous papers; particularly in
Donald (1992).  Much of this discussion remains relevant and is
compatible with the improvements presented here.  Elementary quantum
theory, quantum statistical mechanics, and decoherence theory
produce a picture of the local quantum states of a macroscopic
system as being, to a good approximation, decoherent mixtures of
correlated states weighted by numbers which reflect the
probabilities defined by elementary quantum theory.  The central
purpose of the hypothesis is to analyse the information in such an
approximate decoherent mixture of correlations, and to decompose the
mixture with probabilities determined by the weights.  Thus,
ultimately, the probabilities of elementary quantum mechanics and
the probabilities defined by the hypothesis agree because they all
reduce to weights in quantum mixtures as exemplified by (5.3).

Nevertheless, the introduction of an explicit stochastic process on the space
of switching structures is a significant step which makes plain an
incompleteness in Donald (1992, 1995).  Those papers, I believe, provide a
correct definition of ``a priori probability'' and correctly identify high
probability switching structures and the states to which they are
correlated.  What was lacking, however, was an adequate theory for sets of
observers, and the suggestion in this direction, made in section 9 of Donald
(1995), now strikes me as insufficient and wrong.   

In classical probability theory, if $\Pr(a) = {2\over 3}$ and $\Pr(b) =
{1\over 3}$, then in $N$ independent trials, a string of $N$ $a$'s is more
likely than any other single string, while for $N$ large, most members of
the set of all possible strings, by simple counting, contain about as many
$b$'s as $a$'s.  Probability theory applied to sets of strings tells us, however,
that, what we are likely to see, at least for $N$ large, is  a ``typical'' string,
which will contain about $N/3$ $b$'s.  More precisely, what the
mathematics tells us is that for $N$ large, there is a set of strings, all of
which have about $N/3$ $b$'s, which has probability close to one.  
Similarly in the present theory, what we are likely to see will be
determined by what is ``typical'', and to make sense of this, it is not enough
just to have a definition of the most likely observers, nor is it enough
merely to be able to count observers.  We also have to have a definition for
probabilities of sets of observers.  The stochastic process provides this.

Nevertheless, these objective definitions are only a foundation for
probabilistic reasoning which, in general, will involve both objective
probabilities and probabilities as degrees of rational belief.  When we ask
how likely it is that we will observe some outcome for an observation, we
need to take into account how much knowledge is available to us. 
According to classical physics, outcomes are determined, but we still say
that a fair die will show three with probability ${1 \over 6}$, because we do
not expect to know the exact initial conditions.  The present theory is
indeterministic but our ignorance often goes beyond that indeterminism. 
To discover exactly how likely it is that one should observe some outcome,
one would have to know exactly what switching structure one started from,
and define precisely the set of switching structures corresponding to the
observation of the particular outcome.  As pointed out in section 9 of
Donald (1995), neither of these is achievable.  Moreover, it is usually not
possible to be precise about the extent of our knowledge.  Objective
probabilities have to be precisely definable in order to be objective; if life
is like a game of chance, then the chances must be given in the rules. 
Rational belief, on the other hand, is by its nature a vague concept.  

As a result, there will always be some imprecision in the idea of ``typical'',
even beyond the question of what is sufficient for a probability to be
``close to one''.  But after all, we can only identify an occurrence as ``typical''
if it can occur repeatedly and in a way which is, at least to some extent,
unaffected by circumstances.  For example, we can claim that coins will
typically land heads about as often as they land tails without needing to
specify which coin is tossed, or when, or by whom, or even the currency. 
Similarly, when we discuss the outcomes of observations, we shall refer
only quite vaguely to repeatable physical situations which humans like us
might observe.  This imprecision is not important.  Although the definitions
of the hypothesis must be exact, exactness is irrelevant when it comes to
explaining how those definitions relate to the insubstantial mental
construct which, according to the present theory, is the everyday world. 
Here, all that is needed, and all that is possible, are explanations at a
practical level.

In the next three sections, we shall consider the relationship between what
is typical according to the hypothesis and empirical quantum probabilities. 
Partly because of the imprecision of the idea of ``typical'', and partly
because of the complexity of the theory, it will clearly not be possible to
give an explicit analysis from first principles of the properties of the
manifestions of typical switching structures.  Instead, we shall consider
aspects of the consistency of the theory and the way in which it supports
and completes the intuitions developed by decoherence theory.  On the
basis of arguments already given, it will be assumed that the development
of the hypothesis has been successful and that a valid many-minds
interpretation has been achieved.  The relationship between the
probabilities defined in the hypothesis and those of conventional quantum
theory will then be examined.  The central question in this section is why a
typical modern observer should be aware of a world in which the detailed
theoretical predictions of conventional quantum theory are confirmed. 
This will be explained using conventional quantum techniques, without a
detailed analysis of single events at the neural level.  Subsequently, we
shall consider the probabilistic analysis of individual events; first, in
section 7, with the study of elementary models; and then, in section 8, at a
more realistic level.

It will be useful to express in precise terms some aspects of the
assumptions to be made:

\proclaim{Definition 6.1}  Given an initial state $\omega$ and $\varepsilon
> 0$, an $\varepsilon$-manifestation of \newline $S(M, N, [d, \varphi])$
consists of a manifestation
 $((\sigma_m)^M_{m=1}, W)$ of $S(M, N, [d, \varphi])$, such that, for $m = 1,
\dots, M$,
$$\displaylines{
 |\mathop{\rm app}({\cal N}(W), \B(W), m, \omega) -
\app{\B(W)}{(\sigma_i)^{m}_{i=1}}{\omega}| < \varepsilon \cr
\text{and }  |\app{}{S(M, N, [d, \varphi])}{\omega} - \app{}{W}{\omega}| <
\varepsilon.   }$$ A property which, for all sufficiently small $\varepsilon >
0$, holds for every
$\varepsilon$-manifestation of $S(M, N, [d, \varphi])$ will be referred to as
a necessary property of the physical structure of $S(M, N, [d, \varphi])$.
\endproclaim

The definitions imply that there are $\varepsilon$-manifestations for any
$\varepsilon > 0$.

\proclaim{Assumption 6.2}  For $S(M, N, [d, \varphi])$ corresponding to a
typical human observer, it is a necessary property that its manifestations
 $((\sigma_m)^M_{m=1}, W)$ involve quantum states $\sigma_m$ which, at
least for large $m$, are quantum states describing an appropriate and
persistent human brain.
\endproclaim

\proclaim{Assumption 6.3}  In the circumstances of 6.2, it is also necessary
that the $\sigma_m$ have high probability extensions which describe the
world observed by $S(M, N, [d, \varphi])$.
\endproclaim

Although both assumptions use concepts which need some elucidation,
essentially assumption 6.2 says that the hypothesis is sufficient to
characterize the physical structure of human observers, while assumption
6.3 says that the observer's representation of the world of appearances is
accurate.  Both assumptions are fundamental for a many-minds
interpretation, and arguments which can be seen as explaining and
justifying them are to be found throughout this series of papers.

The consistency arguments which use assumption 6.2 will be based on the
assumption that our own observations, of ourselves and of our society, 
give us an accurate picture of a ``typical human observer'' and of an 
``appropriate'' brain for such an observer.  That brain should be
``persistent'' in the sense that, for all sufficiently small $\varepsilon > 0$, if
$S(M^s, N^s, [d^s, \varphi^s])$ is a significantly probable successor of $S(M,
N, [d,
\varphi])$, then the brain described by an $\varepsilon$-manifestation of
$S(M, N, [d, \varphi])$ should be suitably similar to the relevant part of the
history of the brain described by the $\varepsilon$-manifestations of
$S(M^s, N^s, [d^s, \varphi^s])$.

In referring to ``the world observed by $S(M, N, [d, \varphi])$'' in
assumption 6.3, it is assumed that sufficient structure is contained within
$S(M, N, [d, \varphi])$ for the corresponding observer to be able to
construct a mental representation of a world.

Assumption 6.3 also refers to the idea of a high probability extension of a
state $\sigma_m$.  This was discussed at length in Donald (1992).  For an
example, suppose that the structure $S(M, N, [d, \varphi])$ corresponds to a
particular person who is watching a football match.  According to 6.2,
manifestations $((\sigma_m)^M_{m=1}, W)$ exist for which the sequences of
states $(\sigma_m)^M_{m=1}$ give an accurate description of the spectator's
brain.  Information about the present position of the ball will be contained
both in $S(M, N, [d, \varphi])$ and in $(\sigma_m)^M_{m=1}$.  This means
that, if the theory is consistent, then there should be probable extensions
$\rho'$ of
$\sigma_M$ (the present moment state of the observer's brain) to sets of
observables $\B$ which are contained in some local algebra $\A(\Lambda)$,
where $\Lambda$ contains the observed ball position, such that $\rho'$
restricted to
$\B$ is a possible quantum state for a football.  $\rho'$ will depend on the
initial state $\omega$ in the same way that, in conventional physics, our
personal observations will not be sufficient to determine the exact current
state of a football and for the remaining information we could look to the
initial conditions for the universe.  The following definition provides one
way of defining suitable states $\rho'$:

\proclaim{Definition 6.4}  Given a switching structure $S(M, N, [d,
\varphi])$, a set of operators $\B$, an initial state $\omega$, $\varepsilon >
0$, and an
$\varepsilon$-manifestation $((\sigma_m)^M_{m=1}, W)$ of $S(M, N, [d,
\varphi])$ define a state $\rho'$ on $\B$ to be an $\varepsilon$-prediction
for $S(M, N, [d, \varphi])$ of the current state on $\B$, if  $\rho' =
\sigma_M'|_\B$ for some state $\sigma_M'$ such that there exists a
sequence $(\sigma'_m)^M_{m=1}$ with $\sigma'_0 = \omega$, and
$\sigma'_m|_{\B(W)} = \sigma_m|_{\B(W)}$ with
$$\app{\B_1}{(\sigma'_m)^M_{m=1}}{\omega}
\geq \sup\{ \app{\B_1}{(\sigma''_m)^M_{m=1}}{\omega}
 : \sigma''_0 = \omega, \sigma''_m|_{\B(W)} = \sigma_m|_{\B(W)}\} -
\varepsilon $$ where $\B_1 = \B(W) \cup \{ B C : B \in \B, C \in {\cal
C}(W)\}$.  ($\B_1$ is the minimal set containing $\B$ and $\B(W)$ which
also allows correlations between $\B$ and the switch projections
experienced by the observer to be expressed (\sect 3 of Donald (1992).)

A property which, for all sufficiently small $\varepsilon > 0$, holds for
every
$\varepsilon$-prediction will be referred to as a property predicted by 
$S(M, N, [d, \varphi])$ of the state on $\B$.
\endproclaim 

Although this is a reasonable definition, it is not unmodifiable.  Indeed, an 
alternative definition could be based on postulate eight of Donald (1992). 
However, the idea of a high probability extension is not a fundamental
concept within the theory, and so does not need to be absolutely and
unambiguously defined.  $\B$ and $\Lambda$ in the example are clearly
fuzzy, and, in general, there need not be a unique limiting state $\rho'$ for
$\varepsilon
\rightarrow 0$.  Nevertheless, if the present theory has any validity, then,
in appropriate circumstances, both suitable sequences
$(\sigma_m)^M_{m=1}$ and suitable extensions $\rho'$ of $\sigma_M$ will
exist ``for all practical purposes''.  This will be sufficient as a basis for an
explanation of ordinary observation in the framework of the full theory
which will in turn make the theory plausible.  

Ultimately, extensions as defined by 6.4, or in any other way, are merely
means of expressing the fundamental ``elements of reality'' referred to at
the start of this section.  Such extensions are the best possible theoretical
representations of the world seen by the observer; that is, of the world of
which, through his switching structure, the observer has created a mental
representation.  Our ordinary physical knowledge has been developed to
make predictions about the world we observe.  We can use that knowledge,
both to choose appropriate extensions, given partial information about an
observer's structure; and, given an extension, to make predictions about
the observer's subsequent structure.  Because we understand our own
switching structures through the representations of external worlds that
we construct from them, the easiest way for us to predict the likely
temporal development of any switching structure is, first to construct a
theoretical representation of the corresponding observed world -- in other
words, an extension -- and then use that extension to predict future
observations.  Although this is the easiest way, it is not the fundamental
way, which is, of course, that defined by the hypothesis.

Agreement between the two ways will be a consequence of properties of
the a priori probability function and of the fundamental quantum
dynamics.  In the most likely extended states, the information in a
switching structure is correlated with observables external to the observer. 
The a priori probability function requires the quantum states of the
observer to change as slowly as is possible, given the information
available.  No change in quantum state, in the Heisenberg picture used
throughout this work, would correspond to evolution of observables under
Hamiltonian dynamics.  Our knowledge of conventional quantum theory
allows us to model that evolution on either internal or external
observables, and we can then use the correlations to relate one set to the
other.  Changes in information also can be correlated between internal and
external observables.

Statistical mechanics is one field in which it is particularly valuable to be
able to claim that a definition like 6.4 provides the best possible theoretical
representation of the state of the observed world.  For suitable sets $\B$,
$\rho'$ will be the predicted state of an observed thermal system.  Such
states certainly need not be pure.   A fairly exactly specified non-zero (von
Neumann) entropy may well be a predicted property, in the sense of 6.4,
for a thermal system.  The observer may only have information about
macroscopic parameters, and the states predicted by 6.4 will then be
something like the maximum entropy state given fixed values of those
parameters.  This means that in the present framework, it is not necessary
to rely on ergodic theory in order to explain why, in suitable circumstances,
statistical mechanical systems behave as if they occupied Gibbs equilibrium
states.  In a sense made precise by definition 6.4, the observed state {\sl is}
an ``ensemble''.

One of the major difficulties in the analysis of quantum probability is the
variety of different notions of ``probability'' which are involved.  A
distinction between an ontological idea of probability as objective and an
epistemological idea of probability as degree of rational belief has already
been drawn.  It is now useful to make two further distinctions, between
public and private and theoretical and empirical probabilities.  First is the
theoretical concept of public (observer independent) quantum probability. 
Depending on the circumstances, this may be calculated using the expected
value of a projection, or the square of a transition amplitude, or the
coefficient of a component of the expansion of a density matrix into disjoint
pure states, or an a priori probability defined by the function app of (5.9). 
Sometimes, as in equation (5.3), different calculations will give the same
answer.  The corresponding empirical notion can be expressed by, for
example, the published long-term relative frequencies of particular types
of observational results.  A second theoretical notion is expressed by the
probabilities defined by the stochastic process of the hypothesis.  These are
private in the sense that they apply to an individual observer.  The
corresponding relative frequencies will be experienced in the awareness of
that observer.  According to the present theory, public probabilities are
only shadows of the fundamental private probabilities.

Relations between the public notion of quantum probability and the
private notion are subtle and give rise to two closely-related theses:

\proclaim{Thesis 6.5} A typical modern human observer should be aware
of a world in which quantum theory is accepted and in which its detailed
theoretical predictions are confirmed.
\endproclaim

\proclaim{Thesis 6.6} Under apppropriate circumstances, there is a fairly
direct agreement between the private and public probabilities.
\endproclaim

It is a consequence of thesis 6.5 that the present theory is an interpretation
of quantum theory.  In section 8, a model of the functioning of a human
brain which is consistent with thesis 6.6 will be expounded.  In the
remainder of the present section, the relation between the theses will be
examined, and it will be argued that 6.5 would hold even under conditions
much weaker than 6.6.  6.6 involves the study of precisely-specified
individual experiences, but for 6.5, it is sufficient to consider only
broadly-characterized events with nearly-certain ``typical'' outcomes.

The model of brain functioning in section 8 separates quantum probabilities
from the counting of neural events and geometries and shows that
agreement between private and public probabilities follows from
appropriate assumptions of indifference in the counting between possible
alternative observations.  The possibility of 6.6 failing while 6.5 succeeded
might arise if these indifference assumptions failed to hold for some
particular class of personal experience.  For example, a significant
difference between the hypothesis of this paper and the version of Donald
(1995) is that in the earlier version, the determinations of switch status
were required, in the language of part F, to alternate between ``open'' and
``closed''.  Under this assumption, the number of structures corresponding
to a high level of neural activity was raised relative to the number
corresponding to a low level.  This would breach the indifference
assumption when an observer was presented with a pair of alternatives
which would give rise to significantly different levels of activity.  As an
example, one might consider an observer listening to the sound of a Geiger
counter and presume that more neural activity would follow clicking than
silence.

Because the fundamental probabilities are private, ultimately the only
empirical tests of either 6.5 or 6.6 consist of asking yourself whether it is
in accord with your own personal observations.  As far as 6.5 is concerned,
we do (do we not?) experience a world in which it appears that satisfactory
tests of the public notion of quantum probability abound, and, that so far,
when ``appropriate'' calculations have been used, those tests have been
passed.  This means that we do have tests of 6.5.  Indeed, an explanation of
why the present theory predicts that typical observers will experience a
world in which the details of quantum theory are publicly confirmed will
allow the present theory to annex all the public empirical evidence for
those details.

As far as 6.6 is concerned, suppose once again that it failed, and that the
probability of some possible observation which involved a high level of
neural activity was raised relative to an alternative involving a low level. 
This would lead to a situation in which a typical observer would tend to
find himself seeing more of the high activity alternative than he would
expect; either from quantum mechanical calculation, or from reports from
other laboratories.  Thus, with the Geiger counter example, he might find
that his personal measurements, with enhanced probability of clicking,
corresponded to radioactive lifetimes that were shorter than expected.  If
he performed his experiments in the company of a colleague, then, because
of the most basic quantum correlations, he would see her seeing the same
results, and thus he would find her agreeing with him about the puzzling
discrepancy; but if he asked her to repeat the experiments in his absence,
then he would be most likely to see her reporting to him that the
discrepancy had disappeared.  In such a case, each individual separately
would typically believe that his own personal observations were somewhat
unusual.  The performance of quantum experiments in public would even
allow a public consensus about these anomalies, at least among the
audience for the experiments.

Turning now to sufficient conditions for 6.5, recall that in classical theory, if
we are given a source of strings and we have reason to believe that the
letters are produced independently and with constant probabilities, then
we can estimate those probabilities by assuming that the string we happen
to have been given is typical -- or more precisely that the letters in it
have typical relative frequencies.  The laws of large numbers tell us that
this is a consistent procedure and that the longer our test string, the better
our estimates will be likely to be.  Similarly, in quantum theory, we can
use the assumption that suitable observations have been typical to
estimate the probabilities of simple repeated individual events.  The
present theory will be consistent as an interpretation of quantum theory
only if those estimates are likely to agree with the theoretical predictions
of elementary quantum theory.  Moreover, the classical laws of large
numbers explain how the empirical basis for probabilistic reasoning can be
reduced to the observation of properties with sharply peaked distributions,
and the aim here is to show how the same reduction can be made in
quantum theory. 

Experimental tests of probabilistic theories always depend on the laws of
large numbers to produce significant and near certain answers from noisy
and uninteresting random data.  This is particularly important in the
present theory, because, in situations in which the noisy and uninteresting
random data is not directly observed, only the significant answers affect the
switching structures and the set of manifestations of the observer.  Then,
because of the near certainty of those answers, the details of the definition
of private probabilities are irrelevant.  

In order to be aware of a world in which quantum theory is accepted, it is
not necessary to be aware of the detailed history by which the
experimental evidence for quantum theory has been built up.  It is only
necessary to be aware of a limited range of almost inevitable facts about
the world, involving the sort of summaries of experimental evidence which
we could expect to learn at second-hand, either from colleagues, or from
machines, or even from textbooks. For example, these facts might include
the relation between the observed spectrum of hydrogen and the calculated
eigenvalues of the corresponding Schr\"odinger equation.  In conventional
terms, any account of the observation of a hydrogen spectrum would
ultimately depend on the existence of many individual atomic excitations,
but in the present theory, even direct observation of the spectral lines no
more provides evidence for specific excitations, than observation of a
double slit interference pattern provides evidence for the passage of
electrons through specific slits.  

When we look at an interference pattern formed by an appropriately
prepared stream of electrons hitting a screen, we can almost certainly
expect to see a pattern which will show the wave-like behaviour of the
electron and which may be used to confirm the de Broglie relation between
electron energy and wave-length.  We will not however see any evidence
of the order in which individual electrons have struck the screen.  And
because, according to the present theory, only the observations of an
individual observer are authentic, with each individual considered
separately; this means that, in this situation, there is no actual order in
which individual electrons struck the screen.  The present theory has built
in to it a natural ``coarse graining'' which means that no definite existence
is required beyond the direct personal experience of individual observers.

An observation may have both typical and specific aspects.  In many
circumstances, it is permissible to focus on the typical aspects, because the
classical probabilistic structure defined by the hypothesis allows
us simply to aggregate the specific aspects.  For example, when we
look at a complicated bubble chamber photograph, we do observe
specific quantum events of low a priori probability.  The
corpuscular behaviour that is revealed, however, will be typical of
all such photographs.  When we have looked at the photograph, we
will have seen something which, in its precise details, was of low
probability.  Because those details will have affected our internal
structure, this probability is ``private'', and its analysis will
depend on the validity of thesis 6.6.  Nevertheless, we will be part
of a large set of observers, who in looking at the photograph, both
see evidence for particulate structure and share a common history up
to the moment of starting to examine the photograph.  As a future of
the common history, that set will have had high probability.

Among all our predictable information about quantum theory, there is
much information about the values of quantum probabilities.  Indeed, as
observers  of quantum probabilities, we are, more often than not, in the
position of the gambler's wife who sees only the nightly remorse, and not
the card-by-card ups and downs.

\proclaim{example 6.7}  Let strings ${\bf s} = (s_n)_{n=1}^N$ of $a$'s and
$b$'s be generated by $N$ independent and identically distributed classical
trials, with $\Pr(s_n = a) = p$, $\Pr(s_n = b) = 1 - p$.  The laws of large
numbers tell us that the set of all strings with approximately $pN$ $a$'s has
probability close to one.

For example, for $\delta > 0$ and $\eta > {1\over 2}$, let $X^N_{\delta,
\eta}$ be the set of strings with between $pN - \delta N^{\eta}$ and $pN +
\delta N^{\eta}$ $a$'s.  Then, applying Chebyshev's inequality to the
random variable which gives the relative frequency of $a$ in ${\bf s}$, gives
$$\Pr(X^N_{\delta, \eta}) \geq 1 - \delta^{-2} p(1-p) N^{1-2\eta}$$ and so
$\Pr(X^N_{\delta, \eta}) \rightarrow 1$ as $N \rightarrow \infty$.

Suppose that $\rho$ is a density matrix on a Hilbert space $\H$, that $P$ is
a projection, and that $\rho(P) = p$.  Let $(\H_n, \rho_n, P_n)_{n=1}^N$ be a
sequence of $N$ isomorphic copies of $(\H, \rho, P)$.  Set $Q_n = 1- P_n$. Let
$\H^N = \otimes_{n=1}^N \H_n$ and $\rho^N = \otimes_{n=1}^N \rho_n$.

$\rho^N$ provides a quantum model of the classical distribution with
$s_n = a$ corresponding to $P_n$ and $s_n = b$ corresponding to
$Q_n$, in the sense that the expected value in the state $\rho^N$ of any
projection of the form $\otimes_{n=1}^N R_n$ where $R_n$ is either
$P_n$ or $Q_n$ is the same as the probability of the corresponding string. 
Sums of these projections have expected values which are the same as the
probabilities of the corresponding sets of strings.  For example, there is a
projection $P_{X^N_{\delta, \eta}}$ on $\H^N$ corresponding to the set
$X^N_{\delta,
\eta}$ and $\rho^N(P_{X^N_{\delta, \eta}}) \rightarrow 1$ as $N
\rightarrow \infty$.

The relative frequency operator $F^N$ is defined by
$$F^N = \sum_{M=0}^N \textstyle{M \over N} P_{S^N_M}$$ where
$P_{S^N_M}$ \vadjust{\kern1pt} is the projection corresponding to the
set $S^N_M$ of strings of length $N$ with exactly $M$ $a$'s.  Direct
calculations, or standard results on the binomial distribution,
yield $\rho^N(F^N) = p$ and $\rho^N((F^N - p)^2) = p(1-p)/N$ (cf.
Hartle (1968) and the papers by DeWitt and by Graham in DeWitt and
Graham (1973)).
\endproclaim

For an application of this example, suppose that an observer becomes aware
that many independent repetitions of an appropriate experiment on
identically prepared systems have been performed.  If he has sufficient
information about the systems used, then this awareness would mean that
the existence of the experiments would be a prediction of the observer in
the sense of 6.4. Thus, at the beginning of the experiment, it would be
possible to construct sets $\B_n$ which could be modelled by
$\B(\H_n)$, such that the state predicted by the switching structure of the
observer on $\otimes_{n=1}^N \B_n$ would be a product state which could
be modelled by $\rho^N$.   

The observer can now use example 6.7 to predict that, as far as public
quantum probabilities are concerned, the most likely observed proportion
of experiments with result $a$ will be $p$.   More precisely, at the time that
the experiment is set up, the observer can predict, in the sense of 6.4, that
expectations of the observables modelling $P_{X^N_{\delta, \eta}}$, or
$F^N$, or $(F^N - p)^2$ will be as given by example 6.7, and, through the
analysis of appropriate collapse-free quantum dynamics, he can extend
such expectations to observables modelling public records of the proportion
of experiments with result $a$.

Suppose that subsequently the observer is told in roughly what proportion
of the experiments result $a$ was observed, but that he does not observe
the individual results.  Then his own personal quantum states will become
directly correlated to a limited range of values for
$F^N$.  In theory, to make a complete model of this situation, one would
start from manifestations of a particular switching structure
$S(M, N, [d, \varphi])$.  One would then construct its successors and predict
properties of the manifestations of those successors.  One might do this, for
example, by an explicit dynamical model of a von Neu\-mann-type
measurement of $F^N$, or, perhaps, just by considering chains of
correlation from sums of eigen-projections of $F^N$ (like $P_{X^N_{\delta,
\eta}}$), to possible records of the relative frequency, and thence to possible
neural observations of those records.  

A ``collapse'' process is necessarily involved in this situation, as it is not
inevitable that the observed relative frequency will be close to $p$. 
However, the point of example 6.7 is that far more weight in $\rho^N$ is
attached to such relative frequencies.  If the observer did observe all the
individual results, then thesis 6.6 would be relevant because there would
be the possibility of a systematic bias in the individual observations. 
However, as long as he is only aware of, or influenced by, the overall
proportion of different results and particularly if he is just making a simple
choice between whether the relative frequency is close to $p$ or not, the
large relative difference of weights in $\rho^N$ will continue to dominate in
the probabilistic analysis.

In fact, as will be discussed in section 8, the nature of the human brain, and
the way that it is modelled by the hypothesis, mean that $S(M, N, [d,
\varphi])$ will have a huge number of relevant successors, differing in a
myriad of fine details of neural processing.  Nevertheless, because they are
not correlated to individual results, those fine details will not systematically
affect the observed proportion of $a$ results, and correlation to sums of
eigen-projections of $F^N$ will ensure that, in all but a probabilistically
negligible set of successors, the proportion will be close to $p$.  This implies
that, in this situation, the proportion observed by the ``typical'' observer
will agree with the proportion predicted by elementary quantum theory.

Example 6.7 indicates how the hypothesis can provide a framework in
which the large $N$ mathematics of the quantum laws of large numbers
can be applied without either having to pass to the limit
$N = \infty$ or having to use the multitude of individual systems as our
conceptual basis.  It is not necessary to pass to the limit $N = \infty$ to
derive classical probabilities from example 6.7,  because the hypothesis
defines objective classical probabilities for individual finite observers.  On
the other hand, the elementary projections $\otimes_{n=1}^N R_n$ need not
be considered as being more fundamental than composite projections like
$P_{X^N_{\delta, \eta}}$.  It is the individual observer who is fundamental
and whose structure provides the natural coarse-graining.  Given this, it
becomes possible to generalize the analysis of example 6.7 to a wide
variety of circumstances in which records of predictable quantum
probabilities are observed.  It also becomes possible to apply many other
schemes in which large $N$ mathematics has been used to reveal
quasi-classical structures (e.g.~Hepp (1972), Namiki et al.~(1997)),
and to address the conceptual issues which have made those schemes
problematic (Bell (1975), Giulini et al.~(1996, \sect 9.1 -- 9.3)).

In this section, it has been argued that the hypothesis is compatible with
empirical evidence for quantum theory.  This means that that evidence can
be used in turn as an important part of the justification for the hypothesis. 
This is a consistency argument which starts with the development of the
hypothesis as a way of making sense of the empirical evidence.  Then it is
supposed that the hypothesis does describe quantum states which are
states of a brain processing definite information, and that those states can
be extended to describe systems external to the brain.  Finally, by
considering the properties that such systems could be expected to have,
and how those properties will evolve and correlate with possible future
states of the observer, it is argued that a typical observer will be aware of
empirical evidence for quantum theory.  The consequence of this is that the
overall picture is consistent.  Such consistency between observation and
theory is what it means to have an interpretation.

What remains is to consider whether the details of the hypothesis provide
a plausible theory of individual observations of individual events with
uncertain outcomes.

\proclaim{7. Elementary Models.}
\endproclaim

	There are many interwoven aspects to the hypothesis.  In this section, I
shall attempt to explain some of these aspects by showing how the
hypothesis develops from Everett's original picture.  The consistent
histories theory will also be briefly discussed, and analogies will be drawn
to the present theory.

Following Everett, most elementary versions of the many-minds
interpretation assume that, if we wish to describe the observation of a
quantum experiment, then we can split the Hilbert space $\H$ of the
universe naturally into a tensor product  $\H = \H_O \otimes \H_{ex}$
where $\H_O$ is a space of wave-functions for the observer and
$\H_{ex}$ is a space of wave-functions for his experimental apparatus and
for the world external to him.  Then, it is assumed that the true state of the
universe ($\omega$) is a pure state,
$\omega = |\Psi\>\<\Psi|$  and that $\Psi$ has a decomposition of the form
$$\Psi = \sum_{r=0}^{\infty} \sqrt{p_r} \psi_r \otimes \varphi_r
\eqno{(7.1)}$$ where $(\varphi_r)_{r=0}^{\infty}$ is an orthonormal set of
wave-functions representing the distinct results of the experiment and the
rest of the universe, while $(\psi_r)_{r=0}^{\infty}$ is a sequence of
wave-functions for the observer, with $\psi_r$ representing the observer
observing the result represented by $\varphi_r$.  (7.1) then models
correlations between wave-functions of the observer and wave-functions
of the rest of the universe by proposing that $\psi_r$ is correlated with
$\varphi_r$, and models the probabilities of different observations by
proposing that observation $r$ has probability $p_r$.

	In my earlier papers, I have argued that almost every aspect of this
model is an over-simplification.  Nevertheless, equation (7.1) does
introduce the intuition which powers the many-minds interpretation.  The
hypothesis in the appendix to this paper is no more than an attempt to
solve the problems of (7.1) while preserving the intuition.  

	Even within the framework of (7.1), one important forward step in the
analysis of $\omega$ is possible.  As observers we undoubtedly interact
only with very limited aspects of the entire universe.  Because of this, we
need only consider restrictions of $\omega$ to sets of observables
$\B(t)$ accessible to the observer at time $t$.  Appeal to the existence of
such sets allows us to assume that $\omega$ is a mixture of macroscopically
different ``observer states''.  Such an assumption will not be correct at the
global level of the ``wave-function of the universe'', but, as we are assured
by decoherence theory (Giulini et al.~(1996)), it is almost surely
true for appropriate local restrictions of such a wave-function. 
Let us therefore denote by  $\B(t)$ some suitable choice of local
observables. Working in the Heisenberg picture, the time dependence
of the splitting of $\omega$ can then be expressed in the time
dependence of $\omega|_{\B(t)}$.  Part D of the hypothesis defines a
sophisticated version of $\B(t)$, but, even in the framework of
(7.1), we can take $\B(t) = \B(\H_O) \otimes 1$, where
$\B(\H_O)$ is the set of all bounded operators on the Hilbert space
$\H_O$.  This choice allows us to replace (7.1) by
$$|\Psi\>\<\Psi|\big|_{\B(\H_O) \otimes 1} = \sum_{r=0}^{\infty} p_r
|\psi_r\>\<\psi_r|.
\eqno{(7.2)}$$

In (7.2), only the observer states are considered and correlations with the
observed system are lost.  This is also true in the hypothesis.  With the full
hypothesis applied at the human level, the idea, as discussed in the
previous section, is that the external observed system can be
``reconstructed'' from the internal structure of the observer -- in other
words, we ``observe'' (exist as) not the ``real world'', but (as) that apparent
shadow of the apparently real world formed by the functioning of our
brains.  However it is also useful to note that, to a good approximation,
correlations with the observed system can be restored in (7.2) by
appropriate enlargements of the set of operators considered.  For example,
according to various models of environmental decoherence, it would be not
be unreasonable to write  $\H_{ex} = \H_{sys} \otimes \H_{env}$, where
$\H_{sys}$ represents wave-functions of the system under observation and
$\H_{env}$ represents the rest of the universe including the environment
of that system, and to replace (7.1) by
$$\Psi = \sum_{r=0}^{\infty} \sqrt{p_r} \psi_r \otimes \varphi_r \otimes
\chi_r
\eqno{(7.3)}$$ where $(\varphi_r)_{r=0}^\infty$
(resp.~$(\chi_r)_{r=0}^\infty$) is an orthonormal set in $\H_{sys}$ (resp.
$\H_{env}$) (cf.~(3.15) of Giulini et al.~(1996)).  We can now take
$$\B(t) = \B(\H_O) \otimes \B(\H_{sys}) \otimes 1 = \B(\H_O \otimes
\H_{sys}) \otimes 1,$$ where
$\B(\H_{sys})$ is the set of all bounded operators on the Hilbert space
$\H_s$ and this allows us to replace (7.2) by
$$|\Psi\>\<\Psi|\big|_{\B(\H_O \otimes \H_{sys}) \otimes 1} =
\sum_{r=0}^{\infty} p_r |\psi_r\>\<\psi_r| \otimes
|\varphi_r\>\<\varphi_r|. \eqno{(7.4)}$$

In (7.4), the correlations between wave-functions $\psi_r$ and
$\varphi_r$ continue to be expressed.  Nevertheless, (7.4) is only an
approximation, and there is considerable ambiguity in the choice of the
space $\H_{sys}$.  Similar ambiguities will arise in any analysis of
correlation at the level of the full theory (e.g.~in definition 6.4).

The language of (5.6) applies in this scenario: $|\psi_r\>\<\psi_r| \otimes 1
\otimes 1$  (respectively $1\otimes |\varphi_r\>\<\varphi_r| \otimes 1$, 
$|\psi_r\>\<\psi_r| \otimes |\varphi_r\>\<\varphi_r| \otimes 1$)  is a
decohering projection for $|\Psi\>\<\Psi|$ on $\B(\H_O) \otimes 1 \otimes
1$ (resp. $1
\otimes \B(\H_{sys}) \otimes 1$, $\B(\H_O \otimes \H_{sys}) \otimes 1$,
reducing $|\Psi\>\<\Psi|$ to $|\psi_r\>\<\psi_r|$  (resp.
$|\varphi_r\>\<\varphi_r|$,  $|\psi_r\>\<\psi_r| \otimes
|\varphi_r\>\<\varphi_r|$) with, in all cases, probability $p_r$.

The most fundamental problem with (7.1) -- (7.4) lies in the identification
of ``wave-functions for the observer'' and in the indexing of such
wave-functions.  My proposed solution to this problem is part A of the
hypothesis and the idea of switching structures. Using this idea, suggests a
generalization of (7.2) of the form
$$\omega\big|_{\B(t)} = p_0 \rho\big|_{\B(t)} + \sum_{r = 1}^R p_r
\sigma_{O[r]}\big|_{\B(t)}  \eqno{(7.5)}$$  where $0 \leq p_r \leq 1$ for $r
= 0, 1, \dots, R$, $\sum_{r = 0}^R p_r = 1$, $\rho$ is  an arbitrary quantum
state disjoint from the $\sigma_{O[r]}$, $\{O[1], \dots, O[R]\}$ is a set of
possible observers, and
$\sigma_{O[r]}$ is a quantum state of the universe in which the physical
structure of the observer in question is a manifestation of $O[r]$.  Parts C, E,
and F of the hypothesis are concerned with the mapping from switching
structures $O[r]$ to possible states $\sigma_{O[r]}$.

 Assuming that the $\sigma_{O[r]}$ are disjoint in the sense that there exist
projections $Q[r] \in \B(t)$ such that $\sigma_{O[r]}(Q[s]) = \delta_{r s}$ and
$\rho(Q[s]) = 0$, (7.5) and (5.3) yield 
$$\app{\B(t)}{\sigma_{O[r]}}{\omega} = \omega(Q[r]) = p_r. \eqno{(7.6)}$$
This ``a priori probability'' agrees with the standard probabilistic
interpretation of quantum mixtures applied to equation (7.5).

The next stage in the elaboration of the hypothesis is to consider with more
care the time development of an observer.  This suggests a modification of
(7.5) to read
$$\sigma_{O[r^{m}]}\big|_{\B(m +1)}  = p[r^{m}, 0] \rho_{r^{m},
0}\big|_{\B(m+1)} + \sum_{r^{m+1} = 1}^{R^{m+1}} p[r^{m}, r^{m+1}]
\sigma_{O[r^{m}, r^{m+1}]}\big|_{\B(m+1)}. 
\eqno{(7.7)}$$

In (7.7), $t$ has been replaced by a discrete marker $m$ for the steps of a
developing process and the state of a structure at step $m$ is split into
states for its possible immediate successors at step $m +1$.   (7.6)  becomes
$$\app{\B({m+1})}{\sigma_{O[r^{m}, r^{m+1}]}}{\sigma_{O[r^{m}]}} =
\sigma_{O[r^{m}]}(Q[r^{m+1}]) = p[r^{m}, r^{m+1}].
\eqno{(7.8)}$$

	For notational convenience, (7.7) will be rewritten as
$$\sigma_{O[r^{m}]}\big|_{\B({m+1})} =  \sum_{r^{m+1} = 0}^{R^{m+1}}
p[r^{m}, r^{m+1}]
\sigma_{O[r^{m}, r^{m+1}]}\big|_{\B(m+1)}. 
\eqno{(7.9)}$$ In (7.9), the ``left-over'' state $\rho_{r^{m}, 0}$ has been
absorbed into the main sum by setting $\sigma_{O[r^{m}, 0]} = \rho_{r^{m},
0}$.  Similar remainders will occur frequently below.

When (7.1) or (7.2) is presented in elementary versions of many-minds or
many-worlds interpretations, something closer to (7.9) is often being
invoked, because usually the assumption is made that there is a quantum
measurement being considered and that the apparatus for that
measurement already exists.  These assumptions are false in a universal
quantum theory without collapse.  In such a theory, the universal
uncollapsed quantum state $\omega$ superposes all possible situations that
might occur on planets like ours by this stage in the development of the
universe.  

 (7.7), (7.8), and (7.9) model only one time step but generalize immediately
to a succession of steps.  Thus, set $\sigma_{O[\,]} = \omega$ and, for
$m = 0, 1, \dots, M-1$, consider a sequence of equations analogous to (7.9):
$$\displaylines{
 \sigma_{O[ r^1, \dots, r^{m-1}, r^{m}]}\big|_{\B(m+1)} \hcrh = 
\sum_{r^{m+1} = 0}^{R^{m+1}} p[r^1, \dots, r^{m-1}, r^{m}, r^{m+1}]\,
\sigma_{O[r^1, \dots, r^{m-1}, r^{m}, r^{m+1}]}\big|_{\B(m+1)},  \hfill
\llap{(7.10)}   }$$ where there exist projections $Q[r^1, \dots, r^{m-1},
r^{m}, r^{m+1}] \in
\B(m+1)$ such that
$$\sigma_{O[r^1, \dots, r^{m-1}, r^{m}, r^{m+1}]}(Q[r^1, \dots, r^{m-1},
r^{m}, s^{m+1}])  = \delta_{r^{m+1} s^{m+1}}.  
\eqno{(7.11)}  $$

	(7.10) and (7.11) model a single event in which an observer  
$O[ r^1, \dots, r^{m-1}, r^{m}]$,  defined by his entire history, has the
possibility of jumping to one of a set of possible futures $$\{O[r^1, \dots,
r^{m-1}, r^{m}, r^{m+1}] : r^{m+1} = 0,
\dots, R^{m+1}\}$$ with corresponding probabilities
$p[r^1, \dots, r^{m-1}, r^{m}, r^{m+1}]$.  The possibility $r^{m+1} = 0$
corresponds to extinction.

	(7.10) is identical to (7.9) except that the history of each observer has
been made explicit in the notation.  The next proposal however is not
merely notational.  Suppose that, for $m = 0, 1, \dots, M-1$,
$$\displaylines{
 \sigma_{O[r^1, \dots, r^{m-1}, r^{m}]}\big|_{\B(M)} \hcrh = 
\sum_{r^{m+1} = 0}^{R^{m+1}} p[r^1, \dots, r^{m-1}, r^{m}, r^{m+1}]\,
\sigma_{O[r^1,
\dots, r^{m-1}, r^{m}, r^{m+1}]}\big|_{\B(M)}  \hfill \llap{(7.12)}   }$$
where there exist projections $Q[r^1, \dots, r^{m-1}, r^{m}, r^{m+1}] \in
\B(M)$ such that
$$
\sigma_{O[r^1, \dots, r^{m-1}, r^{m}, r^{m+1}]}(Q[ r^1, \dots, r^{m-1},
r^{m}, s^{m+1}]) = \delta_{r^{m+1} s^{m+1}}.  \eqno{(7.13)}$$

In (7.12) and (7.13), the set of observables $\B(m+1)$ has been replaced
by the set $\B(M)$.  This is an important change, which, in a related form, is
discussed at length in Donald (1992).  All states in the sequence have to be
localized to the same set of observables, if the full influence of earlier
states on later states is to be expressed.  Whether or not is it justifiable to
assume that the states $\sigma_{O[r^1, \dots, r^{m-1}, r^{m}, r^{m+1}]}$ are
decoherent both on $\B(m+1)$ and on $\B(M)$ depends on the precise
definition of these sets and is the reason why the set $\B(W)$ of D2 of the
hypothesis is not taken to be a von Neumann algebra.

In the elementary version of the many-minds interpretation, the extension
from single to multiple observations is made by a replacement of (7.3) by
$$\displaylines{
 \Psi = \sum_{r^1=0}^{\infty}\sum_{r^2=0}^{\infty} \dots
\sum_{r^M=0}^{\infty}
\sqrt{p[r^1] p[r^2]\dots p[r^M]} \hcrh
\psi_{[r^1, r^2, \dots , r^M]} \otimes \varphi_{r^1} \otimes \varphi_{r^2}
\otimes \dots
\otimes \varphi_{r^M} \otimes \chi_{[r^1, r^2, \dots , r^M]},
\hfill \llap{(7.14)}   }$$  where now separate independent Hilbert spaces
$\H_{sys^1}$, $\H_{sys^2}$,
\dots, $\H_{sys^M}$ have been introduced for each successive observation.  
(7.14) is essentially the decomposition of $\Psi$ proposed by Everett in his
treatment of ``memory sequences'', together with an added
result-dependent environmental decoherence term $\chi_{[r^1, r^2, \dots ,
r^M]}$.  This additional term means that (7.4) can be replaced by
$$\displaylines{
 |\Psi\>\<\Psi|\big|_{\B(\H_{O} \otimes \H_{sys^1} \otimes \dots \otimes
\H_{sys^M})
\otimes 1} = \sum_{r^1=0}^{\infty}\sum_{r^2=0}^{\infty} \dots
\sum_{r^M=0}^{\infty} p[r^1] p[r^2]\dots p[r^M] \hcr  |\psi_{[r^1, r^2,
\dots , r^M]}\>\<\psi_{[r^1, r^2, \dots , r^M]}| 
\otimes |\varphi_{r^1}\>\<\varphi_{r^1}| \otimes
|\varphi_{r^2}\>\<\varphi_{r^2}| \otimes
\dots \otimes |\varphi_{r^M}\>\<\varphi_{r^M}|.  \cr
\hfill \llap{(7.15)}   }$$ 

Conceptually, (7.15) is very different from the sequence (7.12).  (7.15)
relies on a set of observables ($\B(\H_{O})$) localized to the present time
with which the observer can describe his current memories of the past,
while in the extension from (7.10) to (7.12) in the theory of this paper, it is
assumed that $\B(1) \subset \B(2) \subset \dots \subset \B(M)$ and that
$\B(m)$ is a set of observables which is localized in the past of the
observer (cf.~part D of the hypothesis).  In many-minds quantum theory,
nothing is definite unless it is part of the structure of an observer.  Thus, in
a theory based on (7.15), an observer's past is not definite, except in as far
as it can be reconstructed from his present time structure.  This is one
source of the problem,  referred to in sections 3 and 4, of identity over
time.  The culmination of the instant-to-instant approach suggested by
(7.15) is the idea (Barbour (1994)) that existence in time is an illusion --
that an observer exists only momentarily and that ``the past'' is only a
representation of instantaneous physical ``memory traces''.  This idea is
rejected in the hypothesis, according to which the past of an observer {\sl
is }part of his structure.  I have underpinned this proposal by arguing (in
Donald (1997)) that the analysis of the relationship between mind and
brain can, at the very least, be considerably simplified if we take our
awareness to be constructed from our past as well as from our present.

As a model for the hypothesis, (7.12) is only a model at the level of the
individual manifestations described in section 4.  For this to be successful
when applied to a switching structure $S(M, N, [d, \varphi])$ modelled by
$O[r^1, \dots, r^{M-1}, r^M]$, we need to invoke assumptions 6.2 and 6.3,
and assume that, for some sufficiently small $\varepsilon > 0$, there is an
$\varepsilon$-manifestation $((\sigma_m)^M_{m=1}, W)$ of $S(M, N, [d,
\varphi])$ such that, for $m = 1, \dots, M$, $\sigma_m$ is given by
$\sigma_{O[r^1, \dots, r^{m-1}, r^m]}$, and $\B(W)$ by $\B(M)$.

If this is possible, then, from definition 6.1, 
$$|\app{}{S(M, N, [d, \varphi])}{\omega}  -
\app{\B(W)}{(\sigma_m)^{M}_{m=1}}{\omega}| \leq 2\varepsilon.
\eqno{(7.16)}$$    If we now assume that we can take $\varepsilon = 0$,
then (7.12), (7.13), (5.3), and (5.9) yield
$$\eqalignno{
 \app{}{S(M, N, [d, \varphi])}{\omega} &=
\app{\B(W)}{(\sigma_m)^{M}_{m=1}}{\omega} 
=\app{\B(M)}{(\sigma_{O[r^1, \dots, r^{m-1}, r^m]})^{M}_{m=1}}{\omega}
 \cr  &= p[r^1, \dots, r^{M-1}, r^M] p[r^1, \dots, r^{M-1}] \dots p[r^1]. 
&{(7.17)} \cr}$$

When we extend this model to the immediate successors of  $S(M, N, [d,
\varphi])$, we will model those successors by $$\{O[r^1, \dots, r^{M-1},
r^{M}, r^{M+1}] : r^{M+1} = 1, \dots, R^{M+1} \}$$ and we will replace
(7.12) and (7.13) with 
$$\sigma_{O[r^1, \dots, r^{m}]}\big|_{\B(M+1)} =  \sum_{r^{m+1} =
0}^{R^{m+1}} p[r^1, \dots, r^{m}, r^{m+1}]\, \sigma_{O[r^1, \dots, r^{m},
r^{m+1}]}\big|_{\B(M+1)}  \eqno{(7.18)}$$
where there exist
projections $Q[r^1, \dots, r^{m-1}, r^{m}, r^{m+1}] \in
\B(M+1)$ such that
$$
\sigma_{O[r^1, \dots, r^{m-1}, r^{m}, r^{m+1}]}(Q[ r^1, \dots, r^{m-1},
r^{m}, s^{m+1}]) = \delta_{r^{m+1} s^{m+1}}.  \eqno{(7.19)}$$ for $m = 0, 1,
\dots, M$.

Using these equations, and assuming that we can continue to set
$\varepsilon = 0$ in expressions analogous to (7.16), the normalization
constant $\xi$ of G8 can, by the same steps that led to (7.17), be calculated
to be
$$\eqalignno{
 \xi &= \sum_{r^{M+1}=1}^{R^{M+1}} p[r^1, \dots, r^{M}, r^{M+1}] p[r^1,
\dots, r^{M}] \dots p[r^1]
\cr &= \sum_{r^{M+1}=1}^{R^{M+1}} p[r^1, \dots, r^{M}, r^{M+1}]
\app{}{S(M, N, [d, \varphi])}{\omega}
\cr &= (1 - p[r^1, \dots, r^{M}, 0]) \app{}{S(M, N, [d, \varphi])}{\omega}.  
\cr}$$ This implies that $\xi \leq \app{}{S(M, N, [d, \varphi])}{\omega}$
and it follows that the probability of moving to successor $O[r^1, \dots,
r^{M-1}, r^{M}, r^{M+1}]$, as given by G8, will be \newline $p[r^1, \dots,
r^{M}, r^{M+1}]$, and the probability of extinction will be $p[r^1, \dots,
r^{M}, 0]$.  These probabilities agree with those given in connection with
(7.10).

This long chain of suppositions shows that the hypothesis is a development
of elementary models like (7.12) and (7.18), and that the probabilities it
produces stem from such elementary models.  However, although it may be
a useful starting point, revealing some of the points at issue and linking
directly to other interpretations, in fact, (7.12) is not a particularly accurate
model of the way in which the hypothesis represents the changing quantum
states of a human brain.  There are two problems with the suggested type
of map $O[r^1, \dots, r^{m-1}, r^m] \rightarrow \sigma_{O[r^1, \dots,
r^{m-1}, r^m]}$ from the set of observers into a set of quantum states.  One
problem, which will be discussed in the next section, is that although such a
map can be part of a good model of the hypothesis, the map concerned may
be effectively many-to-one rather than one-to-one.  The second, and even
more fundamental important problem, is that no explicit definition of any
such map has been suggested, and, in fact, there does not seem to be any
precise choice of state corresponding to a given observer.  This problem is
solved in the hypothesis essentially by calculating probabilities through a
supremum over the entire set of possible observer manifestations.

(7.12) and (7.13) model a theory in which a finite number ($M$) of events
occur, and at event $m$ there are $R^m + 1$ possible distinguishable
outcomes with probabilities which depend on the previous outcomes. 
Combining the whole sequence gives
$$\displaylines{
 \omega\big|_{\B(M)} = 
\sum_{r^M=0}^{R^M} \sum_{r^{M-1}=0}^{R^{M-1}} \dots
\sum_{r^1=0}^{R^1}  p[r^1, \dots, r^{M-1}, r^M] p[r^1, \dots, r^{M-1}] 
\hcrh
\dots p[r^1]
\sigma_{O[r^1, \dots, r^{M-1}, r^M]}\big|_{\B(M)}  \qquad \qquad
\llap{(7.20)}   }$$ and, extending (7.13), we may suppose that there exist
projections \newline $Q[r^1, \dots, r^{M-1}, r^M] \in \B(M)$ such that
$$\sigma_{O[r^1, \dots, r^{M-1}, r^M]}(Q[s^1, \dots, s^{M-1}, s^M]) =
\delta_{r^{1} s^{1}} \dots \delta_{r^{M-1} s^{M-1}} \delta_{r^{M} s^{M}}. 
\hfill
\eqno{(7.21)}$$ 

(7.20) and (7.21) express $\omega$ on $\B(M)$ as a decoherent
decomposition of alternative observer states.  Indeed, in the language of
(5.6), $Q[r^1,
\dots, r^{M-1}, r^M]$ is a decohering projection for $\omega$ on $\B(M)$
which reduces $\omega$ to $\sigma_{O[r^1, \dots, r^{M-1}, r^M]}$ with
probability
$$\eqalignno{
 \omega(Q[r^1, \dots, r^{M-1}, r^M])  &= \app{\B(M)}{\sigma_{O[r^1, \dots,
r^{M-1}, r^M]}}{\omega} \cr &= p[r^1, \dots, r^{M-1}, r^M] p[r^1, \dots,
r^{M-1}] \dots p[r^1]. &{(7.22)} \cr}$$

At the level of the full hypothesis, no such simple decomposition into
different alternatives is assumed -- indeed, the set $\B(W)$ which
corresponds to $\B(M)$ depends on the entire structure of the observer
and on the precise manifestation considered, so that there is no appropriate
single set of observables on which different observer states are defined. 
The full hypothesis also does not identify single projections like $Q[r^1,
\dots, r^{M-1}, r^M]$ which at a stroke reduce $\omega$ to the current
observed state.  (7.20) -- (7.22) are oversimplifications; artifacts of the
model of this section.

Nevertheless, in view of the properties of the function
$\mathop{\rm app}$ mentioned in section 5, it is necessary, if the
hypothesis is to be correct, that, for any individual observer, there should,
for all sufficient small $\varepsilon > 0$, be $\varepsilon$-manifestations
$((\sigma_m)^M_{m=1}, W)$ for which each state $\sigma_{m+1}$ is in
some sense a decoherent part of the prior state $\sigma_{m}$  on $\B(W)$. 
For this, it is sufficient, in the framework of this section, to replace (7.12)
and (7.13) by a set of equations for each $m = 0, \dots, M-1$ and each
$r^1$, \dots,
$r^{m-1}$, $r^{m}$ of the form
$$\displaylines{
 \sigma_{O[r^1, \dots, r^{m-1}, r^{m}]}\big|_{\B(M)}  =  p[r^1, \dots,
r^{m-1}, r^{m}, r^{m+1}]\, \sigma_{O[r^1, \dots, r^{m-1}, r^{m},
r^{m+1}]}\big|_{\B(M)}  \hcrh \hfill  + (1 - p[r^1, \dots, r^{m-1}, r^{m},
r^{m+1}]) \sigma^d_{O[r^1, \dots, r^{m-1}, r^{m}, r^{m+1}]}\big|_{\B(M)}
\qquad \hfill  \llap{(7.23)}   }$$
where there exists a projection $Q[r^1, \dots, r^{m-1}, r^{m}, r^{m+1}] \in
\B(M)$ such that
$$\eqalignno{
 \sigma_{O[r^1, \dots, r^{m-1}, r^{m}, r^{m+1}]}(Q[ r^1, \dots, r^{m-1},
r^{m}, r^{m+1}]) &\sim 1 \cr
\text{and}\quad  \sigma^d_{O[r^1, \dots, r^{m-1}, r^{m}, r^{m+1}]}(Q[ r^1,
\dots, r^{m-1}, r^{m}, r^{m+1}]) &\sim 0. \qquad &{(7.24)}  \cr}$$ 

(7.23), (7.24), and (5.4) yield
$$\displaylines{
\qquad \app{\B(M)}{\sigma_{O[r^1, \dots, r^{m-1}, r^{m},
r^{m+1}]}}{\sigma_{O[r^1,
\dots, r^{m-1}, r^{m}]}}
\hcrh \hfill \qquad \quad
\sim \sigma_{O[r^1, \dots, r^{m-1}, r^{m}]}(Q[r^1, \dots, r^{m-1}, r^{m},
r^{m+1}]) \hcrh
\sim p[r^1, \dots, r^{m-1}, r^{m}, r^{m+1}].  \hfill \llap{(7.25)} }$$

(7.23) and (7.24) differ from (7.12) and (7.13) not only in that (7.24) is 
approximate, but also in that the state $\sigma^d_{O[r^1, \dots, r^{m-1},
r^{m}, r^{m+1}]}$ which expresses \newline $O[r^1, \dots, r^{m-1}, r^{m}]$
not becoming $O[r^1, \dots, r^{m-1}, r^{m}, r^{m+1}]$, is not required to be
decomposable into distinguishable and meaningful alternatives.  Only in
estimating probabilities will it be necessary to consider more than one
alternative simultaneously, but even for this, as we shall see in the next
section, only a limited range of alternatives will be relevant at any moment.

The step-by-step, approximate decoherence of (7.23) and (7.24) is much
more plausible than (7.20) -- (7.22); indeed it is no more than the 
approximate decoherence ubiquitous in localized thermal macroscopic
systems.  Nevertheless, it would be absurd to claim that the mere
invokation of decoherence, and the replacement of ``quantum probabilities''
by ``classical probabilities'' is, by itself, the solution to all the problems of
quantum theory.  The crucial difficulty is not to find structures to which
appropriate ``classical probabilities'' can be assigned, but to decide which of
many such structures are significant and to find a way of precisely
characterizing those structures.

Consider, for example, the theory of consistent histories.  There are several
similarities between that theory and the hypothesis of this paper.   Both
theories present abstract structures which are purported to be sufficient to
characterize a ``classical'', ``quasi-classical'', or ``observed'' world.  Both
theories aim to understand quantum uncertainties in terms of classical
probability theory.  Both theories are developments from conventional
quantum mechanics and are built on sets of projections, thought of as
``yes-no'' questions.  In the hypothesis, these projections are defined in
part F.   In consistent histories, consistent families of projections
$(P^1_{r^1})_{r^1=0}^{R^1}$,
\dots, $(P^M_{r^M})_{r^M=0}^{R^M}$ on $\omega$ are considered.  Such
families are required to satisfy some version of the consistency conditions
$$\displaylines{
\omega(P^1_{r^{1}} \dots P^{M-1}_{r^{M-1}} P^{M}_{r^{M}} P^M_{s^M}
P^{M-1}_{s^{M-1}} \dots  P^1_{s^{1}})  \hcrh = \delta_{r^1, s^1}
\dots \delta_{r^{M-1}, s^{M-1}} \delta_{r^M, s^M} \  
 \omega(P^1_{r^{1}} \dots P^{M-1}_{r^{M-1}} P^{M}_{r^{M}} 
P^{M-1}_{r^{M-1}} \dots  P^1_{r^{1}}). \hfill \llap{(7.26)}   }$$
 
However, in my opinion, consistent history theorists have been far too
easily satisfied with the consistency conditions.  The core problem with the
theory is to explain why any particular set of histories from the continuum
which satisfy (7.26) should be physically natural, or important, or should
apply to us as observers.  The consistency conditions make it
straightforward to define sets of numbers which satisfy the axioms of
classical probability theory, but without some fundamental set of histories,
it is not clear what those numbers mean.  Probabilities can be defined only
after the fundamental entities have been identified.  Griffiths
(1998) attempts to evade this problem by assuming that the observer
is not part of the physical system considered.  Under this
assumption, consistent histories is merely a theory of experimental
observations; telling us nothing about ontology.  In particular, the
problem of understanding the nature of observers is entirely
unaddressed.

In the hypothesis, the fundamental entities are defined as information
processing structures.  Then the continuous variations in the possible
physical manifestations of such structures can be dealt with by dealing with
all possibilities simultaneously, and basing the theory not on individual
possibilities, but on equivalence classes of them.  

\proclaim{8.  A model for private probabilities.}
\endproclaim

In section 6, two theses concerning the relationship between the
probabilities of the hypothesis and those of conventional quantum theory
were introduced and discussed.  It was argued that a modern human
observer, who is ``typical'' in the probability of the hypothesis, should be
aware of a world in which quantum theory is accepted.  In this section, we
shall turn to thesis 6.6, and consider a model for the immediate
observation of an individual event. 

In section 7, it was shown that the hypothesis is a development of
elementary models of state change in observation, like (7.12) and (7.18),
and that it produces satisfactory probabilities for those models.  This is a
preliminary step towards thesis 6.6.  However, it is only preliminary,
because, in fact, the models of section 7 are not adequate as representations
of observation at the neural level.  In this section, a more sophisticated
model will be presented.  This will be a model for a brain acquiring possible
information over a time period short in conventional terms but long
enough to contain many determinations of switch status -- a period long
enough for a single glance, for example.  The goal, in this model, is to argue
for a more sophisticated version of a relationship like (5.4) or (7.25)
between a fundamental (hypothesis-defined) probability and the
``expected value'' of a suitable projection.   This will provide short-term
agreement between the probabilities of the hypothesis and those of
conventional theory.  By taking the long term as a developing succession of
such ``glances'', this agreement can be extended to arbitrary time intervals.

We shall suppose that at some moment an observer predicts, in the sense of
definition 6.4, that the state on some set of observables $\B$ is $\rho$, and
that $\B$ contains  projections $P^{\B}_a$ and $P^{\B}_b$ with $P^{\B}_a +
P^{\B}_b = 1$ such that, according to conventional theory, the outcome of
some observation on $\B$ will be $a$ with probability $\rho(P^{\B}_a)$ and
$b$ with probability $\rho(P^{\B}_b)$.  The aim is to link these numbers
with probabilities defined by the stochastic process of the hypothesis.  We
shall begin by analysing the build-up of information within the brain of the
observer.  This will uncover many complexities.  One result of those
complexities is that any outcome will be be observed in many different
ways -- by many different futures of the observer.  However, the analysis
will indicate that quantum probabilities can be separated from geometrical
and combinatorial complexities.  This will make it reasonable to assume that
the complexities will not bias the outcome.

There are two reasons why single step probabilities in the sto\-chastic
process defined by the hypothesis cannot, in general, be directly equated
with the probabilities assigned to experimental results in elementary
quantum mechanics.  One has to do with the number of distinct successors
at each step, and the other with the amount of information recorded in a
single step.  

There is a radical difference in the counting of successors with the model of
(7.18) and with the application of the hypothesis to more realistic neural
models.  Underlying (7.18) is the elementary assumption that each distinct
successor corresponds to a distinct orthogonal projection.  But, according to
the full version of the hypothesis, successors can differ by only seemingly
inconsequential alterations in the precise geometrical pattern of
determinations of switch status and this may have essentially negligible
effect on the corresponding quantum states.  I shall refer to such
differences as being ``minor''.  They arise because of the abstract way in
which information is defined in the hypothesis.  For example, according to
the hypothesis, two switching structures may differ by only the
time-ordering of a single pair of determinations.  It was proposed in
Donald (1990, 1995), that a status determination corresponds a snapshot of
some property of neural membrane which is linked to neural firing and in
Donald (1995), it was estimated that a human brain might have a switching
rate (roughly equivalent to a rate of status determination) as high as
$10^{15}$ switchings per second.  Then we might consider two
situations such that in situation 1, one knows part of neuron A is
firing at spacetime point $(t_1, {\bf x}_A)$ and part of neuron B is
firing at space-time point $(t_2, {\bf x}_B)$, while in situation 2,
part of neuron A is firing at $(t_2, {\bf x}_A)$ and part of neuron
B is firing at $(t_2, {\bf x}_B)$.  If $c^2(t_1 - t_2)^2 - ({\bf
x}_A - {\bf x}_B)^2 > 0$, we shall have different casual information
in these two situations, but that difference will be of negligible
neurophysiological relevance if $t_1 - t_2 < 10^{-4}$s. 

It will be argued in this section that the existence of a multitude of
successors which vary only by minor differences has two important positive
effects.  One is that the ``probability of extinction'' (defined in G8) can be
expected to be zero during the normal functioning of a human brain
because the inevitable loss of a priori probability due to ``collapse'' to any
specific future switching structure will be outweighed by the existence of a
large number of similar structures.  The other positive effect is that the
concentration of the possibility of minor differences to time intervals
during which many switch status determinations have already been made
allows the existence of a ``present moment'' for an observer to be explained;
despite the fact that the hypothesis allows new determinations to be
included at essentially arbitrary times.

In Donald (1992), another model (``postulate nine'' of that paper) was
presented.  In that model, a priori probabilities were defined using
observables beyond those local to the brain of the observer (cf. Definition
6.4), and it was supposed that when an observer observed, at a
macroscopic level, the outcome of an experiment, that observation would
take place within a single step from (in the notation of section 7), say
$O[r^1, \dots, r^{M-1}, r^{M}]$ to $O[r^1, \dots, r^{M-1}, r^{M}, r^{M+1}]$. 
For a more adequate model, in which attention is restricted to observables
within the brain, we shall have to take account of the fact that only a tiny
amount of information is expressed by each determined switch status in
the brain.  The brain is a very noisy system, in which information is
accumulated in small pieces scattered over many parallel channels. 
Although, because of the number of different channels, the time required
for sufficient accumulation to distinguish between different outcomes may
be quite short, inevitably there will be a background of many essentially
simultaneous determinations which are irrelevant to the particular
observation being studied.

We shall continue to use assumptions 6.2 and 6.3.  Indeed given a switching
structure $S(M, N,  [d, \varphi])$, we shall denote by 
$((\sigma^{\varepsilon}_m)^M_{m=1}, W^{\varepsilon})$ an
$\varepsilon$-manifestation for which $\varepsilon$ is sufficiently small to
have the properties required by those assumptions.  We shall assume a
strong form of persistency (assumption 6.2) by assuming that if
$((\sigma^{\varepsilon}_m)^M_{m=1}, W^{\varepsilon})$ is an
$\varepsilon$-manifestation of $S(M, N,  [d, \varphi])$, then, for any
successor $S(\hat M, \hat N,  [\hat d,
\hat \varphi])$ which is relevant to our calculations, we can (ignoring for
simplicity the re-orderings allowed by B3) find an
$\varepsilon$-manifestation of the form $((\hat
\sigma^{\varepsilon}_m)^{\hat M}_{m=1}, \hat W^{\varepsilon})$ such
that $\hat \sigma^{\varepsilon}_m = \sigma^{\varepsilon}_m$ for $m = 1,
\dots, M$.
 
The stochastic process is built on the elementary step from a structure
\newline $S(M, N,  [d, \varphi])$ to an immediate successor  $S(M', N',  [d',
\varphi'])$.  To calculate the probability of this step explicitly, one would
need to know the normalization constant
$$\displaylines{
 \xi(S(M, N,  [d, \varphi])) = \sum \{\app{}{S(M', N', [d',
\varphi'])}{\omega}: \hcrh S(M', N', [d',
\varphi']) \in
\Xi(M, N,  d, \varphi)\} \qquad \qquad \llap{(8.1)}  }$$  of G8, and so one
would need to be able to list all the immediate successors -- the elements
of $\Xi(M, N,  d, \varphi)$.  If the hypothesis is to be successful, then $\xi$
should be closely approximated by the sum of the a priori probabilities of
successors which, for all sufficiently small $\varepsilon$, have
$\varepsilon$-manifestation brain states $\sigma'{}^{\varepsilon}_{M'}$
which can be analysed in classical terms as processing information which is
a meaningful continuation of the information in
$\varepsilon$-manifestation brain states $\sigma^{\varepsilon}_M$ of
$S(M, N,  [d, \varphi])$.  It is at this point that it is important to be able to
rule out the possibility discussed in section 3 of the future of a human
brain being ``dominated by the large number of possible ways in which
small numbers of new short-term artificial switches could arise''.  Then we
have to consider what sort of classically meaningful continuations arise. 
Here we turn to conventional neurophysiology.  

Suppose, as discussed in Donald (1990, 1995), that switchings correspond
to changes, linked to neural firing, in some property of neural membrane. 
A fully functioning brain state $\sigma^{\varepsilon}_M$ will have a very
large number of possible meaningful continuations.  By part B of the
hypothesis, any continuation will correspond either to a new switch or to a
single new determination of the status of an existing switch.  For simplicity,
and without essential loss of generality, I shall mainly discuss the latter
case, involving the determination at some moment of the status of a patch
of neural membrane.  Where that patch has been, and also where it will be
in the short term future, is largely determined, through part E of the
hypothesis, by the geometrical structure in $W^{\varepsilon}$ and the
quasi-classical neural structure represented by $\sigma_M^{\varepsilon}$. 
Most of the possible variation, for a given switch, therefore lies in the
moment at which the new status is specified, and, of course, in the actual
status.

The hypothesis defines a finite structure.  This implies that there are only a
finite number of possible distinguishable ``moments'' (or equivalence
classes of moments) at which a new switch status can be determined. 
However, the docket $d$ of $S(M, N,  [d, \varphi])$ carries information
which is sufficient to determine a very fine division of times up to the
present moment in the brain.  A different docket is defined whenever the
causal relations between any pair of switch determinations is changed.  A
single new determination during a period when there are $10^{15}$
switchings per second, defines a different docket
$d'$ for every change of about $10^{-15}$ seconds in the instant of that
new determination.  On the other hand, there will only be one docket
$d'$ for a new determination (on a given switch) in the strict causal future
of all the determinations in $d$.

As far as the classical meaning of the firing pattern of a brain is concerned,
a change of say $10^{-5}$ seconds in the instant of a single determination
would clearly be a ``minor'' difference.  As neural firing changes only on a
millisecond timescale, it would also be ``minor'' in relation to the local
quantum state.  Thus many terms in (8.1) of equivalent meaning will
correspond to a new determination during a period when many other
determinations have already been made.  Considered together, this large
set of terms will have far higher probability than the comparatively far
smaller set corresponding to a new determination during a period with few
other determinations.  However, C13 of the hypothesis acts as a constraint
on this process, by requiring that a switch cannot repeat status more
rapidly than it has already changed status.  This constraint means that we
cannot continue indefinitely to add new determinations within a given time
period.  Taken together, these imply that the past of a switch will tend to
``fill'' with determinations until no more can be added because of C13.  As a
result, it is reasonable to assume that a developing structure will have a
fairly precisely defined future edge or ``present moment'' and that there
will be a strong tendency for ensuing determinations to be made close to
that future edge.  This effect will be enhanced by the fact that the more
determinations are made within a given spatial and temporal locality
within the brain, the more the status of any new determination within that
locality will tend to be fixed by what has already been determined.  A
status which is compatible with many neighbouring determinations will
have higher a priori probability than a determination which probes the
undecided.

We shall turn now to consider in detail the observation of some stochastic
outcome; in particular, the observation on a set $\B$ of either $a$ or $b$. 
We are concerned with the initial acquisition by an observer of sufficient
information to determine which outcome will be seen.  That initial
acquisition may require many determinations, but will often appear to take
place very quickly -- in the first glance -- at the level of ``pre-conscious''
processing.  Guessing at some numbers just for the sake of argument, it is, I
think, not unreasonable, given a switching rate of $10^{15}$Hz, to suppose a
``present moment'' defined to within a millisecond, which is certainly
shorter than the ``psychological moment'' -- and then we might consider
information from determinations made within a millisecond, scattered over
10$^5$ neurons (0.1\% of retinal cells), during which perhaps 10$^{12}$
irrelevant determinations would also be made elsewhere in the brain. 

For a model in which the probability of observing of a given outcome can
be estimated, consider first a single fixed sequence of switching structures 
\newline $(S(M, N_M, [d_M, \varphi_M]))_{M = M_0}^{M_S}$ which is
sufficiently long that, by its conclusion, a definite outcome will be known. 
Suppose that that outcome is $a$, which happens with conventional
probability
$\rho(P^{\B}_a)$.   When appropriate,
$((\sigma^{\varepsilon}_m)^{M}_{m=1}, W_{M}^{\varepsilon})$, will denote
an $\varepsilon$-manifestation for $S(M, N_M, [d_M,
\varphi_M])$.  (The persistency assumption is implicit here, in that we do
not write $\sigma^{\varepsilon}_{m, M}$.)

We shall suppose that in this sequence, relevant determinations have been
made at instants $M_s$ for $s = 1, \dots, S$, and that the other
determinations are irrrelevant.   In order to model the build-up of
information to the point at which the observed outcome is definite, we
shall begin by modelling each separate relevant determination using (5.2)
and (5.3).  This requires the existence of projections which, in the sense of
(5.6), are ``decohering'' on to the determined statuses.  

We can expect to be able to find projections to characterize particular
events in any sufficiently large quantum system, because, for example,
such events will be characterized by the values of certain observables, and
we can then specify those values using the projections provided by the
spectral theorem. The present situation will certainly be sufficiently large
in this sense.  The switches discussed in Donald (1990, 1995) have length
scales of order $10^{-9}$m or $10^{-8}$m.  Even a region as small as
$(10^{-9}$m$)^3$ can have of order $100$ thermally-active degrees of
freedom in the warm dense environment of the living brain.  The switch
size is chosen to allow the existence of the projections $P_n$ and $Q_n$ of
C14 which discriminate switch status, but it will be possible to find
projections in  $\B(W_{M_s}^{\varepsilon})$ which carry more information
about the local state of the switch than do those projections, because
$\B(W_{M_s}^{\varepsilon})$ is much larger than the algebra to which
$P_n$ and $Q_n$ are required to belong, and because $P_n$ and $Q_n$ are
required to define properties repeated over the whole lifetime of the
switch.

Suitable interactions with the environment of a system are sufficient to
make a given projection decohering.  In the present situation, the set
$\B(W_{M_s}^\varepsilon)$ specifies states only in limited substructures of
the brain, leaving plenty of local environment into which any coherence will
rapidly disappear.  Moreover, switch status represents local neural firing
status and neural firing works as a method of communicating information
because it is a macroscopic, thermally irreversible, ``all-or-nothing''
process which involves transition from one
macro\-scopically-distin\-guishable metastable status of the local neural
membrane to another.  Because of the metastability, status determination
can be expected to have the property assumed in conventional descriptions
of quantum measurement that the determined status will be ``amplified''
across a wider region.  Such amplification into the environment is more
than enough for the required decoherence.  

This means that it is reasonable to suppose that, for each relevant
determination, we can find a projection $P^s \in
\B(W_{M_s}^{\varepsilon})$, which, in the language of (5.6), is decohering
for $\sigma^{\varepsilon}_{M_s - 1}$ on
$\B(W_{M_s}^{\varepsilon})$, and reduces $\sigma^{\varepsilon}_{M_s -
1}$ to $\sigma^{\varepsilon}_{M_s}$. We shall suppose that $B P^s, P^s B
\in \B(W_{M_s}^\varepsilon)$ for all 
$B \in \B(W_{M_s}^\varepsilon)$, and that the constraints imposed by part
F of the hypothesis can be modelled by supposing that in becoming aware
of the new determination, the state of the observer changes from
$\sigma^{\varepsilon}_{M_s - 1}$ to $\sigma^{\varepsilon}_{M_s}$ where
$$\sigma^{\varepsilon}_{M_s} = P^s \sigma^{\varepsilon}_{M_s - 1} P^s/
\sigma^{\varepsilon}_{M_s -1}(P^s).
\eqno{(8.2)}$$ We shall also suppose that, for $s = 1, \dots, S$, 
$$\sigma^{\varepsilon}_{M_s -1}(P^s B) = \sigma^{\varepsilon}_{M_s -1}(B
P^s)  \quad
\text{for all $B \in \B(W_{M_s}^\varepsilon)$}.
\eqno{(8.3)}$$  

Even if the assumption of many simultaneous parallel channels is inaccurate
for some particular type of observation, the macroscopic and thermal
nature of the brain will make it reasonable to assume decoherence on a
given switch between one determination and the next.  With the same
justification, we shall also assume that, for $s = 1, \dots, S$, the projections
$P^s$ all commute.  

For $s = 1, \dots, S$, set $R_s = P^1 \cdot P^2 \cdots P^s$.  The
commutativity of the $P^s$ implies that the $R_s$ are projections and are
independent of the ordering of their components.

(8.2) and (8.3) imply that, for $B \in \B(W_{M_s}^\varepsilon)$
$$\eqalignno{
\sigma^{\varepsilon}_{M_s -1}(B) &= \sigma^{\varepsilon}_{M_s -1}(P^s
B) +
\sigma^{\varepsilon}_{M_s-1}((1-P^s) B) \cr  &=
\sigma^{\varepsilon}_{M_s -1}(P^s B P^s) +
\sigma^{\varepsilon}_{M_s -1}((1-P^s) B (1-P^s)) \cr  &=
\sigma^{\varepsilon}_{M_s-1}(P^s)\sigma^{\varepsilon}_{M_s}(B) +
\sigma^{\varepsilon}_{M_s -1}((1-P^s) B (1-P^s)).
\cr}$$

By (5.3), this implies that 
$$\app{\B(W_{M_s}^\varepsilon)}{\sigma^{\varepsilon}_{M_s}}{
\sigma^{\varepsilon}_{M_s -1}} =
\sigma^{\varepsilon}_{M_s -1}(P^s).
\eqno{(8.4)}$$

A significant aspect of (8.4) is that the projection $P^s$ is localized to the
region $s$.  Because of this, it is not unreasonable to assume that the
expected value $\sigma^{\varepsilon}_{M_s -1}(P^s)$ does not depend on
the irrelevant determinations.  

The state at the beginning of the ``glance'' is
$\sigma^{\varepsilon}_{M_0}$.  Denote this state by $\rho$.  Set $\rho =
\rho_0$ and, mimicking (8.2), define $$\rho_s = R_s
\rho R_s/ \rho(R_s), \eqno{(8.5)}$$ for $s = 1, \dots, S$.  In (8.5), changes in 
$\sigma^{\varepsilon}_{M_0}$ due to irrelevant determinations are
omitted and only the local changes defined by (8.2) are kept.

The commutativity of the $P^s$ implies that
$$\rho(R_s) = \rho(R_{s-1} P^s) = \rho(R_{s-1} P^s R_{s-1}) =
\rho_{s-1}(P^s) \rho(R_{s-1})$$ and hence, by induction, that 
$$\rho(R_s) = \rho_{s-1}(P^s) \rho_{s-2}(P^{s-1}) \cdots \rho_0(P^1).
\eqno{(8.6)}$$

Now we shall use the locality of the $P^s$ to justify replacing
$\sigma^{\varepsilon}_{M_s -1}(P^s)$ in (8.4) by
$\rho_{s-1}(P^s)$.  Making similar replacements for $i = 1, \dots, s$ in (8.6),
allows us to model the situation by supposing that
$$\displaylines{
\sigma^{\varepsilon}_{M_0}(R_s) = \rho(R_s)  = \sigma^{\varepsilon}_{M_s
-1}(P^s) \sigma^{\varepsilon}_{M_{s-1} -1}(P^{s-1}) \cdots
\sigma^{\varepsilon}_{M_1 -1}(P^1) \hcr =
\app{\B(W_{M_s}^\varepsilon)}{\sigma^{\varepsilon}_{M_s}}{
\sigma^{\varepsilon}_{M_s -1}}
\app{\B(W_{M_{s-1}}^\varepsilon)}{\sigma^{\varepsilon}_{M_{s-1}}}{
\sigma^{\varepsilon}_{M_{s-1} -1}} \cdots
\app{\B(W_{M_1}^\varepsilon)}{\sigma^{\varepsilon}_{M_1}}{
\sigma^{\varepsilon}_{M_0}}.
\cr}$$

Using the  localization and decoherence of determinations yet again, allows
us to change the moment at which the set of observables
$\B(W_{M})$ is defined.  In particular, we may suppose that, for
 $M = M_s, M_s +1, \dots, M_{s+1} - 1$,
$$\displaylines{
\sigma^{\varepsilon}_{M_0}(R_s) =
\app{\B(W_M^\varepsilon)}{\sigma^{\varepsilon}_{M_s}}{
\sigma^{\varepsilon}_{M_s -1}}
\app{\B(W_M^\varepsilon)}{\sigma^{\varepsilon}_{M_{s-1}}}{
\sigma^{\varepsilon}_{M_{s-1} -1}}
\hcrh \cdots
\app{\B(W_M^\varepsilon)}{\sigma^{\varepsilon}_{M_1}}{
\sigma^{\varepsilon}_{M_0}}. \qquad \qquad
\llap{(8.7)} }$$

(8.7) is the conclusion of the first step towards the goal of this section.  It
equates a product of a priori probabilities to the expected value of a
composite projection.  The product is a factor in the a priori probability of
the $\varepsilon$-manifestation $((\sigma^{\varepsilon}_m)^{M}_{m=1},
W_{M}^{\varepsilon})$ of the switching structure $S(M, N_M, [d_M,
\varphi_M])$ for $M = M_s, M_s +1, \dots, M_{s+1} - 1$.  All the other
terms in the a priori probability are independent of the outcome being
observed.  Thus, setting $\varepsilon = 0$,
$$\app{}{S(M, N_M, [d_M, \varphi_M])}{\omega}  =
\sigma^{\varepsilon}_{M_0}(R_s) w^{nd}_1(a, S(M, N_M, [d_M,
\varphi_M]), \omega) \eqno{(8.8)} $$
where $w^{nd}_1$ is some factor which does not depend on
the conventional quantum probability ($\rho(P^{\B}_a)$) of the outcome
($a$) which is observed, although it may depend on the nature of the
outcome; for example, it will depend on the number ($s$) of determinations
already made, and on the length of time over which those determinations
have been made. 

In this situation, we return to the calculation of the normalization constant
$\xi(S(M, N_M, [d_M, \varphi_M]))$ defined by (8.1).  For $M = M_s, M_s +1,
\dots, M_{s+1} - 1$, there are two types of term in the sum.  One type
involves successors $S(M', N', [d', \varphi'])$ in which the new
determinations are irrelevant to the observed outcome.  For these terms an
approximate result of the form
$$\app{}{S(M', N', [d', \varphi'])}{\omega} =
\sigma^{\varepsilon}_{M_0}(R_s) w^{nd}_1(a, S(M', N', [d',\varphi']),
\omega)$$ will continue to hold.

There will also be successors $S(M', N', [d', \varphi'])$ in which the new
determinations are relevant.  Now $\app{}{S(M', N', [d',
\varphi'])}{\omega}$ will have a factor like, for example,
$\sigma^{\varepsilon}_{M_0}(P^{s+1})$ and will not be of the same form. 
However, in the sum \newline $\xi(S(M, N_M, [d_M, \varphi_M]))$ there
will be far more terms of the first type than of the second (for example, a
factor of $10^7$ more with the numbers guessed at above).  This means
that terms of the second type can essentially be ignored, and that it is
reasonable to write
$$\xi(S(M, N_M, [d_M, \varphi_M])) = \sigma^{\varepsilon}_{M_0}(R_s) 
w^{nsd}_2(a, S(M, N_M, [d_M, \varphi_M]), \omega)
\eqno{(8.9)}$$  where $w^{nsd}_2$ is some factor which is not strongly
dependent on the conventional probability.

In the ordinary operation of the brain, the total number of terms in the
sum (8.1) is huge.  The suggested number of determinations made within a
millisecond, already shows this, and in calculating (8.1), one would also
have to take account of the number of minor differences in the way in
which any given determination might arise.  With (8.4) indicating that, for
$S(M', N',  [d', \varphi'])$ an immediate successor of $S(M, N,  [d, \varphi])$,
the ratio
$$\app{}{S(M', N', [d', \varphi'])}{\omega}/\app{}{S(M, N, [d,
\varphi])}{\omega}$$  can be approximated by the conventional probability
of some possible impending neural event, this huge number of terms will
ensure that, except in exceptional circumstances, we shall have
$$\xi(S(M, N,  [d, \varphi])) \geq \app{}{S(M, N, [d, \varphi])}{\omega}
\eqno{(8.10)}$$  with the consequence, according to G8, that there will be
no possibility of extinction.

The sequence $(S(M, N_M, [d_M, \varphi_M]))_{M = M_0}^{M_S}$ is one
possible path in the stochastic process which leads from the original
switching structure to a structure in which a given outcome is known.  The
probability of that path is a product of the probabilities of the individual
steps given by G8:
$$\eqalignno{
 \Pr((&S(M, N_M, [d_M, \varphi_M]))_{M = M_0}^{M_S} | S(M_0, N_{M_0},
[d_{M_0}, \varphi_{M_0}]), \omega) \cr &= \prod_{M = M_0}^{M_S -1}
\app{}{S(M + 1, N_{M+1}, [d_{M+1},
\varphi_{M+1}])}{\omega} / \xi(S(M, N_M, [d_M, \varphi_M])) \cr &=
\sigma^{\varepsilon}_{M_0}(R_S) w^{nsd}_3(a, (S(M, N_M, [d_M,
\varphi_M]))_{M = M_0}^{M_S}, \omega) &{(8.11)}
\cr
& &\hbox{ by induction using (8.8) and (8.9).} \cr}$$ 

According to (8.11), the probability of an individual path is given by the
expectation in the initial state $\sigma^{\varepsilon}_{M_0} = \rho$ of a
compound projection $R_S$ multiplied by a complicated geometrical and
combinatorial factor.  $\rho(R_S)$ is not necessarily equal to
$\rho(P^{\B}_a)$.  However, the projection $R_S$ does express sufficient
information within the brain to determine the outcome of the observation. 
This means that, at least as far as the state $\rho$ is concerned, $R_S$
should be a subprojection of $P^{\B}_a$, or, in mathematical terms, that
$\rho(P^{\B}_a R_S B) =\rho(R_S B)$ for all  $B$ in some set sufficiently
large to contain all the observables relevant to the present situation, for
example,
 $\B(W_M^\varepsilon) \cup \{ A C : A \in \B, C \in {\cal
C}(W_M^\varepsilon) \}$ (cf.~Definition 6.4). 

On such a set of observables, $R_S$ differs from $P^{\B}_a$ in that it fixes
not only the outcome of the observation, but also the precise neural
processing by which that observation is made.  Independence between the
internal processing and the cause of the external event makes it possible to
write 
$$\rho(R_S) = \rho(P^{\B}_a) w^{nd}_4(a, (S(M, N_M, [d_M,
\varphi_M]))_{M = M_0}^{M_S}, \omega) \eqno{(8.12)}$$ where, once
again, $w^{nd}_4$ is a factor which, while it may depend on the nature of
the outcome, does not depend on its conventional quantum probability.

Substituting (8.12) into (8.11) gives
$$\displaylines{ \qquad \quad
\Pr((S(M, N_M, [d_M, \varphi_M]))_{M = M_0}^{M_S} | S(M_0, N_{M_0},
[d_{M_0}, \varphi_{M_0}]), \omega) \hcrh = \rho(P^{\B}_a) w^{nsd}_5(a,
(S(M, N_M, [d_M, \varphi_M]))_{M = M_0}^{M_S}, \omega). \hfill
\llap{(8.13)} }$$
\medskip

There are many many paths like $(S(M, N_M, [d_M, \varphi_M]))_{M =
M_0}^{M_S}$ which lead to structures in which outcome $a$ is known.  To
calculate the total probability \newline $\Pr(a| S(M_0, N_{M_0}, [d_{M_0},
\varphi_{M_0}]), \omega)$ of outcome $a$ given the initial structure, we
must sum over all such distinct paths.  Following (8.13), the result will take
the form
$$ \Pr(a | S(M_0, N_{M_0}, [d_{M_0}, \varphi_{M_0}]), \omega) =
\rho(P^{\B}_a) w_6(a,  S(M_0, N_{M_0}, [d_{M_0}, \varphi_{M_0}]),
\omega). \eqno{(8.14)}$$
This states that the total probability of outcome $a$ is
equal to the conventional probability of that outcome multiplied by some
factor $w_6$  which depends on the number of ways in which that $a$ can
be observed.  $w_6$ combines many terms, none of which is strongly
dependent on the conventional probability of $a$, but this does not allow us
to conclude that the combination also does not depend strongly on the
conventional probability.  Here is a caricature of the situation:

\proclaim{example 8.15}{}

\noindent A) \quad Consider a discrete Markov chain $(X_n)_{n\geq0}$
with three states $a$, $b$, and $o$.  $X_0 = o$.  $a$ and $b$ are sink states on
which the process terminates.  At each $n$, the probability of passing to $a$
(respectively $b$) is proportional to $p$ (resp.~$q$).  The probability of
staying at $o$ is proportional to $x$.  The constant of proportionality is
determined by normalizing the probabilities.

$a$ and $b$ model the outcomes of the observation.  Let $F_n(a)$ (resp.~
$F_n(b)$) be the probability of the process terminating at the
$n^{th}$ step at $a$ (resp.~$b$).   We are interested in calculating the net
probability $F(a) = \sum_{n=1}^\infty F_n(a)$ (resp.~$F(b) =
\sum_{n=1}^\infty F_n(b)$) of the process terminating at $a$ (resp.~$b$).

Staying at $o$ is a caricature of making an irrelevant determination, so we
shall suppose that $x \gg p, q$.
$$F_n(a) = {p \over p + q + x} \left({x \over p + q + x}\right)^{n-1},$$ so
that $F_n(a) \sim p/x$ for $x/(p + q) \rightarrow \infty$.  (This caricatures
(8.11).)

$$F(a) =  \sum_{n=1}^\infty {p \over p + q + x} \left({x \over p + q +
x}\right)^{n-1} = {p \over p + q}.$$

If $p$ corresponds to $\rho(P^{\B}_a)$ and $q$ to $\rho(P^{\B}_b)$ then we
would have $p + q = 1$, and it would be the case that $F(a)$ would be
proportional to $p$ with a constant of proportionality ($1$) independent of
$p$.  We would also have $F(a)/F(b) = p/q$.

A variation in the example, however, shows that this is, in general, too
simple a conclusion.
\smallskip

\noindent B) \quad Suppose that there are two sink states $a_1$ and $a_2$
which model distinct ways in which the outcome $a$ might be observed. 
Suppose that the probability of passing to either of these states is
proportional to $p$ (with the same constant of proportionality)

Now we have $F(a) = \dsize{2p \over 2p + q}$.
In this case,
$F(a)/p = \dsize{2 \over p + 1}$ is strongly dependent on $p$ and 
$F(a)/F(b) = 2 p / q$.
\smallskip

\noindent C) \quad For a slightly more realistic model, suppose that there
are many distinct states $a_i$, $i = 1, \dots, N_a$, corresponding to outcome
$a$ and many distinct states $b_j$, $j = 1, \dots, N_b$, corresponding to
outcome
$b$, and suppose that different weights are allowed for each state, so that
the probability of passing from $o$ to $a_i$ (resp.~to $b_j$, to $o$) is
$w(a_i) p/W$ (resp.~$w(b_j) q/W$, $x/W$) where $W = \sum_i w(a_i) p +
\sum_j w(b_j) q  + x$.

This gives $F(a)/p = \dsize{W_a \over W_a p + W_b q}$ and
$F(a)/F(b) = W_a p / W_b q$ where $W_a = \sum_i w(a_i)$ and $W_b =
\sum_j w(b_j)$.
\hfill $\blacksquare$
\endproclaim
\bigskip

Returning to (8.14), we have
$${ \Pr(a | S(M_0, N_{M_0}, [d_{M_0}, \varphi_{M_0}]),
\omega) \over \Pr(b | S(M_0, N_{M_0}, [d_{M_0}, \varphi_{M_0}]),
\omega)}  =  {\rho(P^{\B}_a) w_6(a,  S(M_0, N_{M_0}, [d_{M_0},
\varphi_{M_0}]), \omega) \over
\rho(P^{\B}_b) w_6(b,  S(M_0, N_{M_0}, [d_{M_0}, \varphi_{M_0}]),
\omega)}.$$

We wish to argue that
$$\displaylines{ { \Pr(a | S(M_0, N_{M_0}, [d_{M_0}, \varphi_{M_0}]),
\omega) \over \Pr(b | S(M_0, N_{M_0}, [d_{M_0}, \varphi_{M_0}]),
\omega)} =  {\rho(P^{\B}_a) \over
\rho(P^{\B}_b)} }$$ and this of course would follow if
$$w_6(a,  S(M_0, N_{M_0}, [d_{M_0}, \varphi_{M_0}]), \omega)  =
w_6(b,  S(M_0, N_{M_0}, [d_{M_0}, \varphi_{M_0}]), \omega). 
\eqno{(8.16)}$$

(8.16) will hold if there is a correspondence between the ways in which
outcome $a$ can be observed, and the ways in which outcome $b$ can be
observed.  For example, a sufficient condition for (8.16) is that there should
be a bijection $\lambda$ between sequences  $(S(M, N_M, [d_M,
\varphi_M]))_{M = M_0}^{M_S}$ leading to observation of $a$ and
sequen\-ces
$\lambda((S(M, N_M, [d_M, \varphi_M]))_{M = M_0}^{M_S})$ leading to
$b$ such that 
$$\displaylines{
 w^{nsd}_5(b, \lambda((S(M, N_M, [d_M, \varphi_M]))_{M = M_0}^{M_S}),
\omega)
\hcrh = w^{nsd}_5(a, (S(M, N_M, [d_M, \varphi_M]))_{M = M_0}^{M_S},
\omega).  }$$ The construction of (8.13) suggests that this is not an
unreasonable assumption; although it will be open to refutation in specific
cases by detailed analysis of the precise modes of observation of the
possible outcomes.  In section 6, it was noted that the hypothesis of this
paper differs from the version of Donald (1995), in a way designed to make
such an ``indifference assumption'' more plausible.

We are investigating a situation which, in conventional terms, has only two
outcomes $a$ and $b$.  This means that, to a high degree of accuracy,
$$\Pr(a | S(M_0, N_{M_0}, [d_{M_0}, \varphi_{M_0}]), \omega) + \Pr(b |
S(M_0, N_{M_0}, [d_{M_0}, \varphi_{M_0}]), \omega) = 1.$$

It follows from this, and the fact that $\rho(P^{\B}_a) + \rho(P^{\B}_b) = 1$,
that, if (8.16) holds, then
$$\displaylines{
 \Pr(a | S(M_0, N_{M_0}, [d_{M_0}, \varphi_{M_0}]), \omega) =
\rho(P^{\B}_a) \, \cr
\rlap{\hbox{and}}\hfill
\Pr(b | S(M_0, N_{M_0}, [d_{M_0}, \varphi_{M_0}]), \omega) =
\rho(P^{\B}_b).
\hfill \llap{(8.17)}  }$$

(8.17) is the result we have been aiming at.  It is an exact form of thesis
6.6.  It shows that under suitable circumstances, the stochastic process
proposed in the hypothesis can define ``private'' probabilities which are
equal to the expected values of appropriate projections, and it illustrates
the claim that the hypothesis is a generalization of conventional quantum
mechanics which provides a complete theory within which the probabilistic
role of such ``public'' expected values can be explained.

However, this section has also shown that the detailed analysis of neural
processing in terms of the hypothesis is far from simple.  As a result, there
are many ways in which (8.17) may break down.  Reactions to such a
breakdown might include abandoning the hypothesis, trying to modify it,
or searching for experimental verification.  And yet even if such
verification were to be found, with the sort of bizarre properties described
in section 6, it would not necessarily be anything more than a
demonstration of an ``observer-effect'' with a quite conventional
interpretation.  For example, no matter how often it has been placed in the
infernal apparatus, no cat will remember having heard the cynanide flask
in Schr\"odinger's experiment break more than once.  Even in conventional
terms, the details of neural processing are extremely complex and
unpredictable.  Subtle observer-effects, explicable in terms of sensory
limitations or attention to the biologically significant, are inevitable. 

As individuals, we rarely conduct our own personal tests of quantum
theory; certainly not in the conditions and with the repetitions required for
the collection of significant statistics.  Thus what may be most important
about the arguments for (8.17) is that they make it plausible that the
hypothesis correctly reflects the rough probabilities of everyday
experience (``X always/ usually/ sometimes/ seldom/ never happens'').  In
particular, this applies to the high probabilities needed in section 6 to
confirm thesis 6.5.

\proclaim{9.  Physics without any physical constants.}
\endproclaim

The hypothesis proposes an abstract characterization of the set of all
possible structures for an observer, together with a definition of a
stochastic process on that set, defined in terms of an a priori probability of
existence for each structure.  In this section, we shall consider the nature of
reality in the light of this proposal, and reflect on the possibility of
changing the definition of a priori probability so as to provide a formalism
for a physics in which physical constants are not fixed a priori, but are
indeterminate until observed.  It will turn out that such a formalism will
only be satisfactory if the initial conditions of the universe can be given a
simple description, but as will be discussed, this requirement does not seem
incompatible with current observational evidence.

  The idea that the physical constants we measure are observed from a
range of possibilities is not new, as can be seen from  discussion and
references in Barrow and Tipler (1986 -- in particular,  \sect 4.6), but the
formalism proposed here gives a detailed technical framework for the idea. 
It has always been the case that quantum theory has suggested ways in
which our beliefs about the nature of reality might be altered.  A central
purpose throughout the present work has been the investigation of the
possibility, at the level of serious theoretical physics, of some of the more
radical of these suggestions.

The fundamental set of entities considered in this paper is the set of
minimal switching structures.  Denote this set by $\SM$.  Any possible
observer at any moment corresponds to a unique element $s = S(M, N, [d,
\varphi]) \in \SM$, and G7 defines the a priori probability
$\app{}{s}{\omega}$ of existence of that element. 

			Any future which can be experienced is the possible future of some
observer, so if we know how to compute the probabilities of all possible
futures of all possible observers, then, at least as far as prediction is
concerned, we have a complete physical theory.  In the present theory,  
conventional physics is certainly involved in the definition of the a priori
probabilities.  The definition of $\app{}{s}{\omega}$ is in terms of the
``universal state'' $\omega$ and fundamental physical quantities such as
Lorentz transformations and local von Neumann algebras.  These are
defined by, or in terms of, some postulated underlying universal quantum
field theory.  If that field theory changes then so will the proposed a priori
probabilities.  For example, the sets of sequences of quantum states which
satisfy the definitions in parts E and F of the hypothesis, will change with
changes in the action of time-translations on those states.  

In this section, the dependence of a priori probability on the underlying
field theory will be made explicit in the notation by writing
$\F$ for a possible field theory and replacing  $\app{}{s}{\omega}$ with
$\app{}{s}{\F, \omega}$.  For the sake of discussion, it will be assumed in
this section that $\F$ is some sort of gauge theory;  something like the
standard model or a grand unified theory, with whatever scalar fields are
required to drive cosmological inflation.  In order to discuss possible
constraints on
$\omega$, cosmological arguments involving curved spacetimes will be
invoked.  Of course this is inconsistent with the assumption, in the
hypothesis, of Minkowski spacetime, but this inconsistency will be ignored
here.  There is no doubt that the present hypothesis is incompatible with
general relativity.  Nevertheless, allowing observer-dependent field theory,
and therefore observer-dependent time propagation may be a first step
towards the possibility of observer-dependent spacetime geometry.  I
have considered ways of modifying the hypothesis in order to allow for
such a possibility, but, for the moment, this is unfinished speculation.  I
expect that we shall have to be able to encompass some such possibility
before we can make sense of the ultimate true theory of quantum gravity. 
Leaving this ambition aside, the framework of this section suggests that,
even in the ultimate theory, it may not be necessary for physical constants
to be determined.

	According to the present theory, everything we believe about the world,
we have deduced from the pattern of information represented in our
unique physical structure $s$.  That pattern contains our entire neural
history.  Everything we have taken in from what we have heard or read
exists within that history, and the meaning which we see in it tells us about
our ``reality''  -- the ``world'' which we ``observe''.  In that world, we see
physical events and repetitions of such events, and we see our colleagues
seeing and observing and reporting on physical events.  From all this
information, it becomes possible for us to estimate probabilities for
individual events -- there will be rain before tomorrow, the Tories will not
win the next election -- and to construct and learn about physical theories
which are compatible with some or most or ultimately all of what we see. 
Nevertheless, according to the theory presented here, the ``world'' is
essentially an illusion; a mental construction.

From the very beginning of quantum theory, the conventional view of
reality has been called into question.  It was learnt that a particle cannot
be seen to have, at the same time, both an exact position and an exact
momentum.  The idea arose that the properties of a particle might depend
on the method by which that particle was observed.  Everett's formalism
went further in denying the conventional view.  It suggested, in the first
place, that, although there is an observer-independent physical reality
subject to observer-indepen\-dent physical laws, that reality was not a
``world'' but a much more complex and abstract structure.  

To see this, recall Everett's original many-worlds argument (DeWitt and
Graham (1973) pp 65--68) in its simplest form (cf.~(7.1)).  Everett
imagines a universe consisting of an observer with wave function $\psi$
observing a system with wave function $\varphi$.  The total wave function
of the universe is then a sum of tensor products of $\psi$'s and
$\varphi$'s.  If, at the beginning of a measurement, with the observer in
some fixed initial wave function $\psi$, the system is in an eigenstate
$\varphi_a$ (respectively $\varphi_b$) of the operator being measured,
then the observer at the end will be in some definite corresponding wave
function $\psi_a$ (resp. $\psi_b$) and the final total wave function will be
$\psi_a \otimes \varphi_a$ (resp. $\psi_b \otimes \varphi_b$).  On the
other hand, simply by the linearity of the Schr\"odinger equation, if the
initial wave function of the system is a superposition $\lambda\varphi_a +
\mu\varphi_b$, then the final total wave function must be     
 $$\Psi = \lambda \psi_a \otimes \varphi_a + \mu \psi_b \otimes
\varphi_b.  	    \eqno(9.1)$$ 
Such a superposition does not represent a unique fixed ``world'' outside the
observer, but two ``worlds'' ($\varphi_a$ and $\varphi_b$) each relative to
possible information gained by the observer.  What is left of
observer-independent ``physical reality'' in this picture is, firstly, the
global Hamiltonian which provides the time-dependence for $\Psi$ -- this
corresponds in the present theory to a choice of global quantum field theory
$\F$.  Secondly, physical reality determines the probability of the observer
seeing ``world'' $\varphi_a$ given that he has seen the experiment set up
(this probability is essentially $|\lambda|^2$).  Finally, the determination of
possible physical structures for observers; in other words, the
determination that $\Psi$ should be split as in equation (9.1), and not by
any other possible splitting, is also part of observer-independent physical
reality.  This aspect was not considered by Everett but is fundamental in
the present theory.

  Once we have begun on this path, it becomes possible to go further,
without needing to deny that our observations are shaped by physical
laws.  One further step has already been taken in the current hypothesis
with the proposal that the possible physical structure of an observer
corresponds at an instant not to a unique wave-function but to a set of
quantum states.  Crucial in taking this step is the idea of defining
probability by taking a supremum over possibilities between which the
observer cannot distinguish.  This idea pervades the definitions of
probability in the present theory, not only in G5 and G7, but also, and most
fundamentally, in G2c, in the very definition of the a priori probability
function.

  Suppose that we accept that the set $SM$ does characterize possible
observers and that our lives are characterized by elements $s \in SM$. 
Suppose also that the laws of physics are given by something like some
sort of gauge quantum field theory; by some $\F$.  The physical constants
which we deduce for such a theory (which we shall take to be part of the
definition of $\F$), and the probabilities that it predicts by  G7 and  G8 are
mutually dependent.   Leaving aside the ambitions of string theorists, there
appear to be many different plausible quantum field theories $\F$ which
would allow $s$ to exist.  For example, it seems possible that the fine
structure constant might be exactly ${1\over 137.03601}$ or that it might
be exactly ${1\over 137.03602}$.  The conventional assumption is that,
whether or not we can discover it, there is a true, absolute, real, and
unique value for the fine structure constant (or for a corresponding free
parameter in the ultimate theory of everything).  We estimate that value by
assuming that our world is reasonably probable.   Ultimately, we attach
most credence to the estimate which is {\it most likely }given all the
available evidence.  However, a revision to the hypothesis will show that
this conventional assumption also can be denied.  The question of whether
the fine structure constant is a rational number, for example, simply may
not have an answer.  We may allow merely that our a priori probabilities
are determined by a class of quantum field theories.   In this case,
$\app{}{s}{\F, \omega}$ is not the absolute a priori probability of the
observer defined by $s$ but instead gives the a priori probability for that
observer to observe the universe with the particular set of physical
constants given by $\F$.  

In order to make the revision, suppose that we can identify some suitable
class of quantum field theories; specified, for example, merely by choice of
gauge group or set of possible gauge groups, but with the constants, and
even perhaps with the number of elementary fields, left free.  Suppose also
that for each element $\F$ of the class, we can choose one or more
appropriate ``universal states'' $\omega$.  Denote the set of such pairs $(\F,
\omega)$ by $\V$.  In the hypothesis, $\A(\Lambda)$, $\tau_{(x,L)}$, ${\cal
C}(W)$, $\B(W)$,
${\cal N}(W, E)$, and ${\cal N}(W)$ will all depend on $\F$ and hence so will
$GSO(M, N, d, \varphi)$ and $\app{}{W}{\omega}$, which we shall write as 
$GSO_{\F}(M, N, d,
\varphi)$ and $\app{}{W}{\F, \omega}$.  Now replace G7 and G8 by
\medskip

\noindent G7$'$)	Define the a priori probability for the minimal switching
structure \newline $S(M, N, [d, \varphi])$,  given the field theory $\F$ and
the universal state $\omega$, to be
$$\displaylines{
 \app{}{S(M, N, [d, \varphi])}{\F, \omega}  = \sup\{ \app{}{W}{\F, \omega} :
 W \in GSO_{\F}(M, N, d', \varphi')
\crh \text{where } SO(M, N, d', \varphi') \in S(M, N, [d, \varphi])\}.   }$$

\noindent G8$'$)	Define the a priori probability for the minimal switching
structure $S(M, N, [d, \varphi])$ given the class $\V$ of field theories and
states, to be
 $$\app{}{S(M, N, [d, \varphi])}{\V}  = 
\sup\{  \app{}{S(M, N, [d, \varphi])}{\F, \omega} : (\F, \omega) \in \V \}.
$$  

\noindent G9$'$) Use G8$'$ to define a classical discrete Markov process on
the space of minimal switching structures $S(M, N,  [d, \varphi])$.
$$\text{Set }\xi = \sum \{\app{}{S(M', N', [d', \varphi'])}{\V}: S(M', N',
[d', \varphi']) \in \Xi(M, N,  d, \varphi)\}.$$ 

Define the probability of moving from $S(M, N,  [d, \varphi])$ to an
immediate successor $S(M', N',  [d', \varphi'])$ 
$$\displaylines{
  \text{to be } \hfill
 \app{}{S(M', N', [d', \varphi'])}{\V}/\xi, 
\hfill \text{if } \xi \geq \app{}{S(M, N, [d, \varphi])}{\V},
\cr \text{and to be } \hcr
\app{}{S(M', N', [d', \varphi'])}{\V}/ \app{}{S(M, N, [d, \varphi])}{\V},  
\hfill \text{ if } \xi < \app{}{S(M, N, [d, \varphi])}{\V}.   }$$

Define the probability of extinction to be $0$ if  $\xi \geq \app{}{S(M, N, [d,
\varphi])}{\V}$, and to be $1 - \xi/\app{}{S(M, N, [d, \varphi])}{\V}$
otherwise.
\medskip

The essence of this modification is to add a choice of $\F$ to the definition
\newline $((\sigma_m)_{m=1}^M, W)$ of an individual manifestation of a
switching structure.  With the modification, the theory answers the
question of what the most likely values are for the coupling constants and
the particle masses, given the observer's experience, in exactly the same
way in which it answers the question of what will be the most likely future
experiences of the observer.   G8$'$ simply broadens the supremum over
observationally-indistinguishable possibilities.  In particular, this means
that there is no need to postulate a prior distribution on $\V$.

	The conceptual advantage to this approach is the potential disappearance
of arbitrary initial conditions and arbitrary parameters.  In a deterministic
theory, all the complexity of everything which will happen has to be
present from the beginning of time.  In the proposed revised theory, it is
possible to imagine that the set $\V$ might have a comparatively simple
description -- for example, that there is a gauge group and a thermal
equilibrium state -- in which case there will be very little constraint on
what could have happened, and the constraints on what can happen will
come almost entirely from what has been experienced.

Without a complete theory of quantum cosmology, it is impossible to be
precise about $\V$.  Nevertheless, $\V$ cannot be arbitrary.  G7$'$, G8$'$,
and G9$'$, and G7 and G8 represent different proposals about the
fundamental physical laws.  Each different choice of $\V$ also corresponds
to a different proposal and these proposals have empirical consequences. 
For a given plausible quantum field theory $\F$, $\V$ cannot contain all
pairs of the form $(\F, \omega)$, because, if it did, the most likely choice for
$\omega$ would be highly time-dependent; highly dependent on the
observations which have been made; and the a priori probabilities defined
by the hypothesis would not agree with those predicted from previous
observations.  This is best demonstrated by an elementary example:

\proclaim{example 9.2}  Suppose that an observation with two possible
outcomes ($a$ and $b$) has been made on many similar systems all of which
we expect to be in identical states, and that we wish to predict probabilities
for the outcomes of the next observation.  Suppose that after $T$
observations (with $T$ large) $a$ has been seen around 0.2T times and $b$
around 0.8T times.  A simple quantum mechanical model proposes that an
operator of the form $u|\varphi^a\>\<\varphi^a| +
v|\varphi^b\>\<\varphi^b|$ with $\varphi^a$ orthogonal to $\varphi^b$
and $u \ne v$ is being measured.  An elementary model for the relevant
aspects of $\omega$ would involve a set of independent Hilbert spaces
$\H_{sys^1}$, $\H_{sys^2}$, \dots $\H_{sys^T}$ external to the observer,
one for each observation, (cf.~(7.15) and example 6.7).  Then we might
expect that
$$\displaylines{
 \omega\big|_{\B(\H_{sys^1}) \otimes \B(\H_{sys^2}) \otimes \dots
\otimes \B(\H_{sys^T})} = \sum_{r^1=1}^{2}\sum_{r^2=1}^{2} \dots
\sum_{r^T=1}^{2} p_{r^1} p_{r^2}\dots p_{r^T} \hcrh 
|\varphi_{r^1}\>\<\varphi_{r^1}| \otimes |\varphi_{r^2}\>\<\varphi_{r^2}|
\otimes \dots
\otimes |\varphi_{r^T}\>\<\varphi_{r^T}| 
\qquad \qquad \llap{(9.3)}   }$$
where, for each $t$,  if $r^t = 1$ then $p_{r^t} = 0.2$
and $\varphi_{r^t}$ corresponds to $\varphi^a$ while if $r^t = 2$ then
$p_{r^t} = 0.8$ and $\varphi_{r^t}$ corresponds to $\varphi^b$, so that, in
the language of example 6.7, the right hand side of (9.3) is equal to
$(0.2|\varphi^a\>\<\varphi^a| + 0.8|\varphi^b\>\<\varphi^b|)^T$.

This is satisfactory both as a model for the first $T$ observations, and, by
extension, for subsequent prediction.  

In the framework of G7$'$ -- G9$'$, a modification of definition 6.1 would
allow us to define an $\varepsilon$-manifestation of a structure  $S(M, N,
[d, \varphi])$ as a sequence \newline $((\sigma^\varepsilon_m)^M_{m=1},
W^\varepsilon, \F^\varepsilon, \omega^\varepsilon)$ which comes
sufficiently close to attaining all the relevant suprema.  If $S(M, N, [d,
\varphi])$ is sufficiently complex that, for all sufficiently small
$\varepsilon$, $\F^\varepsilon$ is well-characterized, and if, given $\F$,
the choice of $\omega$ is highly restricted, then (9.3) may be a plausible
model for restrictions of $\omega^\varepsilon$ to previously entirely
unobserved systems, such as we might imagine sometimes observing in
astronomy.  In these circumstances, the suggestion is that, in the language
of a suitably modified form of 6.4, the predicted state on each system
$\B(\H_{sys^t})$, prior to its observation, will be close to
$\omega^\varepsilon|_{\B(\H_{sys^t})}$, and our suppositions that the
initial states of the systems are all identical and that outcome $a$ has
probability 0.2, would be inductively-confirmed cosmological hypotheses.

However, if, for given $\F$, a supremum over arbitrary $\omega$ were
allowed, then (9.2) could be replaced by
$$\omega\big|_{\B(\H_{sys^1}) \otimes \B(\H_{sys^2}) \otimes \dots
\otimes \B(\H_{sys^T})}  =  |\varphi_{r^1}\>\<\varphi_{r^1}| \otimes
|\varphi_{r^2}\>\<\varphi_{r^2}| \otimes \dots
\otimes |\varphi_{r^T}\>\<\varphi_{r^T}|. 
\eqno{(9.4)}$$
where in this case $\varphi_{r^t}$ would correspond
to $\varphi^a$ if $a$ had been the outcome of observation $t$ and
$\varphi^b$ if $b$ had been the outcome.  If, as in this elementary model,
it is possible for $\omega$ to shadow each new piece of information, then
the supremum in G8$'$ would be unity, independent of the actual
observations.  With the normalization of probabilities, G9$'$ would then
imply that the outcome of the next observation would be $a$ with
probability $0.5$ and $b$ with probability
$0.5$.  This would be absurd. \hfill $\blacksquare$
\endproclaim 

The idea of this section is to introduce a formalism which allows the field
theory $\F$ to adjust to match the observation of physical constants. 
However, each observation of a given constant will be probabilistically
constrained, in a sense precisely defined by G9$'$, by all previous
observations.  Anthropic arguments show how such constants can often
effectively be observed, and to quite surprising degrees of accuracy, by our
mere existence.  On the other hand, example 9.2 indicates that if we were to
permit $\omega$ to adjust to match the result of each new observation,
then G7$'$ -- G9$'$ would  produce probabilities in conflict with observed
statistics.  This problem can be avoided if, for each choice of $\F$, we can
define a suitable limited set of fixed possibilities for $\omega$.  However,
unless this set can be given a simple and satisfying definition, there would
be little point in introducing G7$'$ -- G9$'$ just to  avoid the arbitrariness of
an exact definition of constants.  Simple definitions which avoid the
problem raised by 9.2 will correspond to initial conditions which are by
and large purged of information.  

The most attractive proposal for $\V$ is that $\omega$ should be the (or a)
ground state of $\F$ (Tryon (1973)).  Albert (1988) made such a proposal in
the context of a many-minds interpretation but according to his theory the
proposal was without empirical consequence.  The present interpretation
provides a much more sophisticated analysis of probability, and now the
proposal is an empirical postulate about cosmology.  It is about cosmology,
because to discover $\omega$ we need to undo each observed
``wave-packet collapse''; we need to go back to the state of the universe
prior to any observation.  Although the present interpretation is based on
the idea that only the observations of each separate observer are relevant,
so that $\omega$ is only observed during the lifetime of an observer, this
does not imply that $\omega$ can be modelled by the sort of state which
we would conventionally assign to the universe at the moment of an
observer's birth. Such a conventional assignment refers only to the
``observed world'' constructed by the observer, and many of the observed
events in it, like  supernovae explosions, are represented as being in the
distant past.  To undo the ``collapse'' by which we see a supernova explode
at some particular instant in some particular galaxy, we need to go back in
our observed world to before the observed event occured.  The conclusion
of this process is that the state assigned by cosmologists to the very early
universe should be a satisfactory model for $\omega$.  

At least in inflationary cosmology, the idea that the initial state of fields
other than gravity might be some sort of vacuum is certainly taken very
seriously:   ``a non-vacuum initial state contradicts the whole spirit of the
maximally symmetric initial state of the Universe which lies at the heart of
the inflationary scenario''   (Lesgourgues, Polarski, and Starobinsky
(1997)).  Quite general maximally symmetric states might also be
considered.  For example, assuming a background spacetime with a
sufficiently large symmetry group  (e.g.~de~Sitter space -- Borchers and
Buchholz (1999)), one could consider the class of homogeneous and
isotropic thermal equilibrium states for $\F$.   It is of course not required,
in the present theory, that $\omega$ should be a pure state, nor even that it
should be ``ergodic'' in the sense of quantum statistical mechanics (Ruelle
(1969), \sect 6.3).  On the largest visible scales, the universe does seem to
be remarkably homogeneous.  Indeed, explaining this apparent
homogeneity is one of the central motivations for inflationary cosmology
(Peacock (1999), chapter 11).  

Accepting a homogeneous initial state, however, then calls for an
explanation of the inhomogeneities which are so apparent all around us. 
This can be done on a descending hierarcy of scales.  Scales larger than the
visible universe may be relevant in theories of quantum gravity or of
chaotic inflation.  On these scales, it is possible to assume, if necessary, that
our observations give us a symmetry-broken part of some homogeneous,
or otherwise simple, total state.  On the scale of the visible universe,
inflation attempts to explain the existence of galaxies as quantum vacuum
fluctuations ``frozen in'' by an early exponential expansion.  Finally, on
subgalactic scales, it is permissible to assume homogeneity even at the end
of the inflationary era.  This is because, in a quantum mechanical analysis
without collapse, for almost all plausible states at that time, the current
state of a galaxy, at a fixed radial distance from the centre, will be close to
a homogeneous mixture and will need to be broken by observation into a
specific pattern of stars, planets, life-forms, and events.  This means that
there is no advantage in not assuming that the initial state itself was
homogeneous.

The conventional historical description of the path which has lead from the
early density fluctuation which gave rise to our galaxy to our existence on
this particular planet involves many processes in which observed outcome
prediction would have been impossible under a global quantum theory. 
These include fluid dynamical processes involving fragmentation and shock
waves in unstable gas clouds, nuclear processing in turbulent stellar
interiors, and the dispersal of the resulting nuclei by supernova
explosions.  Global quantum mechanical descriptions of all these
processes would involve local decoherence, and a final state which,
locally, contains almost no trace of any initial state, with the
exception of large-scale thermodynamic parameters and the radial
mass distribution. Implicit in a conventional account is a continual
symmetry-breaking sequence of ``wave-packet collapses'' which keeps
individual atoms well-localized.  Decoherence by itself does not
break symmetry; only decoherence plus collapse or observation. In a
many-minds theory, apparent symmetry breaking can be a result of the
requirement that an observer have a specific type of structure. 
Thus we can suppose that we see the specific constellations that we
do, for the same reason that we see a particle in a bubble chamber
move in a specific direction, even if that particle has arisen in a
spherically-symmetric decay process.

When cosmologists try to calculate whether, for example, observational
evidence about the spectrum of inhomogeneities in the cosmic microwave
background is compatible with particular hypotheses about the quantum
field theory of the universe and its initial state, what they are doing can be
interpreted as attempting to constrain the choice of $\V$.  If cosmology
were to provide clear evidence that a homogeneous initial state was
unlikely to give rise to the sort of observations that we make -- for
example, if the microwave background temperature had turned out to be
strongly direction-dependent -- then it might be difficult to find a choice
of $\V$ which would be both compatible with observation and simple to
describe.  At present however, there does not seem to be any such
evidence.

Given a suitable choice of $\V$, the conceptually radical change from G7 and
G8 to G7$'$, G8$'$, and G9$'$ seems unlikely to have directly observable
consequences. This is because, under both scenarios, we describe our
present observations in terms of the quantum field theory and the initial
conditions which best describe our observations.  In my opinion, G7$'$ --
G9$'$ constitutes the simpler and more attractive theory.

In this section, the choice of $\V$ has been discussed under the assumption
of the sort of fixed and symmetrical background spacetime in which
equilibrium states can be defined.  This assumption would surely fail in a
complete quantum gravity theory.  Nevertheless, as long as cosmology can
provide us with a time in the early universe when the universal state can
be given a simple description, it will remain plausible that a simple
description may also be available for the initial states of any complete
theory.  The possibility of a simple description in the full theory may also
be indicated by the apparent low entropy of the geometry of the early
universe (Penrose (1979)).

If G7$'$, G8$'$, and G9$'$ hold, then the value of, for example, the fine
structure constant is not determined, and, indeed, the fine structure
constant does not have a value.  There is no single unique quantum field
theory $\F$ which governs the physical structure of every observer. 
Quantum field theory is observer-dependent.  In fact, even given an
observer, the supremum in G8$'$ need not be attained at a unique $\F$.   
The physical universe as we know it has disappeared.  It has become a
projection from our experiences, rather than being the arena in which our
experiences occur.  Our experiences are possibilities allowed by a set of
rules.  They still give us a localized view of ``what is really happening'', but
this view is much more parochial than we had led ourselves to believe. 

Not everything is illusory.  A given human has a switching structure $s \in 
\SM$ and the possible futures of that structure are governed by
probabilistic laws.  $\SM$ and the probabilistic laws are discovered by
trying to make sense of the history presented by $s$.  To do this, we have to
assume, as discussed in section 6, that, given the probabilistic laws, $s$ is
``reasonably typical''  as an observer of its type.  Although it is inevitable
that such an assumption cannot be precisely defined, and although it is the
nature of probability that it need not be true; nevertheless, the apparent
success and consistency of the deduced laws of physics in explaining our
observations does indicate why it would be reasonable to take such a
theory seriously.

\proclaim{10.  Conclusion.}
\endproclaim

An attempt has been made in this paper and its predecessors to present a
technically complete and consistent version of a many-minds
interpretation of special-relativistic quantum field theory.  This attempt
has revealed some of the issues which may be relevant to any such
project.  

At present quantum theory would seem to be our best handle on the
weirdness of reality.  It is almost certainly a mistake to expect that that
weirdness to be tamed with yet another search for those hidden variables
which would  return us once and for all to the classical physics of our
innocence.  But it would also be a mistake to assume that reality is so far
beyond our comprehension that all speculation is idle.  Nor should we give
up trying to understand how the world is, merely because we might
be faced with a range of different possibilities, between which we
cannot decide.  It is important to know what the options are and to
test each separate option as far towards its destruction as our
abilities allow.

  The empirical verification of an ever-widening range of aspects of
quantum theory has continued for many decades.  But a theory can be
tested by other means than direct observation.  If, for example, the
incompleteness and inconsistency of Copenhagen quantum mechanics does
not count towards its refutation, then it is hard to see what could.  Of
course we can wait indefinitely for something to turn up to explain the
problems, but then we can also wait indefinitely for an explanation for any
adverse experimental result.

This paper is about progress in a many-minds interpretation.  It
demonstrates that the many-minds idea is not empty metaphysical
speculation, by showing how the careful analysis of a detailed theoretical
program can refine an interpretation and lead to deeper questions and
wider issues.  For example, the refinement of the intrepretation through a
precise definition for individual probabilities requires the discussion of the
many probabilistic notions differentiated in section 6, and leads to the
consideration of the details of neural processing in section 8.  Another
example is the resolution of the trimming problem, referred to in section
3.  This requires the explanation, again in section 8, of the existence of a
``present moment''.  Finally, the attempt, in section 9, to extend the breadth
of the interpretation leads to contact with issues in cosmology.  Many
further issues are bound to arise, before full compatibility with a theory of
quantum gravity can be attained.  However, here the progress may go both
ways, as the development of such a theory will almost certainly depend on
the simultaneous development of a compatible interpretation of quantum
theory.

Everett's many-worlds idea has always been taken fairly seriously because
it seems to fit quite naturally into the mathematics of elementary quantum
mechanics.  The current work tests the idea.  The test has not been failed,
and progress has been made.

\bigskip

\noindent{\bf Acknowledgements } I am grateful to 
Guido Bacciagaluppi, Katherine Brading, Jeremy Butterfield,
Talal Debs, Doreen Fraser, Peter Morgan, Oliver Pooley, Simon Saunders,
and Christopher Timpson for useful conversation and comments.
\bigskip

\vfill \eject

\parindent = 0pt
\def\dent{\leavevmode\hbox to 20pt{}}

\proclaim{Appendix \hfill A Hypothesis. \hfill \hphantom{Appendix}}
\endproclaim
\bigskip
	
\dent The structure of a mind, at a given moment, can be described by a
minimal switching structure $S(M, N, [d, \varphi])$.
\bigskip
	
{\bf A}\quad   A minimal ordered switching structure $SO(M, N, d,
\varphi)$ is given by:
\medskip

A1) Two positive integers $M$ (the number of determinations of switch
status)  and $N$ (the number of switches).

A2) An $M$-component ascending docket  $d$.  (This defines the spacetime
relations between determinations.)

\dent  (A docket is a geometrical structure in spacetime defined as an
equivalence class of ordered sequences $(A_i)_{i=1}^M$ of suitable
spacetime sets.  Two such sequences $(A_i)_{i=1}^M$ and $(B_i)_{i=1}^M$
will have the same docket if they have the same spacetime, or causal,
arrangement -- in other words, if, for every pair $i$, $j$,  $B_i$ is in the past
of/spacelike to/in the future of $B_j$ exactly when $A_i$ is in the past
of/spacelike to/in the future of $A_j$ -- and if one sequence can be
continuously deformed into the other while the arrangement is essentially
unaltered.  A docket $d$ is ``ascending'' if and only if $(A_i)_{i=1}^M \in d$
and $i < j$ implies that $A_i$ is not in the strict timelike future of $A_j$.)

A3) A function   $\varphi$ from $\{1, \dots, M \}$ onto $\{ \pm 1, \dots, 
\pm N\}$.  ($|\varphi(m)| = n$ is to be interpreted as meaning that the
$m^{th}$ determination is a determination of the status of switch $n$.  The
two possible statuses of that switch are represented by the sign of
$\varphi(m)$.) 

A4) Write $|\varphi|^{-1}(n) = \{j_n(k) : k = 1, \dots, K_n\}$, where $j_n(1)
< j_n(2) < \dots < j_n(K_n)$.   (Determination number $j_n(k)$ is the
$k^{th}$ determination of the status of switch $n$.  We shall write
$j(\varphi)_n(k)$ in place of $j_n(k)$ in B to show the dependence on
$\varphi$.)  

A5) For each $n \in \{1, \dots, N\}$, $K_n \geq 4$ and there exist $k_1, k_2,
k_3, k_4$ with
$1 \leq k_1 < k_2 < k_3 < k_4 \leq K_n$ such that
$\varphi(j_n(k_1)) = -\varphi(j_n(k_2)) = \varphi(j_n(k_3)) =
-\varphi(j_n(k_4))$. (A switch must open and close at least twice if all the
constraints imposed below are to be brought into play.)
\bigskip

{\bf B}\quad Given $M$, $N$, $d$, and $\varphi$ as in A, define a minimal
ordered switching structure $SO(M', N', d', \varphi')$ to be an immediate
ordered successor of $SO(M, N, d, \varphi)$ if and only either B1 or B2
holds:

B1) $N' = N$ and $M' = M + 1$ (there is a single new determination of status
on an existing switch).  There exists an order-preserving map $f : \{1, \dots,
M\} \rightarrow \{1, \dots, M'\}$ ($m_1 < m_2 \implies f(m_1) < f(m_2)$)
such that, for
$m \leq M$, $\varphi'(f(m)) = \varphi(m)$ ($f$ preserves switch numbers
and their statuses), and there exists a sequence  $(A_i)_{i=1}^{M+ 1} \in d'$
such that
$(A_{f(i)})_{i=1}^{M} \in d$ (the geometric relations of the existing
determinations are unchanged).

B2) $N' = N +1$, $M' = M + 4$, and there exists an order-preserving map $f :
\{1, \dots, M\}$ $\rightarrow \{1, \dots, M'\}$  and a sequence 
$(A_i)_{i=1}^{M'} \in d'$  such that, for $m \leq M$, $\varphi'(f(m)) =
\varphi(m)$ and $(A_{f(i)})_{i=1}^{M} \in d$. (In this case, a new switch is
introduced.) 

B3) The minimal switching structure  $S(M, N, [d, \varphi])$ is defined by 
$$\displaylines{
 S(M, N, [d, \varphi]) = \{SO(M, N, d', \varphi') : d' \text{ is ascending and
there exist permutations} \hcrh
\text{ $\pi$ on $M$ elements and $\pi'$ on $N$ elements }
\crh \text{ such that $d' = d^\pi$, $\varphi' = \pi' \circ \varphi \circ \pi$, }
\crh \text{ and $\pi(j(\varphi')_{\pi'(n)}(k)) = j(\varphi)_n(k)$ for each $n$
and $k$.} }$$  ($d^\pi$ is the docket defined by $(A_i)_{i=1}^{M} \in d \iff
(A_{\pi(i)})_{i=1}^{M} \in d^\pi$.  B3 identifies structures which differ only
in the labels assigned to switches or to determinations.)

B4) Define a minimal switching structure $S(M', N', [d', \varphi'])$ to be an
immediate successor of  $S(M, N, [d, \varphi])$ if and only if there exist 
$SO(M, N, d, \varphi) \in S(M, N, [d, \varphi])$ and $SO(M', N', d', \varphi')
\in S(M', N', [d',
\varphi'])$ such that $SO(M', N', d', \varphi')$ is an immediate ordered
successor of
$SO(M, N, d, \varphi)$.  

\dent Let $\Xi(M, N, d, \varphi)$ denote the set of immediate successors of
$S(M, N, [d, \varphi])$.
\bigskip

{\bf C}\quad  The geometrical manifestations of $SO(M, N, d, \varphi)$ 
comprise the set \newline  $GSO(M, N, d, \varphi)$ of all sequences 
$$W = (x, \Lambda, \theta, (T_n, (t_{nk})^{K_n}_{k=1}, (t'_{nm})_{m =
m_n^i}^{m_n^f}, x^n(t), L^n(t), P_n, Q_n)^N_{n=1})$$  such that, for
$n = 1, \dots, N$,

C1)   $x \in \Lambda \subset \Minkowski$ (Minkowski space).  $\Lambda$
is a spacetime retract.  (At any moment, any switch occupies a Poincar\'e
transform of the set $\Lambda$.  The requirement that $\Lambda$ be a
spacetime retract is a weak restriction on
$\Lambda$ related to the definition of the docket $d$.)

C2)  $0 \leq t_{n 1} < t_{n2} < \dots < t_{n K_n} \leq T_n$.  We shall write
$S_n$ for $t_{n 1}$.
 (The parameter on the path of switch $n$  runs from $0$ to $T_n$.
$t_{n k}$ is the parameter time on that path of the $k^{th}$ determination
of the switch status.  The switch is only ``active'' from $S_n$ but the path is
extended back to parameter $0$ to allow comparison between switches.)

C3)  $1 \leq m_n^i \leq m_n^f$. $S_n = t'_{n m_n^i} \leq t'_{n (m_n^i+1)}
\leq \dots
\leq t'_{n m_n^f} \leq T_n$. Set $t'_{n(m_n^f + 1)} = T_n$. (For $m = m_n^i ,
\dots, m_n^f$, the state of switch $n$ is $\sigma_{m}$ from parameter time
$t'_{nm}$ until $t'_{n (m +1)})$).

C4)  The $x^n(t)$  are continuous paths in $\Minkowski$ defined for  $t \in
[0, T_n]$  and with   $x^n(0) = x$.  ($x^n(t)$ defines the translational motion
of switch $n$.)

C5)  The $L^n(t)$  are continuous paths in ${\cal L}^\uparrow_+$  (the
restricted Lorentz group) defined for  $t \in [0, T_n]$, having a right
derivative $L^n\,'(t^+)$ for $t \in [0, T_n)$, and with  $L^n(0) = 1$. ($L^n(t)$
defines the rotations and boosts of switch $n$.)

C6)  For $m =  m_n^i , \dots, m_n^f$ and
$t \in [t'_{nm}, t'_{n(m+1)}]$,
$$x^n(t) = x^n(t'_{nm}) + \int_{t'_{nm}}^t L^n(s) u^n(t'_{nm}) ds \text{
where $u^n(t'_{nm})$ is a four-vector.}$$  (This implies that
$\dsize{{dx^n\over dt}(t) = L^n(t) u^n(t'_{nm})}$.) 

C7) The $u^n(t'_{nm})$ are timelike, future directed, and $(u^n(t'_{nm}))^2 =
-1$. (It follows from C6 that  $u^n(t) = \dsize{{dx^n\over dt}(t)}$ has the
same properties and that the path $x^n$ is timelike, future directed, and
parametrized by proper time $t$.)

C8) For $m =  m_n^i , \dots, m_n^f$, $x^n(t'_{n m})$ is in the closure of the
causal future of at least $m$ members of $\{x^{n'}(t_{n'k'}) : n' = 1, \dots N,
k' = 1, \dots, K_{n'}\}$. (The ``collapses'' of the quantum state are ``caused''
by the determinations and ordered by those causes.)

C9)   Set  $\Lambda_n(t) = \{ x^n(t) + L^n(t)(y - x) :  y \in \Lambda \}$ 
for  $t \in [0, T_n]$.   Set   $A_{j_n(k)} = \Lambda_n(t_{nk})$. 

\dent Then $(A_m)^M_{m=1}$ has docket $d$.

($\Lambda_n(t)$ is the spacetime set occupied by switch $n$ at parameter
time $t$, and $A_{j_n(k)}$ is the spacetime set which it occupies at the
moment of its $k^{th}$ determination.)

C10)  For $1 \leq m', m'' \leq M$, if $y' \in A_{m'}$ and $y'' \in A_{m''}$ are
spacelike separated, then there is a spacelike path from $y'$ to $y''$ in $\{ y
\in  \Lambda_{n'}(t) : n' = 1, \dots, N, \  t \in [S_{n'}, T_{n'}] \}$.
(Individuals are spatially connected.)

C11) For any $t \in [S_n, T_n]$, the number of elements of 
$$\{n' : n' \ne n \text{ and for some } t' \in [S_{n'}, T_{n'}],\   \Lambda_n(t)
\cap
\Lambda_{n'}(t') \ne \emptyset \}$$  (the number of switches whose paths
hit $\Lambda_n(t)$) is bounded by a constant $C$ -- the ``contact
number''.  $C = 13$.

C12)  $\theta : \{1, \dots, N\} \rightarrow \{(n', k', k''): n' = 1, \dots, N, 1
\leq k' < k'' \leq K_{n'}\}$ such that writing $\theta(n) = (n', k', k'')$, we
have $n' \ne n$ if $N > 1$, and $\varphi(j_{n'}(k')) = -\varphi(j_{n'}(k''))$. 
Then, for some 
$ t' \in [S_{n'}, T_{n'}]$, $A_{j_{n}(1)}$ is neither in the strict future nor the
strict past of $\Lambda_{n'}(t')$.  (In F5, the states of switch $n$ will be
required to be similar to those of switch $\theta(n)$.)

C13)  If $k_4, k_5 \in \{1, \dots, K_n\}$ with $k_4 < k_5$ and
$\varphi(j_n(k_4)) =
\varphi(j_n(k_5))$ then there exist $k_1, k_2, k_3 \in \{1, \dots, K_n\}$
with
$k_1 < k_2 < k_3$ and $\varphi(j_n(k_1)) = -\varphi(j_n(k_2)) =
\varphi(j_n(k_3))$ such that $t_{n{k_5}} - t_{n{k_4}} \geq  {1\over
2}(t_{n{k_3}} - t_{n{k_1}})$.  (A  status on a given switch cannot be
redetermined until after the elapse of a proper time at least as large as half
the minimum cycle time.)

C14) $P_n$ and $Q_n$ are projections in $\A(\Lambda)$ with $P_n$
orthogonal to $Q_n$. 
\bigskip
	
{\bf D}  

D1)  Let  ${\cal C}(W)$  be the von Neumann algebra generated by 
$$\displaylines{
  \{\tau_{(x^n(t_{n k})-L^n(t_{n k})x, L^n(t_{n k}))}(P_n), 
\tau_{(x^n(t_{n k})-L^n(t_{n k})x, L^n(t_{n k}))}(Q_n) 
\hcrh :	  k = 1, \dots, K_n , n = 1, \dots, N\}.   }$$  (This is the algebra of
correlations of switch projections experienced by the observer.  If $(x, L)$
is a Poincar\'e transformation which acts to send the spacetime set
$\Lambda$ to $(x, L) \Lambda = \{x + Ly : y \in \Lambda\}$, then
$\tau_{(x,L)}$ is the corresponding transformation on observables, so that,
for
$A \in \A(\Lambda)$,
$\tau_{(x,L)}(A) \in \A((x,L)\Lambda)$.  Transformations on states
$\sigma$ will be defined so that $\tau_{(x,L)}(\sigma)(\tau_{(x,L)}(A)) =
\sigma(A)$, implying that if $\sigma$ is a state on $\A((x,L)\Lambda)$,
then $\tau_{(x,L)}^{-1}(\sigma)$ is a state on
$\A(\Lambda)$.)

D2)  Let  $\B(W)$  be the norm closure of the linear span of 
$$\{ A_1 C_1 + C_2 A_2 :  A_1, A_2 \in
\A(\Lambda_n(t)), C_1, C_2 \in {\cal C}(W), t \in [S_n, T_n], n= 1, \dots, 
N\}.
$$ (This is the set of all observables accessible to the observer.  Elements of
${\cal C}(W)$ are correlated with local observables along the paths of the
switches.)

D3)  A quantum state $\rho$ on $\B(\H)$ -- the set of all bounded
operators on a Hilbert space $\H$ -- is a positive linear functional of unit
norm on the set $\B(\H)$.  (Thus $\rho(A A^*) \geq 0$ for all $A \in
\B(\H)$ and $\rho(1) = 1$.  A density matrix $\rho = \sum_{n=1}^\infty r_n
|\psi_n\>\<\psi_n|$ defines a state (referred to as a ``normal'' state) on
$\B(\H)$ by $\rho(A) = \tr(\rho A)  =
\sum_{n=1}^\infty r_n \<\psi_n| A |\psi_n\>$.)  For $\B \subset \B(\H)$, a
state $\rho$ on
$\B$ is the restriction to $\B$ of some state $\rho'$ on $\B(\H)$.  (This is
written $\rho = \rho'|_\B$.)
\bigskip

\dent E defines the set of sequences of states for which $x^n(t)$ is the path
along which change of state is locally minimized.  E1 requires that the
states be such that the initial conditions $u^n(t'_{nm})$ and
$L^n\,'(t'_{nm}{}^+)$ are optimal and E2 requires that the continuation at
parameter $t$ is optimal.
\medskip

{\bf E}\quad   ${\cal N}(W, E)$ is the set of all sequences of restrictions to
$\B(W)$ of sequences of quantum states $(\sigma_m)^M_{m=1}$ which
satisfy the following requirements for each $n \in \{1, \dots, N\}$ and each
$m$ such that $m_n^i \leq m \leq m_n^f$,

E1) Set $$ \eqalign{ X^{nm} =
\{(L,& v) : L  \text{ is a $C^1$ path in ${\cal L}^\uparrow_+$  on some
interval} \cr
		&[t'_{nm}, t'_{nm} + \varepsilon) \text{ with $\varepsilon > 0$ and with
$L(t'_{nm}) = L^n(t'_{nm})$, and} \cr	 &v \text{ is a future-directed
four-vector satisfing } (v)^2   = -1\}.  \cr}$$
 For  $(L, v) \in X^{nm}$,  define   $f_{nm}(s, L, v) =
\tau^{-1}_{(y^{nm}(s, L, v), L(s))}(\sigma_m)|_{\A(\Lambda)}$ where
$$y^{nm}(s, L, v) = x^n(t'_{nm}) + \int_{t'_{nm}}^s L(s') v ds' - L(s) x.$$

Then we require that $f_{nm}(s, L^n, u^n(t'_{nm}))$ has a right derivative
at
$s = t'_{nm}$ and that  
$$\inf\{\ \limsup_{h\rightarrow0^+}||(
 f_{nm}(t'_{nm}+h, L, v) - f_{nm}(t'_{nm}, L, v))/h\ ||  : (L, v) \in X^{nm}
\}$$  is attained when $L'(t'_{nm}{}^+) = L^n\,'(t'_{nm}{}^+)$ and $v =
u^n(t'_{nm})$. 

E2)   For each  $t \in (t'_{nm}, t'_{n(m+1)})$, set 
$$ \displaylines{ X^n_t = \{L : L  \text{ is a $C^1$ path in ${\cal
L}^\uparrow_+$  on some interval $[t, t +\varepsilon)$ with}\cr 
\varepsilon > 0, \text{  and } L(t)  = L^n(t)\}. }$$ 
\dent ($X^n_t$  and $X^{nm}$ could be replaced by finite dimensional sets
defined in terms of the Lie algebra of
${\cal L}^\uparrow_+$.) 

For  $L \in X^n_t$,  define  $f_t(s, L) = \tau^{-1}_{(y^n_t (s, L),
L(s))}(\sigma_{m})|_{\A(\Lambda)}$ where $$y^n_t(s, L) = x^n(t) +
\int_t^s   L(s')u^n(t'_{nm}) ds' - L(s)x.$$  Then we require that $f_t(s, L^n)$
has a right derivative at $s = t$ and that  $$\inf\{\
\limsup_{h\rightarrow0^+} ||(f_t(t+h, L) - f_t(t, L))/h\ || : L \in X^n_t \}$$
is attained when $L'(t^+) = L^n\,'(t^+)$.
\bigskip

\dent F1 -- F4 are the formal expression for the idea that ``A switch is
something spatially localized, the quantum state of which moves between a
set of open states and a set of closed states, such that every open state
differs from every closed state by more than the maximum difference
within any pair of open states or any pair of closed states.''  F5 is a
requirement of ``homogeneity''; allowing only for gradual change between
different switches.
\medskip

{\bf F}\quad  Set
$$\sigma_{nk} = \tau^{-1}_{(x^n(t_{nk})-L^n(t_{nk})x, 
L^n(t_{nk}))}(\sigma^n(t_{nk}))|_{\A(\Lambda)} $$ where $\sigma^n(t)$ is
the state of switch $n$ at parameter time $t$, which is defined to be
$\sigma_{m^n(t)}$ for $m^n(t) = \sup\{m' \leq m_n^f: t \geq t'_{nm'}\}$.

Then ${\cal N}(W)$ is the subset of ${\cal N}(W, E)$ consisting of sequences
$(\sigma_m)^M_{m=1}$ such that,  for each $n \in \{1, \dots, N\}$ and for
$k$, $k' \in \{1, \dots, K_n\}$,
\smallskip 

F1)  $\sigma_{nk}(P_n) > {1\over 2}$ \vadjust{\kern1pt}  for
$\varphi(j_n(k)) > 0$.  (``a set of open states'') 

F2)  $\sigma_{nk}(Q_n) >  {1\over 2}$  for $\varphi(j_n(k)) < 0$.  (``a set of
closed states'') 

F3)  $|\sigma_{nk}(P_n) - \sigma_{nk'}(P_n)| >  {1\over 2}$  and 
$|\sigma_{nk}(Q_n) -
\sigma_{nk'}(Q_n)| >  {1\over 2}$ for all pairs $k$ and $k'$ such that
$\varphi(j_n(k))
\varphi(j_n(k')) < 0$.  (``every open state differs from every closed state'') 

F4)  There is no triple $(P,k,k')$ with $P \in \A(\Lambda)$ a projection and
$k$ and $k'$ satisfying $\varphi(j_n(k)) \varphi(j_n(k')) > 0$ such that
$|\sigma_{nk}(P) -
\sigma_{nk'}(P)| \geq  {1\over 2}$.  (``by more than the maximum
difference within any pair of open states or any pair of closed states.'')

F5)  If $\theta(n) = (n', k', k'')$, then there are no projections $P \in
\A(\Lambda)$ such that either $|\sigma_{n1}(P) - \sigma_{n' k'}(P)| \geq 
{1\over 2}$ or 
$|\sigma_{n2}(P) - \sigma_{n' k''}(P)| \geq  {1\over 2}$.
\bigskip

{\bf G} 

G1)  The set of manifestations of the minimal switching structure $S(M, N,
[d, \varphi])$ is $$\displaylines{
 \{((\sigma_m)^M_{m=1}, W) : W \in GSO(M, N, d',
\varphi') \text{ and } (\sigma_m)^M_{m=1} \in {\cal N}(W), \hcrh
\text{ for }  SO(M, N, d', \varphi') \in S(M, N, [d, \varphi])\}.   }$$

G2) For $\sigma$ and $\rho$ quantum states on a set of operators $\B$, a
function
$\app{\B}{\sigma}{\rho}$ which gives ``the probability, per unit trial of the
information in $\B$, of being able to mistake the state of the world on $\B$
for $\sigma$, despite the fact that it is actually $\rho$'' is defined by 
$$\app{\B}{\sigma}{\rho} = \exp\{ \ent{\B}{\sigma}{\rho} \}$$ where
$\ent{\B}{\sigma}{\rho}$ (the relative entropy of $\sigma$ with respect to
$\rho$ on $\B$) is the unique function satisfying
$$\displaylines{
 \rlap{a)}\hfill  \ent{\B(\H)}{\sigma}{\rho} = \tr(-\sigma \log \sigma +
\sigma \log \rho)  \hfill \text{for  $\sigma$  and  $\rho$  normal states on
$\B(\H)$.} }$$
$$\displaylines{
 \rlap{b)} \hfill  \ent{\B(\H)}{\sigma}{\rho} = \inf\{ F(\sigma, \rho) : F
\text{ is w* upper semicontinuous, concave, and given}
\crh \text{ by a) for $\sigma$ and $\rho$ normal} \}. }$$ (This is a natural
extension of a) to the closure of the set of density matrices in the
$w^*$-topology.) 
$$\ent{\B}{\sigma}{\rho} = \sup\{
\ent{\B(\H)}{\sigma'}{\rho'} :  \sigma'|_\B =
\sigma \text{ and } \rho'|_\B = \rho \}. \leqno{c)}$$

G3)  The a priori probability of a sequence $(\sigma_m)^M_{m=1}$ of states
on a set $\B$, given an initial state $\omega$, is defined by
$$\app{\B}{(\sigma_m)^M_{m=1}}{\omega}  =
{\textstyle\prod\limits^M_{m=1}}
\app{\B}{\sigma_m}{\sigma_{m-1}} \qquad \text{where  $\sigma_0 =
\omega$.}$$

G4)  For $m = 1, \dots, M$,  define 
		$${\cal N}^m(W) = \{(\sigma_i)^m_{i=1} : \exists (\sigma_i)^M_{i=m+1}
\text{ with } (\sigma_i)^M_{i=1} \in  {\cal N}(W)\}.$$ 

G5)   Define, by induction on $m$, the following a priori probabilities.  Start
with $$\mathop{\rm app}({\cal N}(W), \B(W), 1, \omega) = \sup\{
\app{\B(W)}\sigma\omega : \sigma \in {\cal N}^1(W)\}.$$ Then, for $1 <
m+1 \leq M$,  set 
$$\displaylines{
  \mathop{\rm app}({\cal N}(W), \B(W), m+1, \omega) 
	= \sup\{ \limsup_{n\rightarrow\infty}\, \app{\B(W)}{(\sigma^n_i
)^{m+1}_{i=1}}{\omega} : ((\sigma^n_i )^{m+1}_{i=1})_{n\geq1} \text{ is}
\hfill \cr \hfill \text{a sequence of elements of } {\cal N}^{m+1}(W) \text{
and, for }1 \leq k \leq m,  
\cr \hfill \app{\B(W)}{(\sigma^n_i )^{k}_{i=1}}{\omega} \rightarrow
\mathop{\rm app}({\cal N}(W), \B(W), k, \omega) \}.  }$$

G6)  Define the a priori probability  $\app{}{W}{\omega}$  of existence of
an individual geometric manifestation $W \in GSO(M, N, d, \varphi)$ by 
$$\app{}{W}{\omega}  =  \mathop{\rm app}({\cal N}(W), \B(W), M,
\omega).$$

G7)	Define the a priori probability for the minimal switching structure 
$S(M, N, [d, \varphi])$ to be
$$\displaylines{
 \app{}{S(M, N, [d, \varphi])}{\omega}  = \sup\{ \app{}{W}{\omega} : W \in
GSO(M, N, d', \varphi')
\hcrh \text{ where } SO(M, N, d', \varphi') \in S(M, N, [d, \varphi])\}.   }$$
(This takes account of the re-labellings allowed by B.) 

G8) Use G7 to define a classical discrete Markov process on the space of
minimal switching structures $S(M, N,  [d, \varphi])$.
$$\text{Set }\xi = \sum \{\app{}{S(M', N', [d', \varphi'])}{\omega}: S(M', N',
[d', \varphi']) \in
\Xi(M, N,  d, \varphi)\}.$$ 

\dent Define the probability of moving from $S(M, N,  [d, \varphi])$ to an
immediate successor $S(M', N',  [d', \varphi'])$ 
$$\displaylines{
  \text{to be } \hfill
 \app{}{S(M', N', [d', \varphi'])}{\omega}/\xi, 
\hfill \text{if } \xi \geq \app{}{S(M, N, [d, \varphi])}{\omega},
\cr \text{and to be } \hcr
\app{}{S(M', N', [d', \varphi'])}{\omega}/ \app{}{S(M, N, [d,
\varphi])}{\omega},  
\hfill \text{ if } \xi < \app{}{S(M, N, [d, \varphi])}{\omega}.   }$$

\dent Define the probability of extinction to be $0$ if $\xi \geq \app{}{S(M,
N, [d, \varphi])}{\omega}$, and to be $1 - \xi/\app{}{S(M, N, [d,
\varphi])}{\omega}$ otherwise.
 
\bigskip

\vfill \eject

\proclaim{References.}
\endproclaim

\frenchspacing
\parindent=0pt

\everypar={\hangindent=0.75cm \hangafter=1} 

Albert, D.Z. (1988)  ``On the possibility that the present quantum state of
the universe is the vacuum.''   pp 127--133 of {\sl Proceedings of the 1988
Biennial Meeting of the Philosophy of Science Association, Vol. 2,} ed. A.
Fine and J. Leplin. (P.S.A.).

Bacciagaluppi, G., Donald, M.J., Vermaas, P.E. (1995)
 ``Continuity and discontinuity of definite properties in the modal
interpretation.'' {\sl Helv. Phys. Acta \bf 68}, 679--704. 

Barbour, J.B. (1994) ``The timelessness of quantum gravity: I, II.'' {\sl
Class. Quant. Grav. \bf 11},  2853--2897.

Barrow, J.D. and Tipler, F.J.  (1986) {\sl The Anthropic Cosmological 	
Principle.} (Oxford)

Bell, J.S. (1975)  ``On wave packet reduction in the Coleman-Hepp model.'' 
{\sl Helv. Phys. Acta \bf 48}, 93--98.  Reprinted, pp 45--51 of Bell (1987).

Bell, J.S. (1976)  ``The measurement theory of Everett and de Broglie's
pilot wave.''   pp 11--17 of {\sl Quantum Mechanics, Determinism,
Causality, and Particles,} ed. M. Flato et al.~(Reidel).  Reprinted,
pp 93--99 of Bell (1987).

Bell, J.S. (1981)  ``Quantum mechanics for cosmologists.''  pp 611--637 of
{\sl Quantum Gravity 2,} ed. C. Isham et al.~(Oxford).  Reprinted, pp
117--138 of Bell (1987).

Bell, J.S. (1987)  {\sl Speakable and Unspeakable in Quantum Mechanics.}
(Cambridge)

Borchers, H.-J. and Buchholz, D. (1999)  ``Global properties of vacuum
states in de Sitter space.'' {\sl Ann. Inst. H. Poincar\'e, Phys. Th. \bf 70}, 
23--40.   {\sl gr-qc/9803036}

Butterfield, J. (1996) ``Whither the minds?'' {\sl Brit. J. Phil. Sci. \bf 47}, 
200--221.

DeWitt, B.S. and Graham, N. (1973)  {\sl The Many-Worlds Interpretation of
Quantum  Mechanics.}  (Princeton) 

Donald, M.J. (1986)  ``On the relative entropy.''  {\sl Commun. Math. Phys.
}{\bf 105},  13--34.

Donald, M.J. (1990) ``Quantum theory and the brain.''  {\sl Proc.
R. Soc. Lond. \bf A 427},  43--93.

Donald, M.J. (1992) ``A priori probability and localized observers.'' {\sl 
Foundations of Physics, \bf 22}, 1111--1172.

Donald, M.J. (1995) ``A mathematical characterization of the physical
structure of observers.'' {\sl  Foundations of Physics \bf 25},
529--571.

Donald, M.J. (1997)  ``On many-minds interpretations of quantum
theory.'' \newline {\sl quant-ph/9703008}.

Dowker, F. and Kent, A. (1996) ``On the consistent histories approach
to quantum mechanics.'' {\sl   J. Stat. Phys.
\bf 82},  1575-1646.  {\sl gr-qc/9412067}.

Giulini, D., Joos, E., Kiefer, C., Kupsch, J., Stamatescu, I.-O., and Zeh, H.D.
(1996) {\sl Decoherence and the Appearance of a Classical World in
Quantum Theory.}  (Springer)

Griffiths, R.B. (1998) ``Choice of consistent family, and quantum
incompatibility.'' {\sl Phys. Rev. \bf A 57}, 1604.
 {\sl quant-ph/9708028}.

Haag, R. (1992)  {\sl Local Quantum Physics.} (Springer)

Hartle, J.B. (1968) ``Quantum mechanics of individual systems.'' 
{\sl Amer. J. Phys. \bf 36}, 704--712.

Hepp, K.  (1972)  ``Quantum theory of measurement and macroscopic
observables.''  {\sl Helv. Phys. Acta \bf 45}, 237--248.

Lesgourgues, J., Polarski, D., Starobinsky, A.A. (1997)
 ``Quantum-to-classical transition of cosmological perturbations for
non-vacuum initial states.'' {\sl Nucl. Phys. \bf B497}, 479--508. 
{\sl gr-qc/9611019}.

Lockwood et al.~(1996) Symposium: ``Many-minds interpretations of
quantum mechanics.''  {\sl Brit. J. Phil. Sci. \bf 47}, 159--248 and {\sl Brit.
J. Phil. Sci. \bf 47}, 445--461.

Mermin, N.D. (1998) ``What is quantum mechanics trying to tell us?'' 
{\sl Amer. J. Phys. \bf 66}, 753--767. {\sl
quant-ph/9801057}.

Namiki, M., Pascazio, S., and Nakazato, H. (1997) {\sl Decoherence and
Quantum Measurements.}  (World Scientific)

Papineau, D. (1995) ``Probabilities and the many minds interpretation of
quantum mechanics.'' {\sl   Analysis, \bf 55,}  239--246.

Peacock, J.A. (1999) {\sl Cosmological Physics.}  (Cambridge)

Penrose, R. (1979)  ``Singularities and time-asymmetry.''   pp
581--638 of {\sl General Relativity: An Einstein Centenary Survey,}
ed. S.W. Hawking and W. Israel. (Cambridge).

Ruelle, D.  (1969) {\sl Statistical Mechanics.} (Benjamin)

Saunders, S. (1996) ``Reply to Michael Lockwood.'' {\sl Brit. J.
Phil. Sci. \bf 47},   241--248.

Tryon, E.P. (1973)  ``Is the universe a vacuum fluctuation?''  {\sl Nature
}{\bf 246},  396--397.

Wheeler, J.A. (1957)  ``Assessment of Everett's `relative state' formulation
of quantum theory.''  {\sl Rev. Mod. Phys. \bf 29}, 463--465.  Reprinted, pp
151--153 of DeWitt and Graham (1973).

\end